\documentclass[review]{elsarticle}

\usepackage[utf8]{inputenc} % for accents
\usepackage{hyperref}
\usepackage[mathlines,running]{lineno}
\usepackage{amsmath,amssymb,amsfonts}
\usepackage{siunitx}
\usepackage{mathpazo,microtype,nicefrac,graphicx}
\usepackage{bm}
\usepackage{mathtools}
\modulolinenumbers[5]
\usepackage[margin=1in]{geometry}
\usepackage[toc,page,title]{appendix}

\journal{Journal of Computational Physics}
\begin{document}

\begin{frontmatter}

\title{Simulation of crumpled sheets via alternating quasistatic and dynamic representations}

\author[penn,harvard]{Jovana Andrejevic}

\author[wisc,lbl,harvard]{Chris H. Rycroft\corref{corrauthor}}

\cortext[corrauthor]{Corresponding author at: Department of Mathematics, University of Wisconsin--Madison, Madison, WI 53711, United States}
\ead{chr@seas.harvard.edu}

\address[penn]{Department of Physics, University of Pennsylvania, Philadelphia, PA 19104, United States}
\address[harvard]{John A.~Paulson School of Engineering and Applied Sciences, Harvard University, Cambridge, MA 02138, United States}
\address[wisc]{Department of Mathematics, University of Wisconsin--Madison, Madison, WI 53711, United States}
\address[lbl]{Computational Research Division, Lawrence Berkeley Laboratory, Berkeley, CA 94720, United States}

\begin{abstract}
In this work, we present a method for simulating the large-scale deformation and crumpling of thin, elastoplastic sheets. Motivated by the physical behavior of thin sheets during crumpling, two different formulations of the governing equations of motion are used: (1) a quasistatic formulation that effectively describes smooth deformations, and (2) a fully dynamic formulation that captures large changes in the sheet's velocity. The former is a differential-algebraic system of equations integrated implicitly in time, while the latter is a set of ordinary differential equations (ODEs) integrated explicitly. We adopt a hybrid integration scheme to adaptively alternate between the quasistatic and dynamic representations as appropriate. We demonstrate the capacity of this method to effectively simulate a variety of crumpling phenomena. Finally, we show that statistical properties, notably the accumulation of creases under repeated loading, as well as the area distribution of facets, are consistent with experimental observations.
\end{abstract}


\end{frontmatter}

%\linenumbers

\section{Introduction}
Crumpled structures are often regarded as exemplars of complexity in soft matter. Crumpling is found both in nature, from geological deformations~\cite{beloussov1961origin} to the packing of genetic material~\cite{gomes2008geometric,de2016scaling}, and in scientific applications: Crumpled graphene, for instance, has been used to develop high-performance biosensors~\cite{hwang2020ultrasensitive} and electrodes for batteries and supercapacitors~\cite{song2016advanced,zang2013multifunctionality,wen2012crumpled} by leveraging the increased functional surface area of these porous structures. However, crumpling often proceeds in an unpredictable manner: As a thin sheet is confined, stresses spontaneously localize to produce a complex network of creases in the sheet~\cite{wood2002witten,witten2007stress}. To elucidate the mechanisms underlying this complex process, crumpled structures have been investigated at different length scales, from the energy and geometry of individual ridges~\cite{lobkovsky1995scaling,lobkovsky1996boundary,lobkovsky1997properties,croll2019compressive} and vertices~\cite{liang2006spontaneous,cerda1998conical}, to the mechanical response of single folds~\cite{thiria2011relaxation,lechenault2014mechanical,jules2019local,farain2020relaxation}, to the collective properties of disordered ridge networks, such as slow relaxation~\cite{matan2002crumpling,amir2012relaxations,albuquerque2002stress,balankin2015mechanical,balankin2011slow} and non-monotonic aging~\cite{lahini2017nonmonotonic}. However, more remains to be understood about these rich phenomena.

Complementary to these efforts, the work of Gottesman \textit{et al.}~\cite{gottesman2018state} revealed surprising mathematical order hidden behind the disordered crumpling process. Through repeated, uniaxial crumpling of Mylar sheets in a cylindrical setup, the authors found that the accumulated crease length evolves logarithmically in the number of crumpling cycles. In a subsequent study, a physical explanation for this robust logarithmic scaling was proposed by relating crumpling to a fragmentation process that dictates the subdivision of a sheet’s surface into smaller facets over time~\cite{andrejevic2021model}. The statistics of facet area in each crease pattern were consistent with an area-conserving fragmentation rate equation used previously to describe fracture phenomena~\cite{cheng1990kinetics}. However, a generalization of these results for varying materials, thicknesses, and compaction protocols has not yet been established.

In addition to the statistical characterization of damage networks, predicting their local, spatial evolution over time is also a compelling but challenging question, and it echoes a broader interest in how materials gradually accumulate damage and fail. Understanding structural relationships between creases and predicting crease nucleation ``hot spots'' are complex questions that currently lack adequate theory. However, data-driven methods have been successfully applied to recognize structure in high-dimensional data. Using simulation-augmented machine learning, Hoffmann \textit{et al.}~\cite{hoffmann2019machine} found promising evidence that machine learning models can indeed learn underlying geometric patterns of crease networks. The key to successfully applying machine learning methods in this context was using a combination of high quality experimental crease patterns, which are limited in quantity, alongside computer-simulated flat-folding patterns that are simple to generate in large volumes, during the training process. Nevertheless, the lack of adequate experimental data was a critical barrier to achieving higher accuracy in the predictions.

New data-driven studies will require mitigating the limitations of data acquisition. This motivates our development of an efficient computational model for the crumpling of elastoplastic sheets that can further augment data-driven studies. Accurate simulation of crumpling can enable generalization of the logarithmic scaling and facet area statistics across different material properties and compaction protocols, towards a comprehensive understanding of crumpling dynamics. Furthermore, computational models can offer insight to internal properties such as spatial energy variation that are not experimentally accessible.

In this work, we detail the development and implementation of an efficient computational model of elastoplastic sheets towards this aim. As a starting point, we adopt the flexible membrane model of Seung and Nelson~\cite{seung1988defects} to handle in-plane stretching and out-of-plane bending rigidity. This model has been applied in a wide range of relevant studies, from understanding the energy of stretching ridges~\cite{lobkovsky1995scaling}, to the scaling of strain energy in crumpled sheets~\cite{kramer1997stress}, to the effects of self-avoidance during crumpling~\cite{vliegenthart2006forced}. With reference to Wardetzky \textit{et al.}~\cite{wardetzky2007discrete}, we extend the flexible membrane model, detailed for a hexagonal lattice, to handle more general mesh topologies and sheet geometries. We introduce a simple model of plasticity that enters into the bending rigidity similar to a non-zero rest angle in the discrete shells model of Grinspun \textit{et al.}~\cite{grinspun2003discrete}. Plastic deformation accrues proportionally to the elastic part of a dimensionless curvature in excess of a specified yield curvature. A distinct but related approach has been employed previously by Narain \textit{et al.}~\cite{narain2013folding}. We note that other physically-based models have also been previously developed towards understanding the role of plasticity in crumpling, such as by Tallinen \textit{et al.}~\cite{tallinen2009discrete,tallinen2009effect}. Further approaches to modeling plasticity include the dynamic creation of singular points on a piecewise continuous developable surface, which is well-suited for interactive animation~\cite{schreck2015nonsmooth}.

Next, we discuss our numerical approach for simulating crumpling. Numerical methods for the efficient simulation of folding, wrinkling, and crumpling have extensive prior development particularly in the context of cloth simulation. Implicit methods for time integration typically provide greater stability and efficiency over explicit methods due to the stiff nature of the equations of motion for mesh-based models of deformable materials like cloth and paper~\cite{baraff1998large,grinspun2003discrete}. Mixed implicit--explicit methods have also been introduced for handling different subsets of forces within a single integration step~\cite{bridson2005simulation}. A key aspect of our numerical procedure is to adaptively alternate between quasistatic and fully dynamic formulations of the governing equations of motion, integrated implicitly and explicitly, respectively. This is directly motivated by the physical behavior of sheets during crumpling: A thin sheet under confinement largely deforms smoothly such that the local acceleration may be approximated as zero. We thus treat the sheet as approximately quasistatic during smooth deformations and formulate the equations of motion as a differential-algebraic system of equations (DAE), solved implicitly. Applying the quasistatic approximation where valid significantly accelerates integration. However, thin sheets can exhibit intermittent, local buckling events and sudden self-contact that produce large changes in the local velocity, inconsistent with the quasistatic assumption. When the quasistatic approximation breaks down, we switch to a fully differential set of equations, solved explicitly.

In addition to providing qualitatively physical results and motivating examples for more detailed study, we likewise show evidence that statistical properties such as the logarithmic scaling of crease length and distribution of facet area are quantitatively consistent with experimental findings. The following sections first introduce the models for elastic, dissipative, self-contact, and plastic forces that govern the behavior of thin, elastoplastic sheets, then detail the numerical methods used, and finally conclude with selected results.

\section{Methods}
We consider a mass-spring model of thin sheets by discretizing a two-dimensional surface into a triangular mesh of $n$ discrete masses (nodes) joined to their neighbors by springs. The general equations of motion for a node $i$ of mass $m$ are given by
\begin{linenomath}\begin{equation}
\begin{aligned}
    {\dot{x}}_i &= {v}_i \\
    m{\dot{v}}_i &= {F}_i,
\end{aligned}
\end{equation}\end{linenomath}
with ${x}_i$ the three-dimensional position and ${v}_i$ the three-dimensional velocity degrees of freedom. The net force ${F}_i$ is proportional to the sum of gradients of in-plane stretching and out-of-plane bending energies $E_s$ and $E_b$, respectively, as well as isotropic drag ${F}^{\text{iso}}_d$ proportional to node velocity, internal damping ${F}^{\text{int}}_d$ modeled as a dashpot in parallel with each spring, contact forces ${F}_c$ due to self-avoidance, and possibly applied external forces ${F}_a$:
\begin{align}
    {F}_i &= - \sum_{j \in N_{i}} \left(\nabla_{{x}_i} E_s\left(x_i,x_j\right)\right)^T - \sum_{\substack{(i,j,k) \in T,\\(i,k,l) \in T}} \left(\nabla_{{x}_i} E_b \left(\hat{{n}}_{ijk},\hat{{n}}_{ikl}\right)\right)^T \nonumber \\
    &+ {F}_d^{\text{iso}}\left({v}_i\right) + \sum_{j \in N_i}{F}_d^{\text{int}}\left(x_i,x_j\right) + \sum_{j\in{C_i}}{F}_c\left({x}_i,x_j\right) + {F}_a\left({x}_i\right).
\end{align}
In the notation above, $N_i$ denotes the set of nodes joined by springs to node $i$, $T$ the set of all triangles in the mesh, with triangle unit normal vectors $\hat{n}_{ijk}$, and $C_i$ the set of all nodes experiencing a pairwise contact force with node $i$. We follow the convention of Tamstorf \textit{et al.} such that the gradient of a scalar produces a row vector~\cite{tamstorf2013discrete}. In addition, throughout this work, the circumflex symbol is used to denote a vector of unit length; i.e. $\hat{u} = u/\|u\|$. In this section, we outline the discrete formulation of the forces in our model.
\subsection{Elastic model}
\begin{figure}[ht!]
\centering
\includegraphics[width=6in]{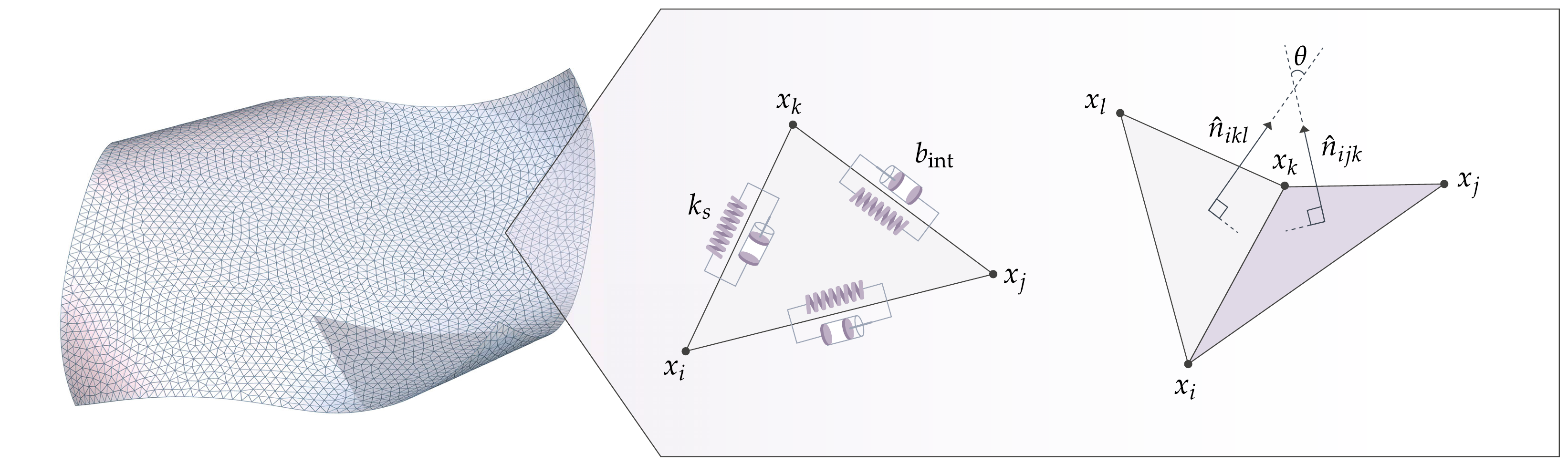}
\caption{\textbf{Microscopic model.} Microscopic model of in-plane springs, internal damping, and bending used to model thin sheet dynamics. In-plane stretching is governed by linear springs joining nearest-neighbor nodes, in parallel with dashpots that dissipate in-plane vibrations. Out-of-plane bending rigidity is modeled by an energy penalty for misalignment in the normal vectors of adjacent triangles in the mesh.}
\label{fig:figure1}
\end{figure}
Discrete in-plane elastic forces that model the behavior of thin, elastic sheets are commonly developed through a correspondence between the continuum and discrete strain energy densities, and have been derived in prior works~\cite{seung1988defects,ostoja2002lattice,kot2015elastic,wang2009hybrid}. Following these results, the energy contribution due to an interaction between a pair of nodes $i$ and $j$ joined by a spring of spring constant $k_s$ and rest length $s_{ij}$ is given by
\begin{linenomath}\begin{equation}
    E_s\left(x_i,x_j\right) = \frac{1}{2} k_s \left(\|r_{ij}\|-s_{ij}\right)^2,
\end{equation}\end{linenomath}
where $r_{ij} = x_i-x_j$ is the vector from node $j$ to node $i$.

Generalizing the bending energy model of Seung and Nelson~\cite{seung1988defects} to triangles of varying symmetry, the discrete bending energy at an edge shared by two triangles $T_1$ and $T_2$ with total area $\bar{A}=A_1 +A_2$, shared edge length $\|e_0\|$, and unit normal vectors $\hat{{n}}_{1}, \hat{{n}}_{2}$ is given by
\begin{linenomath}\begin{equation}
    E_b(\hat{{n}}_1,\hat{{n}}_2) = \frac{1}{2}k_b \frac{\sqrt{3}\|{e}_0\|^2}{2\bar{A}}\|\hat{{n}}_{1} - \hat{{n}}_{2}\|^2,
\end{equation}\end{linenomath}
with bending constant $k_b$. The microscopic model of thin sheet elasticity is depicted in Fig.~\ref{fig:figure1}. A more detailed derivation of these functional forms and clarification on subtle differences from the cited works is provided in~\ref{appendix:elasticity}.

\subsection{Effective thickness}
The bending, or flexural, rigidity $\kappa$ and 3D Young's modulus $Y$ of a thin, elastic sheet are related as~\cite{landau1986theory}
\begin{linenomath}\begin{equation}
    \kappa = \frac{Yh^3}{12(1-\nu^2)},
\end{equation}\end{linenomath}
with $h$ the sheet thickness and $\nu$ the Poisson's ratio. For a 2D model of a thin sheet, the in-plane and out-of-plane elasticity can be considered as imparting an effective thickness on the sheet,
\begin{linenomath}\begin{equation}
    h_{\text{eff}} = \sqrt{\frac{12(1-\nu^2)\kappa}{Y_{2D}}},
\end{equation}\end{linenomath}
where $Y=Y_{2D}/h_{\text{eff}}$. The result for a periodic hexagonal lattice with $Y_{2D}=2k_s/\sqrt{3}$, $\nu=1/3$, and $\kappa=\sqrt{3}k_b/2$ (see~\ref{appendix:elasticity}) simplifies to~\cite{liang2006spontaneous}
\begin{linenomath}\begin{equation}
    h_{\text{eff}} = \sqrt{\frac{8k_b}{k_s}}.
\end{equation}\end{linenomath}
The effective thickness is used to set the minimum allowed distance between nodes for self-avoiding sheets, as well as to determine the degree of refinement needed for contact detection, discussed in Section~\ref{subsec:contact}.

\subsection{Damping}
Physical systems dissipate energy during motion. We introduce two types of dissipative forces: isotropic drag, which acts as an external force on each node proportional to velocity, and internal damping, which may be modeled as a dashpot in parallel to each spring (Fig.~\ref{fig:figure1}). The isotropic drag on a node $i$ with velocity $v_i$ acts as
\begin{linenomath}\begin{equation}
    {F}_d^{\text{iso}}(v_i) = -b_{\text{iso}}{v}_i,
\end{equation}\end{linenomath}
with drag constant $b_{\text{iso}}$. In Section~\ref{sec:numerical_implementation}, we demonstrate the primary role that a small amount of isotropic drag plays in ensuring the numerical stability of linear algebra calculations performed during implicit integration.

For convenience in computation, we select a form for internal damping such that the matrices of in-plane spring coefficients and internal damping coefficients are proportional, \textit{i.e.} proportional damping, which gives the force on node $i$ as
\begin{align}
  {F}_d^{\text{int}}(x_i,x_j) &= -b_{\text{int}}\frac{d}{dt}\left({r}_{ij}-s_{ij}\hat{{r}}_{ij}\right) \nonumber \\
    &= -b_{\text{int}}\left(\left(1-\frac{s_{ij}}{\|r_{ij}\|}\right)\left(v_i-v_j\right)+\frac{s_{ij}}{\|r_{ij}\|^3}\left[r_{ij}\cdot\left(v_i-v_j\right)\right]r_{ij}\right),
\end{align}
with damping constant $b_{\text{int}}$, where ${r}_{ij}={x}_i-{x}_j$ is the vector between nodes $i$ and $j$, and $s_{ij}$ is the rest length of the spring joining nodes $i$ and $j$.

\subsection{Self-contact}\label{subsec:contact}
To prevent unphysical self-intersections of the sheet, a model for contact repulsion is likewise introduced. Literature on contact detection and response is well developed particularly in the area of computer animation of fabric and cloth~\cite{moore1988collision,provot1997collision,bridson2002robust}. Scanning for contacts is typically performed either by traversing an axis-aligned bounding box hierarchy of primitives of the sheet~\cite{bridson2002robust}, or by using a background grid of the simulation domain to reduce the local search space. After testing both methods, we opted for a grid-based approach due to improved performance for our problem. In the grid-based approach, the domain spanned by the extremal values of the node positions is subdivided into blocks. Each node is then assigned to a block based on its position. During contact detection, the nodes contained in all blocks within the interaction range of a particular node are checked for possible contact. The grid is periodically recomputed throughout the course of simulation to maintain a balanced number of nodes per grid block on average.

After detection, the contact response can then be applied in two primary ways: (1) by introducing stiff repulsive forces that act over a short interaction range (\textit{penalty method}), or (2) by obtaining the precise time of contact and correcting node positions and velocities consistent with conservation of momentum (\textit{impulse method})~\cite{bridson2002robust}. Impulse methods typically assume constant velocity over the course of a single timestep for simplicity, which may be lower in order of accuracy than the integration method employed. Another important consideration is the consistency of multiple contacts: The handling of one contact may introduce further self-intersection, and multiple contacts may occur during a single step~\cite{provot1997collision}. On the other hand, penalty methods rely on stiff springs that can significantly reduce the integration step size needed to maintain the desired accuracy~\cite{moore1988collision}. Our numerical integration approach, discussed in Section~\ref{sec:numerical_implementation}, mitigates some of the challenges of the penalty method by adaptively adjusting both the integration step size and the integration method (implicit or explicit); thus, it proved more suitable to our problem over the momentum-conserving approach.

\begin{figure}[ht!]
\centering
\includegraphics[width=6.5in]{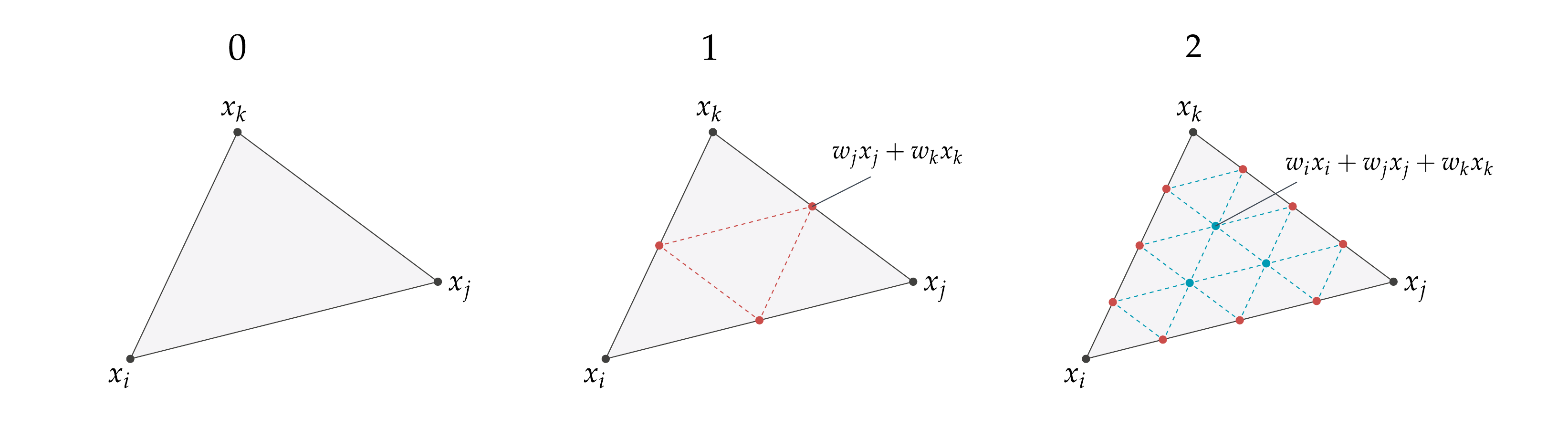}
\caption{\textbf{Contact site refinement.} Additional edge and interior triangle sites introduced for resolving contacts for the first three refinement levels. The position of each edge or interior triangle site may be expressed as a weighted sum of the triangle vertices. These weights determine the proportion of the contact penalty force applied to each vertex during contact response.}
\label{fig:figure2}
\end{figure}

Another important consideration for contact detection is the set of primitives compared. In the simplest approach, checks for pairwise contacts between nodes in the mesh are performed. A more accurate and widely-used method entails checking the proximity of all point--triangle and edge--edge pairs within the mesh, as done by Bridson \textit{et al.}~\cite{bridson2002robust}. In this approach, the closest point of contact between a point and triangle, or between two edges, is computed. If the distance to the contact point is less than a minimum allowed separation, the contact response is applied and distributed to the nodes involved in the interaction, weighted by proximity to the closest point. However, a difficulty with this approach is the presence of discontinuities in the contact force, as nodes may abruptly enter or exit the interaction domains of different primitives. Discontinuities lead to increased error estimates in adaptive step integration schemes, and they can disrupt the accurate prediction of the solution at the next step during implicit integration. To maintain continuity, we use only pairwise contact interaction forces; however, we improve the accuracy of the identified contacts by introducing a finer set of sites on the interior of each triangle and edge, and thus identify contacts with higher resolution than checks between nodes only. Pairwise contacts between the refined sites are checked using the grid-based method described, and the resulting forces remain continuous. As in the point--triangle/edge--edge approach, the contact force is distributed to the nodes nearest to the interpolated sites, weighted by proximity to the site. This refinement procedure is only applied during contact, so the number of nodes and total degrees of freedom in the system remains unchanged. Fig.~\ref{fig:figure2} illustrates the placement of additional sites on the interior of a triangle and its edges for the first three refinement levels. For two interacting sites $1$ and $2$ at positions $x_1=w_ix_i+w_jx_j+w_kx_k$ and $x_2=w_mx_m+w_nx_n+w_ox_o$, with $w_i+w_j+w_k=1$ and $w_m+w_n+w_o=1$, we use the following form for the contact force on site $1$:
\begin{linenomath}\begin{equation}
    F_1 = \begin{cases}
      k_s\frac{\sigma}{\|r_{12}\|}\left(\frac{\sigma}{\|r_{12}\|-\sigma}\right)^2\exp\left({\frac{\sigma}{\|r_{12}\|-\sigma}}\right)r_{12} &\qquad \text{for $\|r_{12}\|<\sigma$,} \\
      0 & \qquad \text{ for $\|r_{12}\| \geq \sigma$,}
    \end{cases}
\end{equation}\end{linenomath}
where $\sigma$ is the interaction range, typically set equal to the effective thickness $h_{\text{eff}}$ of the sheet, and $r_{12}=x_1-x_2$ is the vector between the two interacting sites. Then the contact force on a node $i$ due to node $m$ is given by
\begin{linenomath}\begin{equation}
    F_c(x_i,x_m) = w_iw_mF_1
\end{equation}\end{linenomath}
while the force on $i$ due to the complete interaction with site $2$ is
\begin{linenomath}\begin{equation}
    F_c(x_i,x_2) = w_i\left(w_mF_1 + w_nF_1 + w_oF_1\right)=w_iF_1.
\end{equation}\end{linenomath}

\subsection{Plasticity}
\begin{figure}[ht!]
\centering
\includegraphics[width=6in]{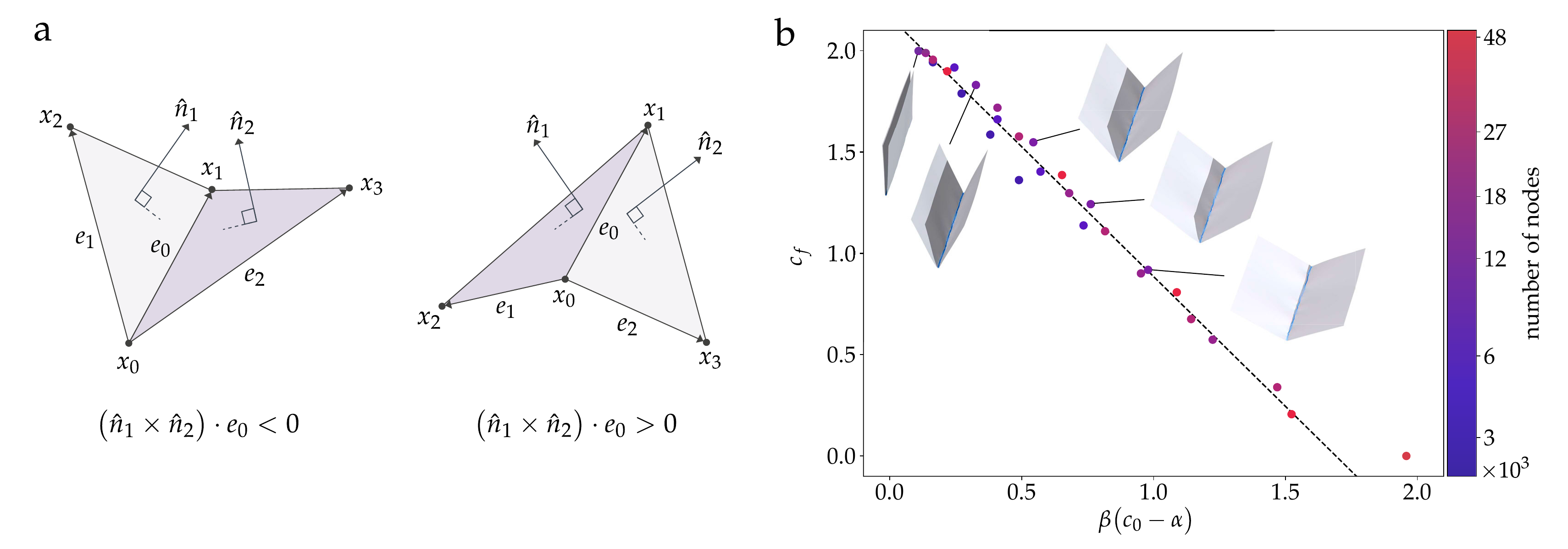}
\caption{\textbf{Curvature and plasticity.} (a) The curvature sign convention between two neighboring triangles, determined by the sign of $\left(\hat{n}_1\times\hat{n}_2\right)\cdot{e}_0$. (b) Relationship between the final fold angle and yield angle for simulations of a single fold performed at different mesh resolutions. When mapped to physical units, each sheet is \SI{10}{\cm} in length and \SI{0.25}{\mm} thick, with a Young's modulus of approximately \SI{1}{\GPa}. For a fold angle $\phi_f$ measured as the angle between the two planes of the sheet, $\theta_f = \pi-\phi_f$ represents the angle between the normals to the two planes, and $c_f = 2\sin\left(\theta_f/2\right)$. $c_0$ is the specified onset of plastic deformation, and $\alpha=0.05$, $\beta=0.04\lambda$ are parameters introduced to achieve an approximate collapse of data across different mesh sizes onto a line, with linear node density $\lambda = \sqrt{n/A}$ for a sheet of area $A$, in simulation units.}
\label{fig:figure3}
\end{figure}

Without plasticity, the bending energy contribution of two adjacent triangles $T_1$ and $T_2$ with unit normals ${\hat{n}}_{1},{\hat{n}}_{2}$ is given by
\begin{linenomath}\begin{equation}
E_b(\hat{n}_1,\hat{n}_2)=\frac{1}{2}k_b\frac{\sqrt{3}\|{e}_0\|^2}{2\bar{A}}\|{\hat{n}}_{1}-{\hat{n}}_{2}\|^2,
\end{equation}\end{linenomath}
as previously presented. This formulation assumes a flat resting state such that ${\hat{n}}_{1}$ and ${\hat{n}}_{2}$ are parallel; in equilibrium, ${\hat{n}}_{1}={\hat{n}}_{2}$. Note that $\|{\hat{n}}_{1}-{\hat{n}}_{2}\|=2\sin(\theta/2)$, where $\theta\in\left[0,\pi\right]$ is the angle between the two normals, measures the deviation from a flat state of the sheet. We refer to this quantity as the magnitude of a dimensionless curvature $c_{12}$ between triangles $T_1$ and $T_2$. To introduce an appropriate sign for $c_{12}$, following Tamstorf \textit{et al.}~\cite{tamstorf2013discrete}, we use the quantity $\left({\hat{n}}_1\times{\hat{n}}_2\right)\cdot{e}_0$ to define a sign convention, as illustrated in Fig.~\ref{fig:figure3}a.
Recognizing that ${n}_1 = {e}_0\times{e}_1$, ${n}_2 = -{e}_0\times{e}_2$ and applying a triple cross product rule, we obtain the following equivalent expression:
\begin{linenomath}\begin{equation}
\left({\hat{n}}_1\times{\hat{n}}_2\right)\cdot{e}_0 = -\frac{\left({\hat{n}}_2\cdot{e}_1\right)}{\|{n}_1\|}\|{e}_0\|^2.
\end{equation}\end{linenomath}
Thus, the same sign information is encoded in the quantity $-({\hat{n}}_2\cdot{e}_1)$. To introduce plasticity, we modify the expression for bending energy to 
\begin{linenomath}\begin{equation}
E_b(\hat{n}_1,\hat{n}_2)=\frac{1}{2}k_b\frac{\sqrt{3}\|{e}_0\|^2}{2\bar{A}} \left(-\frac{\left({\hat{n}}_2\cdot{e}_1\right)}{|{\hat{n}}_2\cdot{e}_1|}\|{\hat{n}}_{1}-{\hat{n}}_{2}\|-p_{12}\right)^2.
\end{equation}\end{linenomath}
The term $p_{12}$ is a signed quantity that assumes the role of an effective plastic deformation, and can be interpreted as having the form $2\sin(\bar{\theta}/2)$ where $\bar{\theta}$ is the rest angle between the two normals. Thus, plastic deformation in this context refers to a local deformation away from the flat state of the sheet, and a nonzero $p_{12}$ implies a nonzero rest angle. This is analogous to the more general model presented by Tamstorf \textit{et al.}~\cite{tamstorf2013discrete},
\begin{linenomath}\begin{equation}
    E_b \sim \frac{\sqrt{3}\|{e}_0\|^2}{\bar{A}}\left(\varphi(\theta)-\varphi(\bar{\theta})\right)^2,
\end{equation}\end{linenomath}
with the choice $\varphi(\theta)=2\sin(\theta/2)$. An advantage of this particular choice of $\varphi$ is that the bending energy, force, and Hessian may be expressed without the use of trigonometric functions, which are comparatively slower to compute numerically.

Next, we outline the update rule for the dimensionless rest curvature $p_{12}$. Within the integration scheme, $p_{12}$ is treated as an additional degree of freedom for each interior edge of the mesh. Let
\begin{linenomath}\begin{equation}
c_{12} = -\frac{\left({\hat{n}}_2\cdot{e}_1\right)}{|{\hat{n}}_2\cdot{e}_1|}\|{\hat{n}}_1-{\hat{n}}_2\|
\end{equation}\end{linenomath}
be the current signed dimensionless curvature; we use the following form for the time derivative of $p_{12}$ to ensure a smooth update rule:
\begin{linenomath}\begin{equation}
\frac{dp_{12}}{dt} = 
\begin{cases} 
  f(c_{12}-p_{12})  & \qquad \text{for $c_{12}-p_{12} > 0$,} \\
  -f(-c_{12}+p_{12}) & \qquad \text{for $c_{12}-p_{12} \leq 0$,} \\
   \end{cases} \\
\end{equation}\end{linenomath}
where
\begin{linenomath}\begin{equation}
    f(\lambda) = \frac{1}{1+\gamma}S\left(\frac{\lambda-c_0}{2-c_0}\right).
\end{equation}\end{linenomath}
Here, $c_0$ is a non-negative yield curvature marking the onset of plastic deformation and $\gamma$ a non-negative constant tuning the rate of damage accumulation. $S(x)$ is the smoothstep function, used to ensure continuity \cite{ebert2003texturing}, with the form
\begin{linenomath}\begin{equation}
    S(x) = 
\begin{cases} 
  0  & \qquad \text{for $x < 0$,} \\
  6x^5 - 15x^4 +10x^3 & \qquad \text{for $0 \leq x \leq 1$,}\\
  1 & \qquad \text{for $x > 1$.}
   \end{cases} \\
\end{equation}\end{linenomath}
Note that the explicit $2$ in the argument $(\lambda-c_0)/(2-c_0)$ of $S$ corresponds to the upper limit of $|c_{12}-p_{12}|$; this is the theoretical maximum for $|c_{12}-p_{12}|$ when $c_{12}$ and $p_{12}$ have the same sign, which may be noted from the equation for $c_{12}$ when ${\hat{n}}_1$ and ${\hat{n}}_2$ are exactly anti-parallel.

In order to calibrate the yield curvature for different mesh resolutions, simulations of a square, elastoplastic sheet folded in half are performed, and the final fold angle $\phi_f$ measured, for varying $c_0$ and mesh size. Fig.~\ref{fig:figure3}b shows the relationship between the final fold angle and onset of plastic deformation. In particular, we obtain a reasonable data collapse onto a single line when $c_f = 2\sin\left(\theta_f/2\right)$, with $\theta_f = \pi - \phi_f$ the angle between the normals of the folded planes, is plotted against $\beta\left(c_0-\alpha\right)$, where $\alpha=0.05$ and $\beta=0.04\lambda$ are fitting parameters, and $\lambda = \sqrt{n/A}$ is the linear node density for a sheet of area $A$ with $n$ nodes. Thus, the onset of plastic deformation may be adjusted in a predictable manner to model sheets with higher or lower propensity for plastic deformation.

To conclude this section, several quantities and parameters relevant to sheet simulation are summarized in Table~\ref{tab:params}. In the following section, we refer to the time over which the sheet is simulated as the \textit{simulation time}, and reserve \textit{computation time} for discussing the duration of program execution.

\begin{table}[h!]
  \begin{linenomath}\begin{equation}
    \begin{matrix*}[l]
    \text{symbol} & \text{parameter} & \text{dimension} \\
    \hline
    n & \text{number of nodes} & 1 \\
    m & \text{mass per node} & M \\
    h_{\text{eff}} & \text{effective thickness} & L \\
    t &  \text{simulation time} & T \\
    k_s & \text{stiffness constant} & M/T^2 \\
    k_b & \text{bending constant} & ML^2/T^2 \\
    b_{\text{iso}} & \text{isotropic drag constant} & M/T \\
    b_{\text{int}} & \text{internal damping constant} & M/T \\
    Y_{2D} & 2D\text{ Young's modulus} & M/T^2 \\
    Y & 3D\text{ Young's modulus} & M/(LT^2) \\
    \nu & \text{Poisson's ratio} & 1 \\
    \kappa & \text{bending rigidity} & M L^2/T^2 \\
    \end{matrix*}
  \end{equation}\end{linenomath}
  \caption{Quantities and parameters relevant to sheet simulation, along with their physical dimensions.\label{tab:params}}
\end{table}

\section{Numerical implementation}\label{sec:numerical_implementation}
This section details the numerical methods used to generate the mesh and integrate the equations of motion in time. In particular, we emphasize the adaptive integration methods used that are responsive to how rapidly the sheet is deforming. The custom code is implemented in C++ and multithreaded using OpenMP~\cite{dagum1998openmp}.

\subsection{Mesh topology}
To generate a mesh, a rectangular domain is partitioned into triangles by a random Delaunay triangulation generated via the Voro++ library~\cite{rycroft2009voro++,rycroft09c}. As Voro++ makes use of a Voronoi tessellation to generate its dual Delaunay triangulation, randomly triangulated sheets are first regularized using Lloyd's algorithm~\cite{lloyd1982least}, which brings the arrangement of nodes closer to a centroidal Voronoi tessellation with cells of higher uniformity in shape and size. This produces more uniform triangle shapes and sizes in the corresponding Delaunay triangulation. The random Delaunay triangulation is also used to mesh a circular sheet geometry demonstrated in later examples.

\subsection{Quasistatic formulation}
The behavior of thin, elastoplastic sheets during crumpling is characterized by slower, large scale deformations alongside localized buckling events and snap-throughs that take place on smaller time and length scales. These intermittent events are experimentally recognized, for example, by measuring the acoustic emissions during crumpling~\cite{kramer1996universal,houle1996acoustic,abobaker2015avalanche,mendes2010earthquake}. In an effort to accurately model and simulate crumpling dynamics, we address this disparity in temporal scales by implementing a hybrid integration scheme that adaptively switches between a quasistatic formulation of the equations of motion, solved implicitly, and a dynamic formulation, solved explicitly, outlined next.

In the quasistatic formulation, the equations of motion may be expressed in the semi-explicit form
\begin{linenomath}\begin{equation}
\begin{aligned}
    {\dot{x}} &= {v}, \\
    0 &= {F}.
\end{aligned}
\end{equation}\end{linenomath}
The above equations omit subscripts denoting a specific node, and use ${x}$, ${v}$, and ${F}$ to refer to the complete vector of positions, velocities, and forces, respectively, each of length $3n$. The acceleration of all nodes is taken to be approximately zero at each step, thereby implying zero net force. The combination of both differential equations and algebraic constraints results in a differential-algebraic system of equations, or DAE~\cite{ascher1998computer}. Here, node positions $x$ represent the differential degrees of freedom, and node velocities $v$ the algebraic degrees of freedom. In order to simultaneously satisfy all algebraic constraints, DAEs are typically solved using implicit methods. We use a backward differentiation formula (BDF) of order $s$ to discretize the set of differential equations as
\begin{linenomath}\begin{equation}
    0 = {x}^{(k+1)} + \sum_{j=0}^{s-1}\left(\alpha_j{x}^{(k-j)}\right) - \beta h{v}^{(k+1)}.
\end{equation}\end{linenomath}
Here, $\alpha$ and $\beta$ are coefficients that depend on the history of integration steps and order $s$ of the BDF method used, $h$ is the current integration step size, and superscripts are used to denote the discrete timestep corresponding to each set of positions and velocities. To allow for adaptive step size control, orders $s=3$ and $s=4$ are used, with step size-dependent coefficients $\alpha$ and $\beta$ as given in Refs.~\citenum{dolejvsi2008adaptive} \& \citenum{decaria2021variable}. The higher order method provides a higher accuracy reference solution from which a local error estimate of the lower order solution may be computed. Switching the sign of ${F}$ for later convenience, the set of discretized differential and algebraic equations now reads
\begin{align}
  0 &= {x}^{(k+1)} + \sum_{j=0}^{s-1}\alpha_j{x}^{(k-j)} - \beta h{v}^{(k+1)} = {r}^{(k+1)}_{\text{diff}}, \\
  0 &= -{F}^{(k+1)} = {r}^{(k+1)}_{\text{alg}}.
\end{align}
The equations are solved iteratively via Newton's method for all degrees of freedom ${q} = ({x},{v})$ of the system using
\begin{linenomath}\begin{equation}
    J({q}^{(k+1)}_j)\Delta{q}^{(k+1)}_j = -{r}({q}^{(k+1)}_j),
\end{equation}\end{linenomath}
where ${r}({q}^{(k+1)}_j) = ({r}^{(k+1)}_{\text{diff}}, {r}^{(k+1)}_{\text{alg}})$ is the residual of the system, and $J$ the Jacobian of ${r}$. The subscripts $j$ denote Newton iterations, and the above linear system is solved iteratively until the norm of the residual falls below a specified threshold. The Jacobian $J({q}) = \nabla_{{q}}{r}$ takes the form of a $2\times 2$ block matrix
\begin{linenomath}\begin{equation}
    J({q}) =
    \begin{pmatrix}
    I & -\beta hI \\
    -\nabla_{{x}}{F} & -\nabla_{{v}}{F}
    \end{pmatrix} = 
    \begin{pmatrix}
    I & -\beta hI \\
    H_{{x}} & H_{{v}}
    \end{pmatrix},
\end{equation}\end{linenomath}
where the notation $H_{{x}}=-\nabla_x{F}$ and $H_{{v}}=-\nabla_v{F}$ is used in reference to the Hessian of the sheet energy (though dissipative forces without a corresponding energy function are also included). The subscript $j$ and superscript ${(k+1)}$ have been omitted for clarity. The solution to the $2\times 2$ block system at each Newton iteration may be computed in two steps:
\begin{enumerate}
  \item \label{step1} a linear system solve for the update to the position degrees of freedom $\Delta{{x}}$,
    \begin{linenomath}\begin{equation}
      \left(H_{{x}}+\frac{1}{\beta h}H_{{v}}\right)\Delta{x} = {F} - \frac{1}{\beta{h}}H_{{v}}{r}_{\text{diff}},
    \end{equation}\end{linenomath}
  \item an explicit calculation of the update to the velocity degrees of freedom $\Delta{{v}}$,
    \begin{linenomath}\begin{equation}
      \Delta{{v}} = \frac{1}{\beta{h}}\left({r}_{\text{diff}}+\Delta{x}\right).
    \end{equation}\end{linenomath}
\end{enumerate}
Recalling our formulation of damping forces such that the damping coefficient matrix is proportional to the in-plane stiffness coefficients, the term $H_{{v}}/\beta h$ may be decomposed into
\begin{linenomath}\begin{equation}
    \frac{1}{\beta h}H_{{v}} = \frac{b_{\text{int}}}{k_s\beta h}H^{\text{in-plane}}_{{x}} + \frac{b_{\text{iso}}}{\beta h}I.
\end{equation}\end{linenomath}
We can now directly observe that isotropic drag offers numerical stability through the term $b_{\text{iso}}I/\beta h$, and the condition number of the corresponding matrix in step \ref{step1} may furthermore be tuned by adjusting the integration step size $h$.

Plasticity introduces an additional set of differential degrees of freedom corresponding to the dimensionless rest curvature of each interior edge. To simplify notation, let $g= dp/dt$; then
\begin{align}
  0 &= {x}^{(k+1)} + \sum_{j=0}^{s-1}\left(\alpha_j{x}^{(k-j)}\right) - \beta h{v}^{(k+1)} = {r}^{(k+1)}_{\text{diff,}x}, \\
  0 &= {p}^{(k+1)} + \sum_{j=0}^{s-1}\left(\alpha_j{p}^{(k-j)}\right) - \beta h{g}^{(k+1)} = {r}^{(k+1)}_{\text{diff,}p}, \\
  0 &= -{F}^{(k+1)} = {r}^{(k+1)}_{\text{alg}}.
\end{align}
Here, $p$ represents the vector of effective plastic deformation at all interior edges of the mesh. For $q=\left(x,v,p\right)$ this results in a $3\times 3$ block matrix structure of the Jacobian:
\begin{linenomath}\begin{equation}
J(q) = \begin{pmatrix}
    I & -\beta{h}I & 0 \\
    -\beta{h}\nabla_xg & 0 & I - \beta{h}\nabla_pg \\
    -\nabla_x{F} & -\nabla_v{F} & -\nabla_p{F}
    \end{pmatrix} =
    \begin{pmatrix}
    I & -\beta{h}I & 0 \\
    \beta{h}G_x & 0 & I + \beta{h}G_p \\
    H_x & H_v & H_p
    \end{pmatrix},
\end{equation}\end{linenomath}
where we have defined $G_x=-\nabla_x{g}$, $G_p=-\nabla_p{g}$, and $H_p=-\nabla_p{F}$.
The resulting problem at each Newton iteration may likewise be reduced to one linear system solve for $\Delta{x}$, followed by two explicit calculations of $\Delta{p}$ and $\Delta{v}$:
\begin{linenomath}\begin{equation}
    \begin{aligned}
        (1) & \quad \left(H_{{x}}+\frac{1}{\beta h}H_{{v}}-\beta{h}H_p\left(I+\beta{h}G_p\right)^{-1}G_x\right)\Delta{x} = {F} - \frac{1}{\beta{h}}H_{{v}}{r}_{\text{diff,}x} + H_p\left(I+\beta{h}G_p\right)^{-1}{r}_{\text{diff,}p}, \\
        (2) & \quad \Delta{{p}} = -\left(I+\beta{h}G_p\right)^{-1}\left({r}_{\text{diff,}p}+\beta{h}G_x\Delta{x}\right), \\
        (3) & \quad \Delta{{v}} = \frac{1}{\beta{h}}\left({r}_{\text{diff,}x}+\Delta{x}\right).
    \end{aligned}
\end{equation}\end{linenomath}
By our choice of plastic update rule, $G_p$ is diagonal, which simplifies intermediate calculations such as $\left(I+\beta{h}G_p\right)^{-1}.$

Sparse matrix storage is handled using the Eigen library for numerical linear algebra~\cite{eigenweb}. All entries of the Jacobian are analytically derived so that an exact Jacobian may be used, and are provided in~\ref{appendix:jac}. This eliminates the need for numerical differentiation and further improves the performance of the implicit scheme. In addition, prior to the Newton iterations, an improved guess for the solution is made using a third-order accurate Lagrange interpolant constructed from prior integration steps~\cite{zahr2013performance}. This improves convergence to approximately two Newton iterations per integration step.

\subsection{Preconditioning}
\begin{figure}[ht!]
\centering
\includegraphics[width=6in]{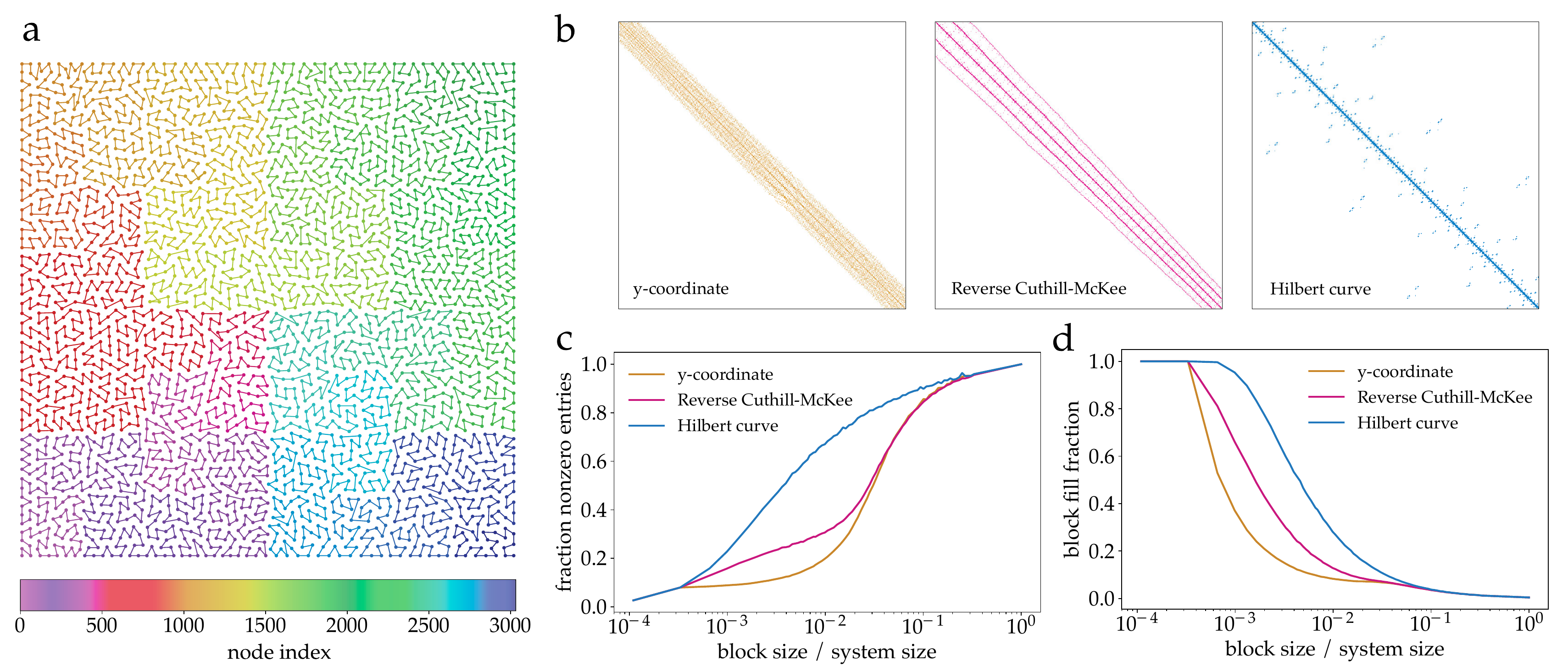}
\caption{\textbf{Node re-indexing for matrix preconditioning.} (a) Order of node indexing based on the Hilbert space-filling curve for a typical random mesh of approximately 3000 nodes. We see that this type of ordering maintains the locality of indices for spatially proximal nodes. (b) A magnified central inset of the sparsity pattern of the matrix $H_x= -\nabla_x{F}$, considering only contributions from internal stretching and bending forces, for three node ordering schemes: $y$-coordinate, in which nodes are sorted in ascending order of their $y$-coordinate, reverse Cuthill--McKee (RCM), and Hilbert curve. The sparsity patterns shown are representative of the matrix at each linear solve prior to self-contact. While the overall matrix bandwidth is smaller for the $y$-coordinate and RCM schemes, the nonzero entries cluster more densely near the diagonal for the Hilbert curve-based approach. (c) The fraction of nonzero entries considered as a function of normalized block size when a block Jacobi preconditioner is applied to solve a linear system with the three sparsity patterns above. The Hilbert curve-based reordering captures the greatest information contained in the matrix for a majority of block sizes. (d) In addition, the Hilbert curve-based reordering provides the most efficient blocks, with the highest fill fraction of the three methods.}
\label{fig:figure4}
\end{figure}
\begin{figure}[ht!]
\centering
\includegraphics[width=4.7in]{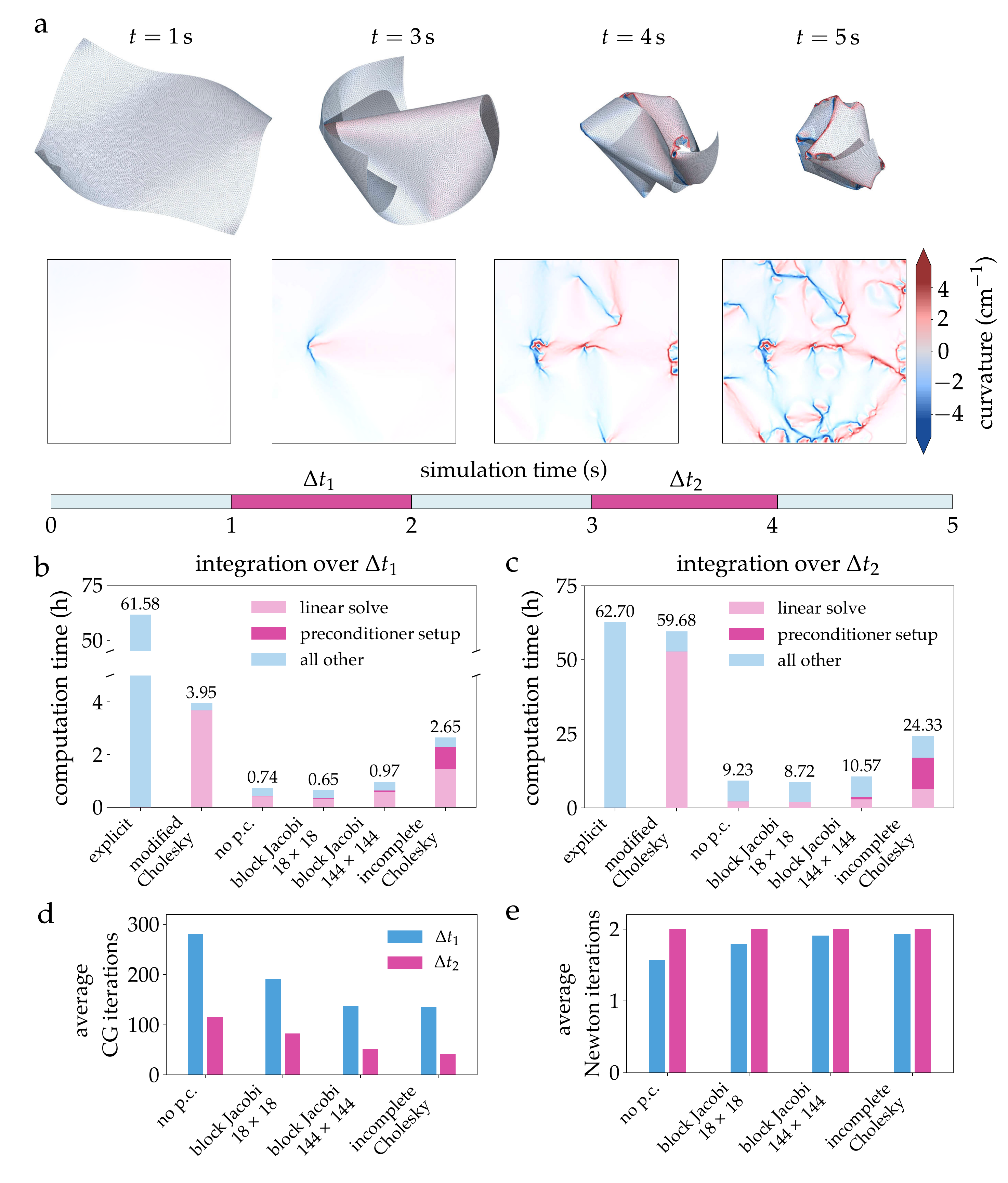}
\caption{\textbf{Integrator and preconditioner performance.} (a) Selected snapshots of a simulated sheet under radial confinement and corresponding projected crease patterns, visualized as the map of mean curvature of the sheet. The mesh contains $12000$ nodes, and simulation parameters correspond to a $\SI{10}{\cm} \times \SI{10}{\cm}$ sheet \SI{0.25}{\mm} thick with a Young's modulus of approximately \SI{1}{\GPa}. The sheet is compressed radially by applying a radially inward force to all nodes exceeding a cutoff radius from the origin. The cutoff radius decreases linearly at a rate of \SI{1}{\cm/\second} for \SI{5}{\s}. (b) Comparison of the required computation time to integrate over a selected interval $\Delta{t}_1$ from time $t=\SI{1}{\s}$ to $t=\SI{2}{\s}$ using several different integration approaches. \textit{Explicit} denotes the use of the fully dynamic formulation throughout; all others use the alternating quasistatic and dynamic approach and are distinguished by the method used for the linear solve: \textit{modified Cholesky} makes use of a direct modified Cholesky solver, \textit{no p.c.} makes use of conjugate gradient without preconditioning, \textit{block Jacobi} uses block Jacobi-preconditioned conjugate gradient with two different block sizes, $18\times{18}$ and $144\times{144}$, and Hilbert curve-based reindexing, and \textit{incomplete Cholesky} uses incomplete Cholesky-preconditioned conjugate gradient. The modified Cholesky solve, Cholesky factorization of matrix sub-blocks, and incomplete Cholesky factorization are computed via the Eigen library~\cite{eigenweb}. Each simulation is performed using $8$ threads. We see that the alternating quasistatic and dynamic approach provides considerable speedup and performs best when using conjugate gradient for the linear solve. (c) The same  comparison as (b) for a second interval $\Delta{t}_2$ from simulation time $t=\SI{3}{\s}$ to $t=\SI{4}{\s}$, during which the sheet undergoes greater self-contact. (d) The average number of conjugate gradient iterations per Newton step and (e) the average number of Newton iterations per implicit integration step for the conjugate gradient-based methods. While Newton iterations remain roughly the same, preconditioning decreases the number of required CG iterations up to approximately 50\% for the surveyed preconditioners.}
\label{fig:figure5}
\end{figure}

The linear system at each Newton step is large, sparse, symmetric, and typically positive-definite, and thereby well-suited for solving via the conjugate gradient method, one of a family of Krylov subspace methods for iteratively solving linear systems. We accelerate the linear system solve by applying block Jacobi preconditioning~\cite{hegland1992block}. This approach has the advantage of being highly parallelizable, as sub-blocks may be factored simultaneously in parallel. The sub-block factorizations are computed directly using Cholesky factorization, suitable for symmetric, positive-definite matrices, via the Eigen library~\cite{eigenweb}. We note that if a bad attempt at solution is made resulting in a matrix that is not positive-definite, the integration step is rejected and repeated with reduced step size. We find this occurs very rarely in practice.
Additionally, we aim to improve the quality of the block Jacobi preconditioner by first re-indexing each degree of freedom to produce a matrix sparsity pattern that is dense along the diagonal. This is related to the task of bandwidth reduction, which seeks to minimize the bandwidth, or furthest distance of non-zero entries from the diagonal, of a matrix. Testing a classic bandwidth reduction algorithm, Reverse Cuthill--McKee (RCM)~\cite{cuthill1969reducing}, as well as a re-indexing scheme based on the Hilbert space-filling fractal~\cite{moon2001analysis}, we find that while RCM achieves a smaller bandwidth, the Hilbert curve better preserves the locality of nodes and produces a sparsity pattern that is denser along the diagonal, as illustrated in Fig.~\ref{fig:figure4}. Thus, the Hilbert curve-based re-indexing is used to accelerate the linear solve alongside block Jacobi preconditioning.
Fig.~\ref{fig:figure5} compares the performance of several different integration and preconditioning approaches for a sample crumpling simulation. Direct and iterative linear solvers are considered, with different preconditioners. We find the best performance using the conjugate gradient method and block Jacobi preconditioning with a moderate block size; larger block sizes present a greater overhead cost for the factorization of diagonal blocks, and thus reduce the computational benefit past a certain size.

\subsection{Dynamic formulation}
\begin{figure}[ht!]
\centering
\includegraphics[width=6in]{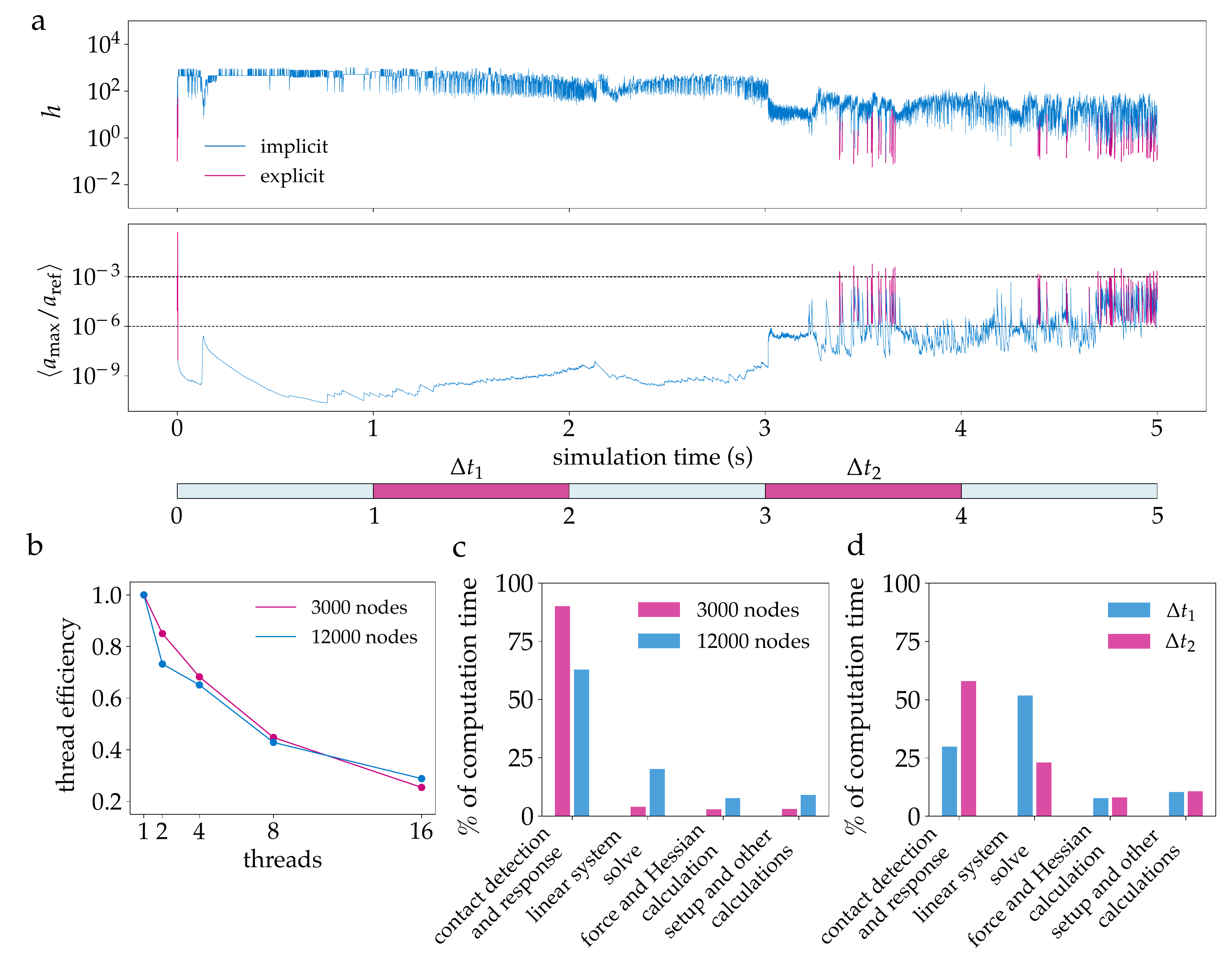}
\caption{\textbf{Adaptive switching and multithreading performance.} (a) The step size $h$ and average scaled maximum acceleration $\left<a_{\text{max}}/a_{\text{ref}}\right>$ as a function of simulation time, for the simulation presented in Fig.~\ref{fig:figure5}a. Segments are colored by the integration scheme used, implicit or explicit, illustrating the adaptive switching between methods. Dashed horizontal lines in the plot of $\left<a_{\text{max}}/a_{\text{ref}}\right>$ denote the switching thresholds: Falling below the lower threshold at $10^{-6}$ prompts a switch to the implicit scheme, while rising above the upper threshold at $10^{-3}$ prompts a switch to the explicit scheme. (b) The thread efficiency for simulations of the radial confinement example using a coarser (3000 nodes) and finer (12000 nodes) mesh as a function of CPU threads used, computed as $T(1)/nT(n)$ where $n$ is the number of threads and $T(n)$ the computation time with $n$ threads. (c) For simulations using $8$ threads, the percent of computation time spent on selected key components of the simulation using 3000 and 1200 nodes, and (d) the same for the two integration intervals $\Delta{t}_1$ from $t=\SI{1}{\s}$ to $t=\SI{2}{\s}$ and $\Delta{t}_2$ from $t=\SI{3}{\s}$ to $t=\SI{4}{\s}$ as in Fig.~\ref{fig:figure5} (12000 nodes). Prior to significant self-contact of the sheet, as in $\Delta{t}_1$, the linear system solve accounts for the majority of the computation time. However, contact detection and response becomes most expensive during later stages of simulation, and dominates the computation time overall.}
\label{fig:figure6}
\end{figure}
As earlier noted, the smooth deformation of the sheet may be interrupted by sudden, large changes in local velocity, for example by elastic rearrangement of ridges and vertices, or by abrupt self-contact. In such cases, the quasistatic approximation no longer holds, and we switch to the fully dynamic formulation of the equations of motion, solved explicitly. The breakdown of the quasistatic approximation is detected by calculating the maximum rate of change of the algebraic (velocity) degrees of freedom, $a_{\text{max}}$, and identifying if it exceeds a specified threshold relative to a reference acceleration $a_{\text{ref}}$. We select $a_{\text{ref}}=k_s/m\lambda$, where $m$ is the mass of a node, $k_s$ the spring constant, and $\lambda$ a linear node density. In practice, a moving average $\left<a_{\text{max}}/a_{\text{ref}}\right>$ is used to discourage frequent fluctuation between the quasistatic and dynamic formulations, with $\left<\cdot\right>$ denoting an average over a specified number of prior steps. In addition, switching is disallowed if too few consecutive steps have been attempted with the implicit method. Fig.~\ref{fig:figure6}a tracks the integration step size and average scaled maximum acceleration over the course of a single simulation and illustrates the alternating behavior of the integrator. The explicit solver uses a standard fourth-order accurate Runge-Kutta method and likewise employs adaptive step control. In this case, an embedded third-order method is compared against a fourth-order candidate solution for step size selection.

Fig.~\ref{fig:figure6}b shows the computation time for a typical crumpling simulation while varying the number of threads. In Fig.~\ref{fig:figure6}c--d, we analyze how computation time is spent for two different mesh resolutions, and for the two distinct integration intervals of Fig.~\ref{fig:figure5}. When significant self-contact is present, the majority of computation time is spent resolving contacts. However, we note that as smaller meshes require a higher refinement level to resolve the same interaction range, there is an advantage to increasing the mesh size and correspondingly decreasing refinement without incurring a proportionally larger penalty.

\clearpage
\section{Results}
\begin{figure}[ht!]
\centering
\includegraphics[width=\textwidth]{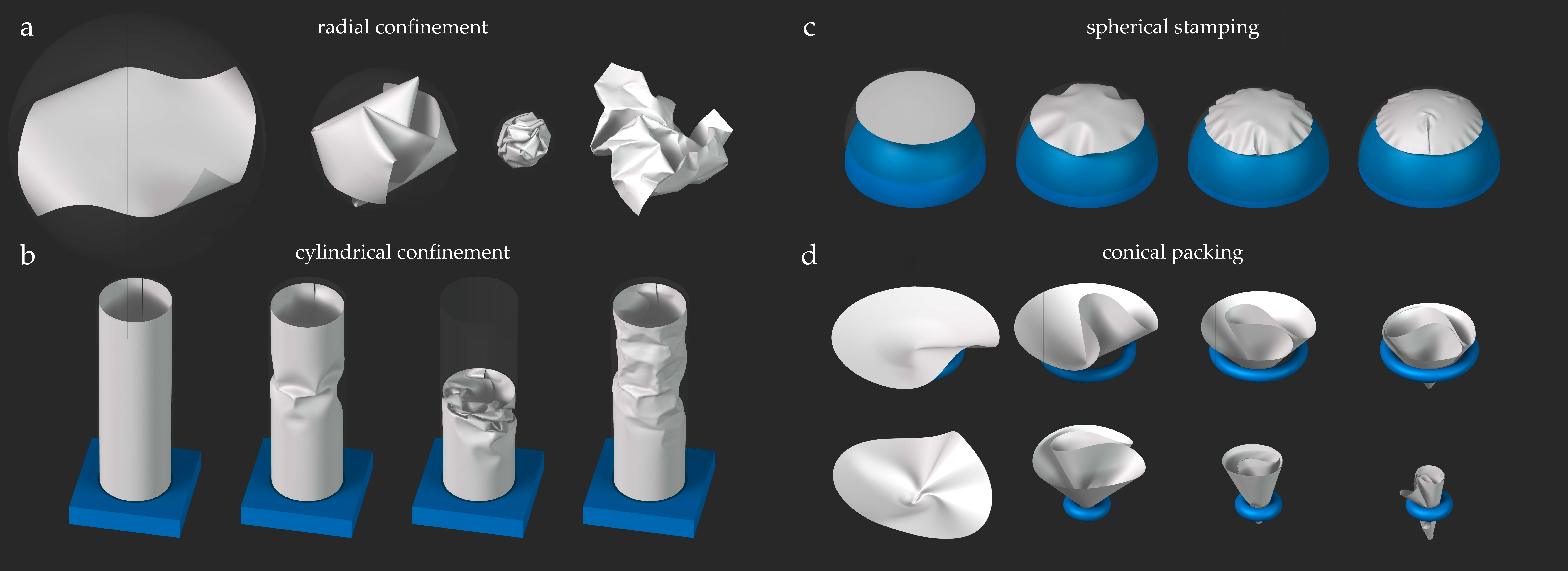}
\caption{\textbf{Crumpling examples gallery.} Simulations of an elastoplastic sheet under radial confinement, cylindrical confinement, spherical stamping, and conical packing (low and high confinement), illustrating a variety of mechanisms by which sheets may wrinkle, fold, and crumple.}
\label{fig:figure7}
\end{figure}
\subsection{Representative examples}
We now demonstrate some of the capabilities of the computational model. We begin with a small gallery of examples displaying a variety of mechanisms by which wrinkling, folding, and crumpling can occur. Motivated by experimental and theoretical work in these areas, Fig.~\ref{fig:figure7} demonstrates radial confinement~\cite{cambou2011three}, cylindrical confinement~\cite{gottesman2018state,cambou2015orientational}, spherical stamping~\cite{hure2012stamping,davidovitch2019geometrically}, and conical packing~\cite{boue2006spiral,adda2010statistical,mellado2011mechanical} of elastoplastic sheets. When mapped to physical units, the sheets are \SI{10}{\cm} in length (diameter) with a Young's modulus of approximately \SI{1}{\GPa}. The radially confined sheet has an effective thickness of \SI{0.125}{\mm}, while the remaining examples are \SI{0.25}{\mm}. Experimental studies of repeated cylindrical confinement have revealed remarkably robust properties in the accumulation of total crease length and the distribution of facet sizes, as previously described~\cite{gottesman2018state,andrejevic2021model}. A possible application of numerical simulation would be to verify whether these statistical properties are universal to different modes of crumpling such as radial confinement, which may otherwise pose experimental challenges.

Spherical stamping of a flat sheet provides a model example of wrinkles localizing into sharp folds with increased confinement. This wrinkle-to-crumple transition has been studied experimentally~\cite{timounay2020crumples}, and detailed simulations could further shed light on stress localization in this process. Furthermore, the general framework for modeling nonzero rest curvature enables us to explore the complementary system of a curved shell pressed to a flat substrate, which has seen recent experimental and theoretical development~\cite{albarran2018curvature,tobasco2020exact,tobasco2021curvature}. Inspired by these studies, wrinkle and crumple morphologies could be surveyed for varying intrinsic curvatures, thicknesses, and geometries.

Finally, the elastic properties of a thin sheet pulled through a small opening (conical packing) have been studied from the emergence of the initial crescent singularity~\cite{cerda1998conical} to the formation of subsequent conical dislocations~\cite{mellado2011mechanical}. It is potentially interesting to consider the role of plasticity and properties of radially propagating damage networks in relation to other failure processes such as crack formation, inspired by recent connections between crumpling and fragmentation~\cite{andrejevic2021model}.

\begin{figure}[ht!]
\centering
\includegraphics[width=\textwidth]{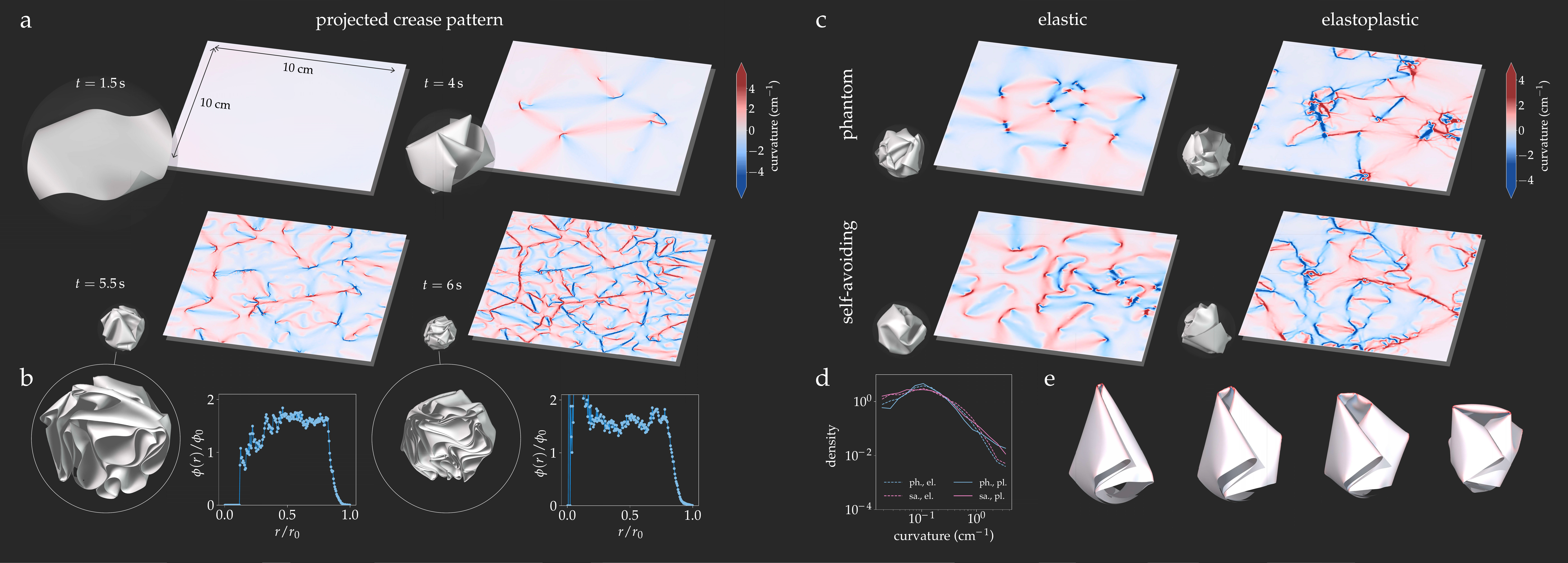}
\caption{\textbf{Crumpled morphologies of sheets with varying properties.} (a) Selected snapshots of an elastoplastic sheet under radial confinement, and corresponding crease patterns. (b) Cross-sections of the crumpled sheet at two different compaction times $t=\SI{5.5}{\s}$ and $t=\SI{6}{\s}$, revealing stacking in the layers of the sheet. For each of the two compaction times, the scaled volume fraction $\phi(r)/\phi_0$ of the three-dimensional structure as a function of the scaled radius $r/r_0$ is also shown, with $r_0$ the maximum radius of the crumpled ball, and $\phi_0$ the volume fraction of the sheet contained in the sphere of radius $r_0$ with origin at the sheet's center of mass. The volume fraction $\phi(r)$ is computed as the volume of sheet contained in a spherical shell of inner radius $r$ and outer radius $r + dr$, divided by the shell volume. The sheet volume is determined as the area of the triangles of the underlying mesh whose center of mass is contained in the shell, times effective thickness. (c) Comparison of the crease pattern for phantom (self-intersecting) elastic and elastoplastic sheets, and self-avoiding elastic and elastoplastic sheets. Qualitative differences in the localization of curvature and homogeneity of the deformations can be observed. (d) Curvature distributions for the four crease patterns in (c). At low curvature, the distributions are most influenced by the topology of the sheet, phantom (ph.) or self-avoiding (sa.). However at high curvature, the distributions mostly reflect the elastic properties, elastic (el.) or elastoplastic (pl.), with elastoplastic sheets displaying higher frequency of large curvatures. (e) Selected snapshots of an elastic, self-avoiding sheet displaying elastic rearrangements of a ridge and vertices during crumpling.}
\label{fig:figure8}
\end{figure}

Fig.~\ref{fig:figure8}a shows the evolution of the projected crease pattern of a radially confined sheet. As the sheet becomes increasingly confined, we observe a high degree of stacking in cross-sections of the sheet, as seen in Fig.~\ref{fig:figure8}b.
Detailed statistical properties of crumpled sheets have been investigated experimentally, for example in Refs.~\cite{deboeuf2013compaction, deboeuf2013comparative, andresen2007ridge, blair2005geometry, balankin2006intrinsically, balankin2008entropic, balankin2006intrinsically, balankin2013fractal,cambou2011three,martoia2017crumpled}. Inspired by experiments of Cambou and Menon~\cite{cambou2011three}, we compute the volume fraction of the three-dimensional structure as a function of radius, shown in Fig.~\ref{fig:figure8}b. In addition to supplementing data-driven studies, we believe a key application of the simulation is to enable measurement of quantities that are challenging or impossible to measure experimentally, such as local stresses in the sheet.

Furthermore, numerical experiments allow properties such as self-avoidance and plasticity to be studied systematically. Fig.~\ref{fig:figure8}c shows the crease patterns of phantom (self-intersecting) elastic and elastoplastic sheets, alongside self-avoiding examples. Qualitative observations apparent from the crease patterns are increased localization of high curvature for elastoplastic sheets, and overall greater homogeneity in the case of self-avoiding sheets for the same degree of confinement. Inspired by the detailed analyses of Marto{\"\i}a \textit{et al}. on experimental sheets~\cite{martoia2017crumpled}, we compute the curvature distributions for all four crease patterns from Fig.~\ref{fig:figure8}c, shown in Fig.~\ref{fig:figure8}d. Interestingly, we observe that at low curvature, the distributions are most influenced by the topology of the sheet, phantom or self-avoiding. However at high curvature, the distributions mostly reflect the elastic properties, elastic or elastoplastic, with elastoplastic sheets displaying higher frequency of large curvatures. These properties could be investigated in more detail to better understand the role of self-avoidance and plasticity in the localization of energy during crumpling. While theoretical work has predicted the scaling of energy in a single ridge,  the scaling of energy localization in crease networks is inherently more complex~\cite{kramer1997stress}.

In addition, the numerical setting is useful for investigating the dynamics of crumpling. In Fig.~\ref{fig:figure8}d, we isolate an example of ridge and vertex rearrangement in the confinement of a purely elastic, self-avoiding sheet. Network rearrangements may explain the discrepancy in predicted and measured accumulation of elastic energy in a confined sheet~\cite{aharoni2010direct}, and could be directly studied using our model.

\subsection{Comparison to experiment}
\begin{figure}[ht!]
\centering
\includegraphics[width=6in]{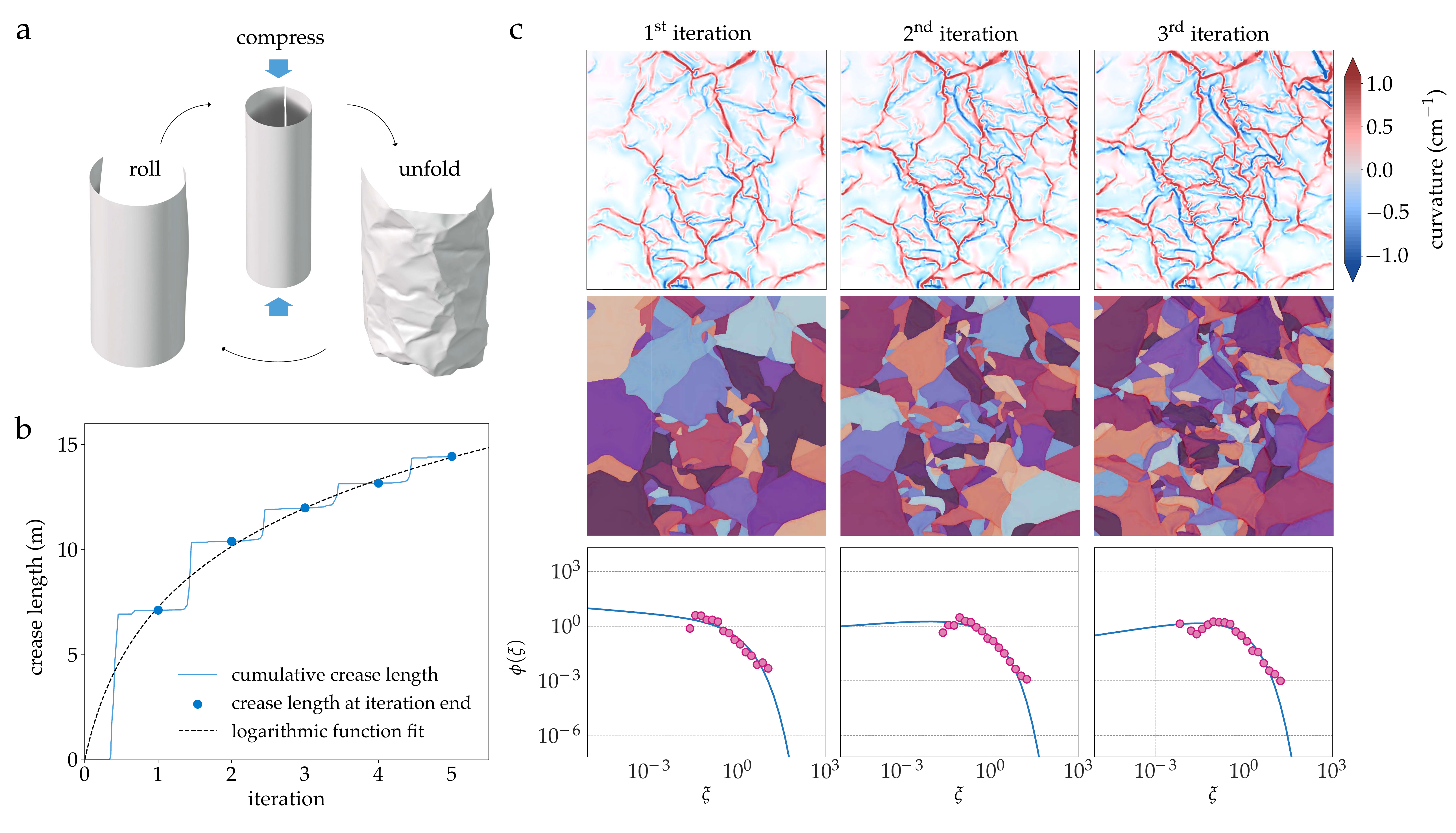}
\caption{\textbf{Numerical experiment of axial crumpling.} (a) Schematic of a repeated axial crumpling simulation. During each iteration, the sheet is rolled into a cylinder, uniaxially compressed under cylindrical confinement, then unfolded, modeled after the experiments by Gottesman \textit{et al.}~\cite{gottesman2018state}. (b) The total crease length as a function of iteration count, measured as the total length of all edges that have undergone plastic deformation. The dashed line demonstrates logarithmic growth in total crease length qualitatively consistent with experimental observations~\cite{gottesman2018state}. (c) The crease pattern (top), facet segmentation (middle), and scaled facet area distribution (bottom) of the sheet after each of the first three crumpling iterations. Facets are colored randomly for visual distinction. In the distribution plots, markers denote the measured density $\phi(\xi)$ of facets of scaled area $\xi$, while the solid line shows the theoretical prediction based on the fragmentation model presented by Andrejevic \textit{et al.}~\cite{andrejevic2021model}, which exhibits reasonable agreement with the simulated results.}
\label{fig:figure9}
\end{figure}
In order to make quantitative comparisons, we also perform a repeated axial crumpling test following the experiments by Gottesman \textit{et al.}~\cite{gottesman2018state}. At each iteration, the simulated sheet is shaped into a cylinder, compressed uniaxially along $z$ while confined radially in $x$ and $y$ directions, and unfolded. This process is repeated five times in the example presented in Fig.~\ref{fig:figure9}. The total length of edge connections in the mesh that undergo plastic deformation serves as an intrinsic measure of the total crease length. We observe growth in the total crease length consistent with a logarithm in iteration number, in agreement with experiment~\cite{gottesman2018state}, as shown in Fig.~\ref{fig:figure9}b. We further perform manual facet segmentation of the crease patterns produced at the end of the first three crumpling iterations and plot the corresponding distributions of facet area (Fig.~\ref{fig:figure9}c). While the present simulation is at a much lower resolution than experimental imaging, we see evidence that the facet area distribution is in fair agreement with the theoretical prediction obtained by the fragmentation model of Andrejevic \textit{et al.}~\cite{andrejevic2021model}. Tuning the material parameters and simulating larger sheets will refine these results.

\section{Discussion}
We have presented a computational model for simulating the large-scale deformation of thin sheets that can serve as an effective tool for data-driven studies of crumpling dynamics. By defining both a quasistatic and dynamic formulation of the equations of motion for the simulated sheet, we have implemented an adaptive integration scheme that uses an efficient implicit method during smooth deformation, and resolves large local changes in velocity explicitly. Our model can flexibly simulate a variety of confinement conditions and demonstrates strong qualitative agreement with physical examples of crumpling. Moreover, we see quantitative evidence that key phenomena such as the logarithmic growth of total crease length and characteristic facet size distribution are likewise reproduced. To conclude, we draw attention to a few current limitations of the model and directions for future work.

With regard to the physical modeling of crumpled sheets, an accurate mechanism for plastic deformation still remains poorly understood. While our current model for plasticity reproduces the qualitative behavior of crumpled sheets, improvements are needed to accurately capture the characteristic slow relaxation of folds and crumpled structures seen in experiment~\cite{matan2002crumpling,amir2012relaxations,albuquerque2002stress,balankin2015mechanical,balankin2011slow}. On a microscopic level, constitutive models for in-plane yield stress relaxation and creep have been proposed for polymer-like networks such as paper fibers and offer an analytical route to slow relaxation~\cite{coffin2009developing}. Future work could entail developing an improved model for plasticity based on these studies. In addition, friction between the layers of a crumpled sheet or with the enclosing container has not been considered in this work, but likely plays an important role in the relaxation of crumpled structures~\cite{matan2002crumpling,balankin2013statistics}.

With regard to numerical methods, contact detection and response remains a critical bottleneck of large-scale simulations of dense crumpling. In addition to the computationally costly collision checks, sudden self-contact leads to large local accelerations, restricting the integration step size and triggering longer periods of slow explicit solves. To mitigate the expensive collision checks for all refined sites in our contact framework, an adaptive refinement scheme could be used, and additional contact sites introduced only when different parts of the sheet become close. However, this would not fully resolve the challenges for densely crumpled sheets. An alternative to explicit integration would be to integrate the fully dynamic equations of motion implicitly as well, and thereby avoid strict limitations on step size required for numerical stability. However, further work is needed to optimize the conjugate gradient solver so that implicit integration can become more computationally efficient even for large accelerations and smaller step sizes. In light of this, we have begun investigating other preconditioning strategies including algebraic multigrid~\cite{brandt1984algebraic} and sparse approximate inverse~\cite{benson1973iterative,benzi2002preconditioning} preconditioners. These approaches may be more promising with further tuning to our problem.

\section*{Acknowledgement}
\noindent This research was partially supported by the National Science Foundation through the Harvard University Materials Research Science and Engineering Center under grant nos.\@ DMR-1420570 and DMR-2011754. Some numerical simulations were performed using resources of the National Energy Research Scientific Computing Center under NERSC award ASCR-ERCAP-0018643. J.A. acknowledges support from the National Science Foundation Graduate Research Fellowship Program under grant no.\@ DGE-1745303. C.H.R. was partially supported by the Applied Mathematics Program of the U.S. DOE Office of Science Advanced Scientific Computing Research under contract number DE-AC02-05CH11231.

\section*{Author contributions}
\noindent Jovana Andrejevic: Software, Validation, Visualization, Writing - Original draft \\
Chris H. Rycroft: Conceptualization, Methodology, Software, Writing - Review and Editing

\section*{Competing interest}
\noindent The authors declare no competing interests.

\bibliography{main}
\clearpage
\appendix

\section{Elastic model}\label{appendix:elasticity}
\renewcommand\thefigure{A.\arabic{figure}}
\setcounter{figure}{0}
\subsection{In-plane elasticity}

For a continuum sheet, the strain energy density is given by
\begin{linenomath}\begin{equation}
    u_s = \frac{1}{2}{\epsilon}\cdot{C}\cdot{\epsilon},
\end{equation}\end{linenomath}
with ${C}$ the stiffness tensor and ${\epsilon}$ the in-plane strain tensor. The components of the stiffness tensor of an isotropic two-dimensional (2D) material can be represented in the form
\begin{linenomath}\begin{equation}
    \begin{bmatrix}
    C_{1111} & C_{1122} & C_{1112} \\
    C_{2211} & C_{2222} & C_{2212} \\
    C_{1211} & C_{1222} & C_{1212} \\
    \end{bmatrix}
    = \begin{bmatrix}
    \lambda + 2\mu & \lambda & 0 \\
    \lambda & \lambda + 2\mu & 0 \\
    0 & 0 & \mu \\
    \end{bmatrix}
\end{equation}\end{linenomath}
where $\lambda$ and $\mu$ are the Lam\'e constants, from which the 2D Young's modulus $Y_{2D}$ and Poisson's ratio $\nu$ may be computed~\cite{seung1988defects}:
\begin{linenomath}\begin{equation}
Y_{2D} = \frac{4\mu(\mu+\lambda)}{2\mu+\lambda}, \quad \nu = \frac{\lambda}{2\mu+\lambda}.
\end{equation}\end{linenomath}
We note that the 2D Young's modulus has dimensions of force per unit length. In the discrete setting, the energy contribution due to an interaction between a pair of nodes $i$ and $j$ joined by a spring of spring constant $k_s$ and rest length $s$ is given by
% Note that for ||, there's a specific LaTeX command of \|, which has slightly tighter spacing 
\begin{linenomath}\begin{equation}
    E_s\left(x_i,x_j\right) = \frac{1}{2} k_s \left(\|r_{ij}\|-s\right)^2,
\end{equation}\end{linenomath}
where $r_{ij}=x_i-x_j$ is the vector from node $j$ to node $i$.
\begin{figure}[ht!]
\centering
\includegraphics[width=0.7\textwidth]{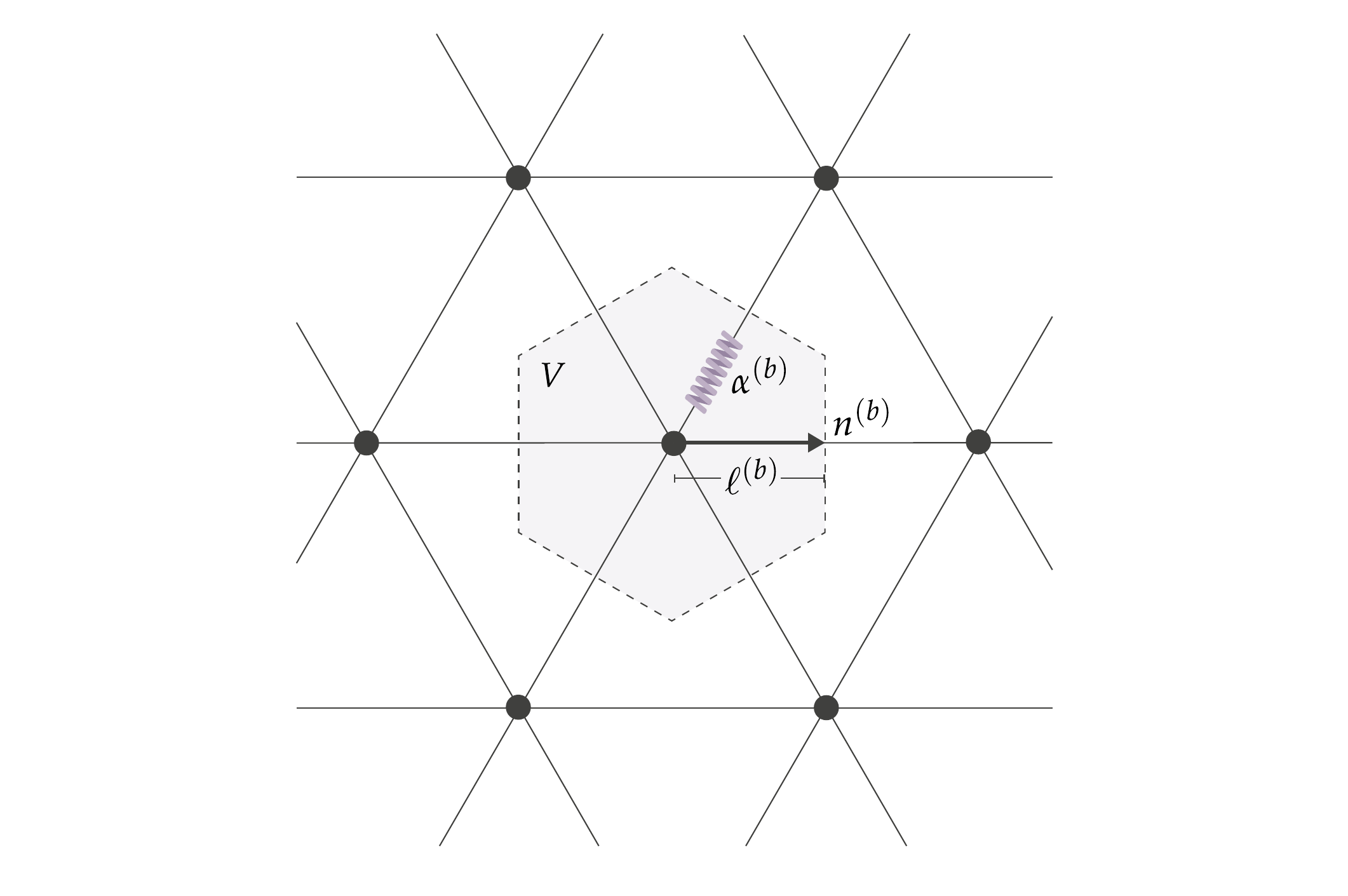}
\caption{\textbf{Energy density for a periodic lattice.} Schematic of the relevant quantities for computing the energy density of a periodic arrangement of sites, as given by Eq.~\ref{eq:en_dens} based on Ref.~\cite{ostoja2002lattice}. A unit cell of volume $V$ is denoted by the dotted lines. Each bond $b$ is of length $\ell^{(b)}$ equal to half the lattice constant, and has stiffness $\alpha^{(b)}$. The unit vector $n^{(b)}$ denotes the direction of the bond.}
\label{fig:figureA1}
\end{figure}
For a periodic arrangement of sites, the energy density can be expressed as \cite{ostoja2002lattice}
\begin{linenomath}\begin{equation}
    u_s = \frac{1}{2V}\sum_b\alpha^{(b)}{(\ell^{(b)})}^{2}n_i^{(b)}n_j^{(b)}n_k^{(b)}n_m^{(b)}\epsilon_{ij}\epsilon_{km}, \label{eq:en_dens}
\end{equation}\end{linenomath}
assuming Einstein summation notation, where $V$ is the volume (area) of the unit cell, $b$ the index of a bond, $\alpha$ the spring constant of a half-length bond, $\ell$ the rest length of a half-length bond, $n_i^{(b)}$ the unit vector component parallel to bond $b$ with component $i\in\{1,2\}$, and $\epsilon_{ij}$ the strain tensor component with directions $i,j\in\{1,2\}$ (Fig.~\ref{fig:figureA1}). The components of the stiffness tensor are then given by
\begin{linenomath}\begin{equation}
    C_{ijkm} = \frac{1}{V}\sum_b\alpha^{(b)}{(\ell^{(b)})}^{2}n_i^{(b)}n_j^{(b)}n_k^{(b)}n_m^{(b)}.
\end{equation}\end{linenomath}
In the case of a hexagonal lattice with identical springs of spring constant $k_s=\alpha/2$, rest length $s=2\ell$, and cell volume $V=2\sqrt{3}\ell^2$, the continuum and discrete expressions for the stiffness tensor components may be equated to obtain~\cite{seung1988defects,ostoja2002lattice}
\begin{linenomath}\begin{equation}
    \begin{aligned}
    \lambda=\mu=\frac{\sqrt{3}}{4}k_s.
    \end{aligned}
\end{equation}\end{linenomath}
Consequently, the 2D Young's modulus and Poisson's ratio for a hexagonal lattice are~\cite{seung1988defects}
\begin{linenomath}\begin{equation}
    Y_{2D} = \frac{2}{\sqrt{3}}k_s, \quad \nu = \frac{1}{3}.
\end{equation}\end{linenomath}
For other periodic or non-periodic mesh topologies, the in-plane stretching interactions can still be modeled as springs between adjacent nodes $i$ and $j$ as depicted in Fig.~\ref{fig:figure1} of the main text, with energy contribution
\begin{linenomath}\begin{equation}
    E_s\left(x_i,x_j\right) = \frac{1}{2} k_s \left(\|r_{ij}\|-s_{ij}\right)^2,
\end{equation}\end{linenomath}
where $s_{ij}$ is a rest length unique to each spring. However, the choice of mesh topology will affect the isotropy of the elastic response. To avoid strong deviation from the theoretical results on a hexagonal lattice, we work with a regularized random mesh, as detailed in Section~\ref{sec:numerical_implementation} of the main text.

\subsection{Out-of-plane elasticity}
The total continuum bending energy of a developable surface $S$ embedded in $\mathbb{R}^3$ is given by~\cite{seung1988defects}
\begin{linenomath}\begin{equation}
    U_b = \frac{1}{2}\kappa\int_S H^2dA,
\end{equation}\end{linenomath}
with $\kappa$ the bending rigidity, $H=c_1+c_2$ the local mean curvature (equal to the sum of principal curvatures $c_1$ and $c_2$ rather than the mean, by the convention of Ref.~\cite{seung1988defects}), and $dA$ an area element of the reference surface $S$. For small deformations $f$ of the surface, the mean curvature may be approximated by the Laplacian of $f$, $H\approx\nabla^2{f}$. In the discrete model of elastic membranes introduced by Seung and Nelson~\cite{seung1988defects}, the surface is partitioned into equilateral triangles, and the contribution to the bending energy at an edge shared by two triangles $T_1$ and $T_2$ with unit normal vectors $\hat{{n}}_{1}, \hat{{n}}_{2}$, respectively, is given by
\begin{linenomath}\begin{equation}
    E_b \left(\hat{{n}}_{1}, \hat{{n}}_{2}\right) = \frac{1}{2} k_b \|\hat{{n}}_{1} - \hat{{n}}_{2}\|^2,
\end{equation}\end{linenomath}
where the bending constant $k_b$ is related to the bending rigidity by~\cite{seung1988defects}
\begin{linenomath}\begin{equation}
    \kappa = \frac{\sqrt{3}}{2}k_b.
\end{equation}\end{linenomath}
We see that the discrete bending energy arises as a penalty for misalignment in the normal vectors of neighboring triangles in the mesh, as shown in Fig.~\ref{fig:figure1} of the main text. As presented, this formulation for bending energy is valid for equilateral triangles; however, a generalization can be made for triangles of varying symmetry. Wardetzky \textit{et al.}~\cite{wardetzky2007discrete} present a discrete formulation of bending energy at an edge as 
\begin{linenomath}\begin{equation}
    E_b({x}) \propto \frac{1}{2}{x}^T(L^TM^{-1}L){x},
\end{equation}\end{linenomath}
where ${x}$ is the embedding ${x}:S\rightarrow\mathbb{R}^3$ of the surface $S$, $L$ a discrete intrinsic Laplace operator, and $M$ a diagonal mass matrix whose diagonal entries equal one third the sum of areas of the two incident triangles; i.e. $m_{ii}=\bar{A}/3$ with $\bar{A}= A_1+A_2$. We note the right-hand expression is dimensionless, and equivalence is achieved by multiplying a bending constant with dimensions of energy to the right-hand side. For a choice of discretization of the Laplace operator consistent with an edge-based linear finite element model, the bending energy takes the form
\begin{linenomath}\begin{equation}
    E_b(\theta) \propto \frac{3\|{e}_0\|^2}{2\bar{A}}\left(2\sin\frac{\theta}{2}\right)^2,
\end{equation}\end{linenomath}
with $\theta$ the angle between the normals of the two triangles incident at an edge defined by vector ${e}_0$. Re-expressed in terms of the triangle unit normals,
\begin{linenomath}\begin{equation}
    \left(2\sin\frac{\theta}{2}\right)^2=2\left(1-\cos\theta\right)=2\left(1-\hat{{n}}_1\cdot\hat{{n}}_2\right)=\|\hat{{n}}_{1} - \hat{{n}}_{2}\|^2.
\end{equation}\end{linenomath}
We can reconcile the resulting model with that of Seung and Nelson~\cite{seung1988defects} by introducing a prefactor of $k_b/\left(2\sqrt{3}\right)$:
\begin{linenomath}\begin{equation}
    E_b(\hat{{n}}_1,\hat{{n}}_2) = \frac{1}{2}k_b \frac{\sqrt{3}\|{e}_0\|^2}{2\bar{A}}\|\hat{{n}}_{1} - \hat{{n}}_{2}\|^2.
\end{equation}\end{linenomath}
For equilateral triangles, $\bar{A}=\|{e}_0\|^2\sqrt{3}/2$, and the expression simplifies to
\begin{linenomath}\begin{equation}
    E_b(\hat{{n}}_1,\hat{{n}}_2) = \frac{1}{2}k_b\|\hat{{n}}_{1} - \hat{{n}}_{2}\|^2,
\end{equation}\end{linenomath}
consistent with Seung and Nelson~\cite{seung1988defects}. Thus we adopt the form
\begin{linenomath}\begin{equation}
    E_b(\hat{{n}}_1,\hat{{n}}_2) = \frac{1}{2}k_b \frac{\sqrt{3}\|{e}_0\|^2}{2\bar{A}}\|\hat{{n}}_{1} - \hat{{n}}_{2}\|^2,
\end{equation}\end{linenomath}
generalized for all types of triangles. For physical thin sheets of interest, the energetic cost of in-plane stretching is much larger than out-of-plane bending, resulting in only small deformations of triangle edge lengths and areas. Thus, the prefactor $\|{e}_0\|^2/\bar{A}$ is treated as constant, evaluated at the equilibrium state of the triangulated sheet.

\section{Derivation of Jacobian terms}\label{appendix:jac}
\renewcommand\thefigure{B.\arabic{figure}}
\setcounter{figure}{0}

Here we detail some of the key derivative calculations necessary for constructing the Jacobian during implicit integration.

\subsection{In-plane elasticity}
The force on a node $i$ at position $x_i$ due to in-plane spring interactions with node $j$ at position $x_j$ is given by
\begin{linenomath}\begin{equation}
    F_s(x_i,x_j) = -\left(\nabla_{x_{i}}E_s(x_i,x_j)\right)^T = k_s\left(\frac{s_{ij}}{\|r_{ij}\|}-1\right)r_{ij},
\end{equation}\end{linenomath}
where $r_{ij}=x_i-x_j$ and $s_{ij}$ is the rest length of the joining spring. Applying the product rule for the gradient of a scalar $\psi$ times a vector $v$,
\begin{linenomath}\begin{equation}
    \nabla\left(\psi{v}\right) = \psi\nabla{v} + v\nabla\psi,
\end{equation}\end{linenomath}
where $v\nabla\psi$ is an outer product of the column vectors $v$ and $\left(\nabla\psi\right)^T$, the Hessian of the in-plane spring energy is given by
\begin{linenomath}\begin{equation}
    -\nabla_{x_k}F_s(x_i,x_j) = \begin{cases}
      k_s\left(\left(1-\frac{s_{ij}}{\|r_{ij}\|}\right)I + \frac{s_{ij}}{\|r_{ij}\|^3}r_{ij}r_{ij}^T\right) & \qquad \text{for $k=i$,}\\
      -k_s\left(\left(1-\frac{s_{ij}}{\|r_{ij}\|}\right)I + \frac{s_{ij}}{\|r_{ij}\|^3}r_{ij}r_{ij}^T\right) & \qquad \text{for $k=j$,}\\
        0 & \qquad \text{otherwise}.
    \end{cases}
\end{equation}\end{linenomath}
Here $r_{ij}r_{ij}^T$ is likewise an outer product, and we have used the fact that
\begin{linenomath}\begin{equation}
    \nabla_{x_k}r_{ij} = \begin{cases}
      I & \qquad \text{for $k=i$,}\\
      -I & \qquad \text{for $k=j$,}\\
    0, & \qquad \text{otherwise.}
    \end{cases}
\end{equation}\end{linenomath}

\subsection{Internal damping}
As provided in the main text, the internal damping force on node $i$ connected to node $j$ via a dashpot is given by
\begin{linenomath}\begin{equation}
  {F}_d^{\text{int}}(x_i,x_j) = -b_{\text{int}}\left(\left(1-\frac{s_{ij}}{\|r_{ij}\|}\right)\left(v_i-v_j\right)+\frac{s_{ij}}{\|r_{ij}\|^3}\left[r_{ij}\cdot\left(v_i-v_j\right)\right]r_{ij}\right).
\end{equation}\end{linenomath}
Working out the contributions to the Jacobian, we obtain
\begin{linenomath}\begin{equation}
    -\nabla_{x_k}F_d^{\text{int}}(x_i,x_j) = \begin{cases}
        b_{\text{int}}\left(\frac{s_{ij}}{\|r_{ij}\|^3}\left(r_{ij}\cdot{u_{ij}}\right)I + \frac{s_{ij}}{\|r_{ij}\|^3}\left(r_{ij}u_{ij}^T + u_{ij}r_{ij}^T - 3\frac{\left(r_{ij}
        \cdot{u_{ij}}\right)}{\|r_{ij}^2\|}r_{ij}r_{ij}^T\right)\right) & \qquad \text{for $k=i$,}\\
        -b_{\text{int}}\left(\frac{s_{ij}}{\|r_{ij}\|^3}\left(r_{ij}\cdot{u_{ij}}\right)I + \frac{s_{ij}}{\|r_{ij}\|^3}\left(r_{ij}u_{ij}^T + u_{ij}r_{ij}^T - 3\frac{\left(r_{ij}
\cdot{u_{ij}}\right)}{\|r_{ij}^2\|}r_{ij}r_{ij}^T\right)\right) & \qquad \text{for $k=j$,}\\
        0 & \qquad \text{otherwise,}
    \end{cases}
\end{equation}\end{linenomath}
and
\begin{linenomath}\begin{equation}
    -\nabla_{v_k}F_d^{\text{int}}(x_i,x_j) = \begin{cases}
      b_{\text{int}}\left(\left(1-\frac{s_{ij}}{\|r_{ij}\|}\right)I + \frac{s_{ij}}{\|r_{ij}\|^3}r_{ij}r_{ij}^T\right) & \qquad \text{for $k=i$,}\\
      -b_{\text{int}}\left(\left(1-\frac{s_{ij}}{\|r_{ij}\|}\right)I + \frac{s_{ij}}{\|r_{ij}\|^3}r_{ij}r_{ij}^T\right) & \qquad \text{for $k=j$,}\\
        0 & \qquad \text{otherwise,}
    \end{cases}
\end{equation}\end{linenomath}
where we have introduced $u_{ij}=v_i-v_j$ to simplify notation.

\subsection{Self-contact}
As given in the main text, the contact force on node $i$ belonging to a contact site at $x_1$, due to interaction with a contact site at position $x_2$ within its interaction range, is given by
\begin{linenomath}\begin{equation}
    F_c(x_i,x_2) = w_iF_1
\end{equation}\end{linenomath}
where
\begin{linenomath}\begin{equation}
    F_1 = k_s\frac{\sigma}{\|r_{12}\|}\left(\frac{\sigma}{\|r_{12}\|-\sigma}\right)^2\exp\left({\frac{\sigma}{\|r_{12}\|-\sigma}}\right)r_{12},
\end{equation}\end{linenomath}
with $r_{12}=x_1-x_2$ and $x_1=w_ix_i+w_jx_j+w_kx_k$, $x_2=w_mx_m+w_nx_n+w_ox_o$ the positions of the refined contact sites. The corresponding Jacobian terms are of the form
\begin{align}
    -\nabla_{x_m}F_c(x_i,x_2) &= -w_iw_m\nabla_{x_2}F_1 \nonumber \\
    &= w_iw_mk_s\frac{\sigma}{\|r_{12}\|}\rho_{12}^2\exp\left(\rho_{12}\right)\left(-I+\frac{1}{\sigma\|r_{12}\|}\left(\rho_{12}^2+2\rho_{12}+\frac{\sigma}{\|r_{12}\|}\right)r_{12}r_{12}^T\right),
\end{align}
where we have introduced $\rho_{12}=\sigma/\left(\|r_{12}\|-\sigma\right)$ to simplify notation.

\subsection{Out-of-plane elasticity}
Following the notation in Fig.~\ref{fig:figure3} of the main text, the force on a node $i$ due to bending at two adjacent triangles is given by
\begin{linenomath}\begin{equation}
    F_b(x_i;\hat{n}_1,\hat{n}_2) = -\left(\nabla_{x_{i}}E_b(\hat{n}_1,\hat{n}_2)\right)^T = k_b \frac{\sqrt{3}\|{e}_0\|^2}{2\bar{A}}\left(\nabla_{x_i}\left(\hat{n}_1\cdot\hat{n}_2\right)\right)^T.
\end{equation}\end{linenomath}
This is an interaction of four nodes; thus the gradient $\nabla_{x_i}$ is computed with respect to $x_i\in\{x_0,x_1,x_2,x_3\}$, again following the notation of Fig.~\ref{fig:figure3}. We note that the relevant edge vectors are defined as
\begin{align}
    e_0 &= x_1-x_0, \\
    e_1 &= x_2-x_0, \\
    e_2 &= x_3-x_0,
\end{align}
and the normal vectors are
\begin{align}
    n_1 &= e_0\times e_1, \\
    n_2 &= -e_0\times e_2,
\end{align}
with $\hat{n}_1= n_1/\|n_1\|$ denoting the unit normal. Here, we consider the case without plasticity; however, the components derived in the elastic case are reused in the plastic case, with a few additional terms. Thus, we focus our attention to deriving these shared contributions in detail. The gradients with respect to the individual position degrees of freedom may be expressed as
\begin{align}
   \nabla_{{x}_0}\left({\hat{n}}_1\cdot{\hat{n}}_2\right) &= \frac{\left[{\hat{n}}_1,{\hat{n}}_2,{e}_1\right]{\hat{n}}_1^T-\left[{\hat{n}}_1,{\hat{n}}_2,{e}_0\right]{\hat{n}}_1^T}{\|{n}_1\|} +\frac{\left[{\hat{n}}_1,{\hat{n}}_2,{e}_2\right]{\hat{n}}_2^T-\left[{\hat{n}}_1,{\hat{n}}_2,{e}_0\right]{\hat{n}}_2^T}{\|{n}_2\|}, \\
\nabla_{{x}_1}\left({\hat{n}}_1\cdot{\hat{n}}_2\right) &= -\frac{\left[{\hat{n}}_1,{\hat{n}}_2,{e}_1\right]{\hat{n}}_1^T}{\|{n}_1\|} -\frac{\left[{\hat{n}}_1,{\hat{n}}_2,{e}_2\right]{\hat{n}}_2^T}{\|{n}_2\|}, \\
\nabla_{{x}_2}\left({\hat{n}}_1\cdot{\hat{n}}_2\right) &= \frac{\left[{\hat{n}}_1,{\hat{n}}_2,{e}_0\right]{\hat{n}}_1^T}{\|{n}_1\|}, \\
\nabla_{{x}_3}\left({\hat{n}}_1\cdot{\hat{n}}_2\right) &= \frac{\left[{\hat{n}}_1,{\hat{n}}_2,{e}_0\right]{\hat{n}}_2^T}{\|{n}_2\|},
\end{align}
where, following from Tamstorf \textit{et al.}~\cite{tamstorf2013discrete}, we have used the shorthand
\begin{linenomath}\begin{equation}
    \left[\hat{n}_1,\hat{n}_2,e_i\right] = \left(\hat{n}_1\times\hat{n}_2\right)\cdot{e_i}.
\end{equation}\end{linenomath}
The above gradients can be written concisely using terms of the form
\begin{linenomath}\begin{equation}
    C_{i\alpha} = \frac{\left[{\hat{n}}_1,{\hat{n}}_2,{e}_i\right]{\hat{n}}_{\alpha}}{\|{n}_{\alpha}\|}
\end{equation}\end{linenomath}
with $i\in\{0,1,2\}$ and $\alpha\in\{1,2\}$. Thus, we have
\begin{align}
   \left(\nabla_{{x}_0}\left({\hat{n}}_1\cdot{\hat{n}}_2\right)\right)^T &= C_{11}-C_{01}+C_{22}-C_{02}, \\
\left(\nabla_{{x}_1}\left({\hat{n}}_1\cdot{\hat{n}}_2\right)\right)^T &= -C_{11}-C_{22}, \\
\left(\nabla_{{x}_2}\left({\hat{n}}_1\cdot{\hat{n}}_2\right)\right)^T &= C_{01}, \\
\left(\nabla_{{x}_3}\left({\hat{n}}_1\cdot{\hat{n}}_2\right)\right)^T &= C_{02}.
\end{align}
To compute the Hessian of the bending energy, we thus require the following set of gradients:
\begin{align}
    \nabla_{{x}_0}\left(\nabla_{{x}_0}\left({\hat{n}}_1\cdot{\hat{n}}_2\right)\right)^T &= \nabla_{{x}_0}\left({C}_{11}-{C}_{01}+{C}_{22}-{C}_{02}\right), \\
\nabla_{{x}_1}\left(\nabla_{{x}_0}\left({\hat{n}}_1\cdot{\hat{n}}_2\right)\right)^T &= \nabla_{{x}_1}\left({C}_{11}-{C}_{01}+{C}_{22}-{C}_{02}\right), \\
\nabla_{{x}_2}\left(\nabla_{{x}_0}\left({\hat{n}}_1\cdot{\hat{n}}_2\right)\right)^T &= \nabla_{{x}_2}\left({C}_{11}-{C}_{01}+{C}_{22}-{C}_{02}\right), \\
\nabla_{{x}_3}\left(\nabla_{{x}_0}\left({\hat{n}}_1\cdot{\hat{n}}_2\right)\right)^T &= \nabla_{{x}_3}\left({C}_{11}-{C}_{01}+{C}_{22}-{C}_{02}\right), \\
\nabla_{{x}_0}\left(\nabla_{{x}_1}\left({\hat{n}}_1\cdot{\hat{n}}_2\right)\right)^T &= \nabla_{{x}_0}\left(-{C}_{11}-{C}_{22}\right), \\
\nabla_{{x}_1}\left(\nabla_{{x}_1}\left({\hat{n}}_1\cdot{\hat{n}}_2\right)\right)^T &= \nabla_{{x}_1}\left(-{C}_{11}-{C}_{22}\right), \\
\nabla_{{x}_2}\left(\nabla_{{x}_1}\left({\hat{n}}_1\cdot{\hat{n}}_2\right)\right)^T &= \nabla_{{x}_2}\left(-{C}_{11}-{C}_{22}\right), \\
\nabla_{{x}_3}\left(\nabla_{{x}_1}\left({\hat{n}}_1\cdot{\hat{n}}_2\right)\right)^T &= \nabla_{{x}_3}\left(-{C}_{11}-{C}_{22}\right), \\
\nabla_{{x}_0}\left(\nabla_{{x}_2}\left({\hat{n}}_1\cdot{\hat{n}}_2\right)\right)^T &= \nabla_{{x}_0}{C}_{01}, \\
\nabla_{{x}_1}\left(\nabla_{{x}_2}\left({\hat{n}}_1\cdot{\hat{n}}_2\right)\right)^T &= \nabla_{{x}_1}{C}_{01}, \\
\nabla_{{x}_2}\left(\nabla_{{x}_2}\left({\hat{n}}_1\cdot{\hat{n}}_2\right)\right)^T &= \nabla_{{x}_2}{C}_{01}, \\
\nabla_{{x}_3}\left(\nabla_{{x}_2}\left({\hat{n}}_1\cdot{\hat{n}}_2\right)\right)^T &= \nabla_{{x}_3}{C}_{01}, \\
\nabla_{{x}_0}\left(\nabla_{{x}_3}\left({\hat{n}}_1\cdot{\hat{n}}_2\right)\right)^T &= \nabla_{{x}_0}{C}_{02}, \\
\nabla_{{x}_1}\left(\nabla_{{x}_3}\left({\hat{n}}_1\cdot{\hat{n}}_2\right)\right)^T &= \nabla_{{x}_1}{C}_{02}, \\
\nabla_{{x}_2}\left(\nabla_{{x}_3}\left({\hat{n}}_1\cdot{\hat{n}}_2\right)\right)^T &= \nabla_{{x}_2}{C}_{02}, \\
\nabla_{{x}_3}\left(\nabla_{{x}_3}\left({\hat{n}}_1\cdot{\hat{n}}_2\right)\right)^T &= \nabla_{{x}_3}{C}_{02},
\end{align}
satisfying the symmetry relations
\begin{align}
    \nabla_{{x}_0}\left({C}_{11}-{C}_{01}+{C}_{22}-{C}_{02}\right) &= \left(\nabla_{{x}_0}\left({C}_{11}-{C}_{01}+{C}_{22}-{C}_{02}\right) \right)^T, \\
\nabla_{{x}_1}\left({C}_{11}-{C}_{01}+{C}_{22}-{C}_{02}\right) &= \left(\nabla_{{x}_0}\left(-{C}_{11}-{C}_{22}\right)\right)^T, \\
\nabla_{{x}_2}\left({C}_{11}-{C}_{01}+{C}_{22}-{C}_{02}\right) &= \left(\nabla_{{x}_0}{C}_{01}\right)^T, \\
\nabla_{{x}_3}\left({C}_{11}-{C}_{01}+{C}_{22}-{C}_{02}\right) &= \left(\nabla_{{x}_0}{C}_{02}\right)^T, \\
\nabla_{{x}_1}\left(-{C}_{11}-{C}_{22}\right) &= \left(\nabla_{{x}_1}\left(-{C}_{11}-{C}_{22}\right)\right)^T, \\
\nabla_{{x}_2}\left(-{C}_{11}-{C}_{22}\right) &= \left(\nabla_{{x}_1}{C}_{01}\right)^T, \\
\nabla_{{x}_3}\left(-{C}_{11}-{C}_{22}\right) &= \left(\nabla_{{x}_1}{C}_{02}\right)^T, \\
\nabla_{{x}_2}{C}_{01} &= \left(\nabla_{{x}_2}{C}_{01}\right)^T, \\
\nabla_{{x}_3}{C}_{01} &= \left(\nabla_{{x}_2}{C}_{02}\right)^T, \\
\nabla_{{x}_3}{C}_{02} &= \left(\nabla_{{x}_3}{C}_{02}\right)^T.
\end{align}
We can decompose the gradient of $C_{i\alpha}$ through successive applications of the product rule:
\begin{align}
\nabla{C_{i\alpha}} &= \frac{\left[{\hat{n}}_1,{\hat{n}}_2,{e}_i\right]}{\|{n}_\alpha\|}\nabla{\hat{n}}_\alpha + {\hat{n}}_\alpha\nabla\left(\frac{\left[{\hat{n}}_1,{\hat{n}}_2,{e}_i\right]}{\|{n}_\alpha\|}\right) \nonumber \\
&= \frac{\left[{\hat{n}}_1,{\hat{n}}_2,{e}_i\right]}{\|{n}_\alpha\|}\nabla{\hat{n}}_\alpha + \left[{\hat{n}}_1,{\hat{n}}_2,{e}_i\right]{\hat{n}}_\alpha\nabla\left(\frac{1}{\|{n}_\alpha\|}\right) + \frac{{\hat{n}}_\alpha}{\|{n}_\alpha\|}\nabla\left(\left[{\hat{n}}_1,{\hat{n}}_2,{e}_i\right]\right).
\end{align}
Through vector calculus it can be shown that
\begin{linenomath}\begin{equation}
    {\hat{n}}_\alpha\nabla\left(\frac{1}{\|{n}_\alpha\|}\right) = \frac{1}{\|{n}_\alpha\|}\left(\nabla{\hat{n}_\alpha}\right)^T,
\end{equation}\end{linenomath}
which allows us to rewrite $\nabla{C_{i\alpha}}$ as
\begin{linenomath}\begin{equation}
    \nabla{C_{i\alpha}} = \frac{\left[{\hat{n}}_1,{\hat{n}}_2,{e}_i\right]}{\|{n}_\alpha\|}\left(\nabla{\hat{n}}_\alpha + \left(\nabla{\hat{n}_\alpha}\right)^T\right) + \frac{{\hat{n}}_\alpha}{\|{n}_\alpha\|}\nabla\left(\left[{\hat{n}}_1,{\hat{n}}_2,{e}_i\right]\right).
\end{equation}\end{linenomath}
We can observe that the first term is always symmetric due to the sum $\nabla{\hat{n}}_\alpha + \left(\nabla{\hat{n}_\alpha}\right)^T$---the sum of a matrix and its transpose. This is useful in recognizing when the required symmetry relationships are satisfied in order for the overall Hessian to be symmetric. We can take one more step in expanding our expression for $\nabla{C_{i\alpha}}$:
\begin{align}
    \nabla{C_{i\alpha}} &= \frac{\left[{\hat{n}}_1,{\hat{n}}_2,{e}_i\right]}{\|{n}_\alpha\|}\left(\nabla{\hat{n}}_\alpha + \left(\nabla{\hat{n}}_\alpha\right)^T\right) + \frac{{\hat{n}}_\alpha}{\|{n}_\alpha\|}{e}_i^T\nabla\left({\hat{n}}_1\times{\hat{n}}_2\right)+\frac{{\hat{n}}_\alpha}{\|{n}_\alpha\|}\left({\hat{n}}_1\times{\hat{n}}_2\right)^T\nabla{e}_i \nonumber \\
&= \frac{\left[{\hat{n}}_1,{\hat{n}}_2,{e}_i\right]}{\|{n}_\alpha\|}\left(\nabla{\hat{n}}_\alpha + \left(\nabla{\hat{n}}_\alpha\right)^T\right) + \frac{{\hat{n}}_\alpha}{\|{n}_\alpha\|}{e}_i^T\left({\hat{n}}_1\times\nabla{\hat{n}}_2-{\hat{n}}_2\times\nabla{\hat{n}}_1\right)+\frac{{\hat{n}}_\alpha}{\|{n}_\alpha\|}\left({\hat{n}}_1\times{\hat{n}}_2\right)^T\nabla{e}_i.
\end{align}
We next outline a set of vector relationships that are useful in simplifying the expressions for the components of the Hessian, as well as verifying that the symmetry relationships hold:
\begin{align}
    \left[{\hat{n}}_1,{\hat{n}}_2,{e}_0\right] &= \left({\hat{n}}_1\times{\hat{n}}_2\right)\cdot{e}_0 = -\frac{\left({\hat{n}}_2\cdot{e}_1\right)}{\|{n}_1\|}\|{e}_0\|^2 = -\frac{\left({\hat{n}}_1\cdot{e}_2\right)}{\|{n}_2\|}\|{e}_0\|^2, \\
\left[{\hat{n}}_1,{\hat{n}}_2,{e}_1\right] &= \left({\hat{n}}_1\times{\hat{n}}_2\right)\cdot{e}_1 = \frac{\left({e}_0\cdot{e}_1\right)}{\|{e}_0\|^2}\left[{\hat{n}}_1,{\hat{n}}_2,{e}_0\right], \\
\left[{\hat{n}}_1,{\hat{n}}_2,{e}_2\right] &= \left({\hat{n}}_1\times{\hat{n}}_2\right)\cdot{e}_2 = \frac{\left({e}_0\cdot{e}_2\right)}{\|{e}_0\|^2}\left[{\hat{n}}_1,{\hat{n}}_2,{e}_0\right], \\
\left({\hat{n}}_1\times{\hat{n}}_2\right) &= \frac{\left({e}_0\times{e}_1\right)}{\|{n}_1\|}\times{\hat{n}}_2 = -\frac{\left({\hat{n}}_2\cdot{e}_1\right)}{\|{n}_1\|}{e}_0 = \frac{\left[{\hat{n}}_1,{\hat{n}}_2,{e}_0\right]}{\|{e}_0\|^2}{e}_0, \\
&= {\hat{n}}_1\times\frac{\left({e}_2\times{e}_0\right)}{\|{n}_2\|} = -\frac{\left({\hat{n}}_1\cdot{e}_2\right)}{\|{n}_2\|}{e}_0 = \frac{\left[{\hat{n}}_1,{\hat{n}}_2,{e}_0\right]}{\|{e}_0\|^2}{e}_0, \\
\left({e}_0\times{\hat{n}}_1\right) &= \frac{{e}_0}{\|{n}_1\|}\times\left({e}_0\times{e}_1\right) = \frac{{e}_0\left({e}_0\cdot{e}_1\right)-{e}_1\|{e}_0\|^2}{\|{n}_1\|}, \\
\left({e}_1\times{\hat{n}}_1\right) &= \frac{{e}_1}{\|{n}_1\|}\times\left({e}_0\times{e}_1\right) = \frac{{e}_0\|{e}_1\|^2-{e}_1\left({e}_0\cdot{e}_1\right)}{\|{n}_1\|}, \\
\left({e}_0\times{\hat{n}}_2\right) &= \frac{{e}_0}{\|{n}_2\|}\times\left({e}_2\times{e}_0\right) = \frac{{e}_2\|{e}_0\|^2-{e}_0\left({e}_0\cdot{e}_2\right)}{\|{n}_2\|}, \\
\left({e}_2\times{\hat{n}}_2\right) &= \frac{{e}_2}{\|{n}_2\|}\times\left({e}_2\times{e}_0\right) = \frac{{e}_2\left({e}_0\cdot{e}_2\right)-{e}_0\|{e}_2\|^2}{\|{n}_2\|}, \\
\|{n}_1\|^2 &= \left({e}_0\times{e}_1\right)\cdot\left({e}_0\times{e}_1\right) = \|{e}_0\|^2\|{e}_1\|^2-\left({e}_0\cdot{e}_1\right)^2, \\
\|{n}_2\|^2 &= \left({e}_2\times{e}_0\right)\cdot\left({e}_2\times{e}_0\right) = \|{e}_0\|^2\|{e}_2\|^2-\left({e}_0\cdot{e}_2\right)^2.
\end{align}
Now that some simplifying properties have been outlined, we derive the contributions to the Hessian of the bending energy in full. For clarity, we detail key steps in the derivation for the first few calculations, then present only the final results later on, as the derivations follow from similar principles.

\subsubsection{\texorpdfstring{$\nabla_{{x}_0}\big({C}_{11}-{C}_{01}+{C}_{22}-{C}_{02}\big)$}{A1}}
Substituting in expressions for each term of the form $C_{i\alpha}$, we have that
\begin{align*}
&\nabla_{{x}_0}\big({C}_{11}-{C}_{01}+{C}_{22}-{C}_{02}\big) \\
&= \frac{\big[{\hat{n}}_1,{\hat{n}}_2,{e}_1\big]}{\|{n}_1\|}\Big(\nabla_{{x}_0}{\hat{n}}_1 + \big[\nabla_{{x}_0}{\hat{n}}_1\big]^T\Big)+\frac{{\hat{n}}_1}{\|{n}_1\|}{e}_1^T\big({\hat{n}}_1\times\nabla_{{x}_0}{\hat{n}}_2-{\hat{n}}_2\times\nabla_{{x}_0}{\hat{n}}_1\big)+\frac{{\hat{n}}_1}{\|{n}_1\|}\big({\hat{n}}_1\times{\hat{n}}_2\big)^T\nabla_{{x}_0}{e}_1 \\
&- \frac{\big[{\hat{n}}_1,{\hat{n}}_2,{e}_0\big]}{\|{n}_1\|}\Big(\nabla_{{x}_0}{\hat{n}}_1 + \big[\nabla_{{x}_0}{\hat{n}}_1\big]^T\Big)-\frac{{\hat{n}}_1}{\|{n}_1\|}{e}_0^T\big({\hat{n}}_1\times\nabla_{{x}_0}{\hat{n}}_2-{\hat{n}}_2\times\nabla_{{x}_0}{\hat{n}}_1\big)-\frac{{\hat{n}}_1}{\|{n}_1\|}\big({\hat{n}}_1\times{\hat{n}}_2\big)^T\nabla_{{x}_0}{e}_0 \\
&+ \frac{\big[{\hat{n}}_1,{\hat{n}}_2,{e}_2\big]}{\|{n}_2\|}\Big(\nabla_{{x}_0}{\hat{n}}_2 + \big[\nabla_{{x}_0}{\hat{n}}_2\big]^T\Big)+\frac{{\hat{n}}_2}{\|{n}_2\|}{e}_2^T\big({\hat{n}}_1\times\nabla_{{x}_0}{\hat{n}}_2-{\hat{n}}_2\times\nabla_{{x}_0}{\hat{n}}_1\big)+\frac{{\hat{n}}_2}{\|{n}_2\|}\big({\hat{n}}_1\times{\hat{n}}_2\big)^T\nabla_{{x}_0}{e}_2 \\
&- \frac{\big[{\hat{n}}_1,{\hat{n}}_2,{e}_0\big]}{\|{n}_2\|}\Big(\nabla_{{x}_0}{\hat{n}}_2 + \big[\nabla_{{x}_0}{\hat{n}}_2\big]^T\Big)-\frac{{\hat{n}}_2}{\|{n}_2\|}{e}_0^T\big({\hat{n}}_1\times\nabla_{{x}_0}{\hat{n}}_2-{\hat{n}}_2\times\nabla_{{x}_0}{\hat{n}}_1\big)-\frac{{\hat{n}}_2}{\|{n}_2\|}\big({\hat{n}}_1\times{\hat{n}}_2\big)^T\nabla_{{x}_0}{e}_0. 
\end{align*}
We can introduce our known expressions for the gradients of ${\hat{n}}_1$, ${\hat{n}}_2$, and all edge vectors:
\begin{align*}
&\nabla_{{x}_0}\big({C}_{11}-{C}_{01}+{C}_{22}-{C}_{02}\big) \\
&= \Bigg(\frac{\big[{\hat{n}}_1,{\hat{n}}_2,{e}_1\big]-\big[{\hat{n}}_1,{\hat{n}}_2,{e}_0\big]}{\|{n}_1\|}\Bigg)\Big(\nabla_{{x}_0}{\hat{n}}_1 + \big[\nabla_{{x}_0}{\hat{n}}_1\big]^T\Big) + \Bigg(\frac{\big[{\hat{n}}_1,{\hat{n}}_2,{e}_2\big]-\big[{\hat{n}}_1,{\hat{n}}_2,{e}_0\big]}{\|{n}_2\|}\Bigg)\Big(\nabla_{{x}_0}{\hat{n}}_2 + \big[\nabla_{{x}_0}{\hat{n}}_2\big]^T\Big) \\
&+ \frac{{\hat{n}}_1}{\|{n}_1\|}{e}_1^T\Bigg(\frac{{\hat{n}}_1\times\big({e}_0\times{\hat{n}}_2\big){\hat{n}}_2^T}{\|{n}_2\|}-\frac{{\hat{n}}_1\times\big({e}_2\times{\hat{n}}_2\big){\hat{n}}_2^T}{\|{n}_2\|}+\frac{{\hat{n}}_2\times\big({e}_0\times{\hat{n}}_1\big){\hat{n}}_1^T}{\|{n}_1\|}-\frac{{\hat{n}}_2\times\big({e}_1\times{\hat{n}}_1\big){\hat{n}}_1^T}{\|{n}_1\|}\Bigg) \\
&- \frac{{\hat{n}}_1}{\|{n}_1\|}\big({\hat{n}}_1\times{\hat{n}}_2\big)^T \\
&- \frac{{\hat{n}}_1}{\|{n}_1\|}{e}_0^T\Bigg(\frac{{\hat{n}}_1\times\big({e}_0\times{\hat{n}}_2\big){\hat{n}}_2^T}{\|{n}_2\|}-\frac{{\hat{n}}_1\times\big({e}_2\times{\hat{n}}_2\big){\hat{n}}_2^T}{\|{n}_2\|}+\frac{{\hat{n}}_2\times\big({e}_0\times{\hat{n}}_1\big){\hat{n}}_1^T}{\|{n}_1\|}-\frac{{\hat{n}}_2\times\big({e}_1\times{\hat{n}}_1\big){\hat{n}}_1^T}{\|{n}_1\|}\Bigg) \\
&+ \frac{{\hat{n}}_1}{\|{n}_1\|}\big({\hat{n}}_1\times{\hat{n}}_2\big)^T \\
&+ \frac{{\hat{n}}_2}{\|{n}_2\|}{e}_2^T\Bigg(\frac{{\hat{n}}_1\times\big({e}_0\times{\hat{n}}_2\big){\hat{n}}_2^T}{\|{n}_2\|}-\frac{{\hat{n}}_1\times\big({e}_2\times{\hat{n}}_2\big){\hat{n}}_2^T}{\|{n}_2\|}+\frac{{\hat{n}}_2\times\big({e}_0\times{\hat{n}}_1\big){\hat{n}}_1^T}{\|{n}_1\|}-\frac{{\hat{n}}_2\times\big({e}_1\times{\hat{n}}_1\big){\hat{n}}_1^T}{\|{n}_1\|}\Bigg) \\
&- \frac{{\hat{n}}_2}{\|{n}_2\|}\big({\hat{n}}_1\times{\hat{n}}_2\big)^T \\
&- \frac{{\hat{n}}_2}{\|{n}_2\|}{e}_0^T\Bigg(\frac{{\hat{n}}_1\times\big({e}_0\times{\hat{n}}_2\big){\hat{n}}_2^T}{\|{n}_2\|}-\frac{{\hat{n}}_1\times\big({e}_2\times{\hat{n}}_2\big){\hat{n}}_2^T}{\|{n}_2\|}+\frac{{\hat{n}}_2\times\big({e}_0\times{\hat{n}}_1\big){\hat{n}}_1^T}{\|{n}_1\|}-\frac{{\hat{n}}_2\times\big({e}_1\times{\hat{n}}_1\big){\hat{n}}_1^T}{\|{n}_1\|}\Bigg) \\
&+ \frac{{\hat{n}}_2}{\|{n}_2\|}\big({\hat{n}}_1\times{\hat{n}}_2\big)^T.
\end{align*}
We apply the vector triple product
\begin{linenomath}\begin{equation}
{a}\times\big({b}\times{c}\big) = {b}\big({a}\cdot{c}\big)-{c}\big({a}\cdot{b}\big)
\end{equation}\end{linenomath}
and eliminate dot products such as ${\hat{n}}_1\cdot{e}_0$ (or ${e}_0^T{\hat{n}}_1$), ${\hat{n}}_1\cdot{e}_1$ (or ${e}_1^T{\hat{n}}_1$), ${\hat{n}}_2\cdot{e}_0$ (or ${e}_0^T{\hat{n}}_2$), and
${\hat{n}}_2\cdot{e}_2$ (or ${e}_2^T{\hat{n}}_2$), which we recognize must be zero by the orthogonality of vectors resulting from a cross product, as $n_1={e}_0\times{e}_1$ and $n_2=-{e}_0\times{e}_2$. This gives
\begin{align*}
&\nabla_{{x}_0}\big({C}_{11}-{C}_{01}+{C}_{22}-{C}_{02}\big) \\
&= \Bigg(\frac{\big[{\hat{n}}_1,{\hat{n}}_2,{e}_1\big]-\big[{\hat{n}}_1,{\hat{n}}_2,{e}_0\big]}{\|{n}_1\|}\Bigg)\Big(\nabla_{{x}_0}{\hat{n}}_1 + \big[\nabla_{{x}_0}{\hat{n}}_1\big]^T\Big) + \Bigg(\frac{\big[{\hat{n}}_1,{\hat{n}}_2,{e}_2\big]-\big[{\hat{n}}_1,{\hat{n}}_2,{e}_0\big]}{\|{n}_2\|}\Bigg)\Big(\nabla_{{x}_0}{\hat{n}}_2 + \big[\nabla_{{x}_0}{\hat{n}}_2\big]^T\Big) \\
&+ \frac{\big({e}_0\cdot{e}_1\big)\big({\hat{n}}_1\cdot{\hat{n}}_2\big)}{\|{n}_1\|\|{n}_2\|}{\hat{n}}_1{\hat{n}}_2^T - \frac{\big({e}_1\cdot{e}_2\big)\big({\hat{n}}_1\cdot{\hat{n}}_2\big)}{\|{n}_1\|\|{n}_2\|}{\hat{n}}_1{\hat{n}}_2^T + \frac{\big({\hat{n}}_2\cdot{e}_1\big)\big({\hat{n}}_1\cdot{e}_2\big)}{\|{n}_1\|\|{n}_2\|}{\hat{n}}_1{\hat{n}}_2^T \\
&+ \frac{\big({e}_0\cdot{e}_1\big)\big({\hat{n}}_1\cdot{\hat{n}}_2\big)}{\|{n}_1\|^2}{\hat{n}}_1{\hat{n}}_1^T - \frac{\|{e}_1\|^2\big({\hat{n}}_1\cdot{\hat{n}}_2\big)}{\|{n}_1\|^2}{\hat{n}}_1{\hat{n}}_1^T - \frac{\|{e}_0\|^2\big({\hat{n}}_1\cdot{\hat{n}}_2\big)}{\|{n}_1\|\|{n}_2\|}{\hat{n}}_1{\hat{n}}_2^T \\
&+ \frac{\big({e}_0\cdot{e}_2\big)\big({\hat{n}}_1\cdot{\hat{n}}_2\big)}{\|{n}_1\|\|{n}_2\|}{\hat{n}}_1{\hat{n}}_2^T - \frac{\|{e}_0\|^2\big({\hat{n}}_1\cdot{\hat{n}}_2\big)}{\|{n}_1\|^2}{\hat{n}}_1{\hat{n}}_1^T + \frac{\big({e}_0\cdot{e}_1\big)\big({\hat{n}}_1\cdot{\hat{n}}_2\big)}{\|{n}_1\|^2}{\hat{n}}_1{\hat{n}}_1^T \\
&+ \frac{\big({e}_0\cdot{e}_2\big)\big({\hat{n}}_1\cdot{\hat{n}}_2\big)}{\|{n}_2\|^2}{\hat{n}}_2{\hat{n}}_2^T - \frac{\|{e}_2\|^2\big({\hat{n}}_1\cdot{\hat{n}}_2\big)}{\|{n}_2\|^2}{\hat{n}}_2{\hat{n}}_2^T + \frac{\big({e}_0\cdot{e}_2\big)\big({\hat{n}}_1\cdot{\hat{n}}_2\big)}{\|{n}_1\|\|{n}_2\|}{\hat{n}}_2{\hat{n}}_1^T \\
&- \frac{\big({e}_1\cdot{e}_2\big)\big({\hat{n}}_1\cdot{\hat{n}}_2\big)}{\|{n}_1\|\|{n}_2\|}{\hat{n}}_2{\hat{n}}_1^T + \frac{\big({\hat{n}}_1\cdot{e}_2\big)\big({\hat{n}}_2\cdot{e}_1\big)}{\|{n}_1\|\|{n}_2\|}{\hat{n}}_2{\hat{n}}_1^T - \frac{\|{e}_0\|^2\big({\hat{n}}_1\cdot{\hat{n}}_2\big)}{\|{n}_2\|^2}{\hat{n}}_2{\hat{n}}_2^T \\
&+ \frac{\big({e}_0\cdot{e}_2\big)\big({\hat{n}}_1\cdot{\hat{n}}_2\big)}{\|{n}_2\|^2}{\hat{n}}_2{\hat{n}}_2^T - \frac{\|{e}_0\|^2\big({\hat{n}}_1\cdot{\hat{n}}_2\big)}{\|{n}_1\|\|{n}_2\|}{\hat{n}}_2{\hat{n}}_1^T + \frac{\big({e}_0\cdot{e}_1\big)\big({\hat{n}}_1\cdot{\hat{n}}_2\big)}{\|{n}_1\|\|{n}_2\|}{\hat{n}}_2{\hat{n}}_1^T
\end{align*}
\begin{align*}
&= \Bigg(\frac{\big[{\hat{n}}_1,{\hat{n}}_2,{e}_1\big]-\big[{\hat{n}}_1,{\hat{n}}_2,{e}_0\big]}{\|{n}_1\|}\Bigg)\Big(\nabla_{{x}_0}{\hat{n}}_1 + \big[\nabla_{{x}_0}{\hat{n}}_1\big]^T\Big) + \Bigg(\frac{\big[{\hat{n}}_1,{\hat{n}}_2,{e}_2\big]-\big[{\hat{n}}_1,{\hat{n}}_2,{e}_0\big]}{\|{n}_2\|}\Bigg)\Big(\nabla_{{x}_0}{\hat{n}}_2 + \big[\nabla_{{x}_0}{\hat{n}}_2\big]^T\Big) \\
&+ \Bigg(\frac{2\big({e}_0\cdot{e}_1\big)\big({\hat{n}}_1\cdot{\hat{n}}_2\big)-\|{e}_0\|^2\big({\hat{n}}_1\cdot{\hat{n}}_2\big)-\|{e}_1\|^2\big({\hat{n}}_1\cdot{\hat{n}}_2\big)}{\|{n}_1\|^2}\Bigg){\hat{n}}_1{\hat{n}}_1^T \\
&+ \Bigg(\frac{2\big({e}_0\cdot{e}_2\big)\big({\hat{n}}_1\cdot{\hat{n}}_2\big)-\|{e}_0\|^2\big({\hat{n}}_1\cdot{\hat{n}}_2\big)-\|{e}_2\|^2\big({\hat{n}}_1\cdot{\hat{n}}_2\big)}{\|{n}_2\|^2}\Bigg){\hat{n}}_2{\hat{n}}_2^T \\
&+ \Bigg(\frac{\big({\hat{n}}_2\cdot{e}_1\big)\big({\hat{n}}_1\cdot{e}_2\big)-\|{e}_0\|^2\big({\hat{n}}_1\cdot{\hat{n}}_2\big)}{\|{n}_1\|\|{n}_2\|} \\
&+ \frac{\big({e}_0\cdot{e}_1\big)\big({\hat{n}}_1\cdot{\hat{n}}_2\big) + \big({e}_0\cdot{e}_2\big)\big({\hat{n}}_1\cdot{\hat{n}}_2\big) - \big({e}_1\cdot{e}_2\big)\big({\hat{n}}_1\cdot{\hat{n}}_2\big)}{\|{n}_1\|\|{n}_2\|}\Bigg)\Big({\hat{n}}_1{\hat{n}}_2^T+{\hat{n}}_2{\hat{n}}_1^T\Big)
\end{align*}
\begin{align*}
&= \frac{\big[{\hat{n}}_1,{\hat{n}}_2,{e}_0\big]}{\|{n}_1\|}\Bigg(\frac{\big({e}_0\cdot{e}_1\big)}{\|{e}_0\|^2}-1\Bigg)\Big(\nabla_{{x}_0}{\hat{n}}_1 + \big[\nabla_{{x}_0}{\hat{n}}_1\big]^T\Big) + \frac{\big[{\hat{n}}_1,{\hat{n}}_2,{e}_0\big]}{\|{n}_2\|}\Bigg(\frac{\big({e}_0\cdot{e}_2\big)}{\|{e}_0\|^2}-1\Bigg)\Big(\nabla_{{x}_0}{\hat{n}}_2 + \big[\nabla_{{x}_0}{\hat{n}}_2\big]^T\Big) \\
&+ \Bigg(\frac{2\big({e}_0\cdot{e}_1\big)\big({\hat{n}}_1\cdot{\hat{n}}_2\big)-\|{e}_0\|^2\big({\hat{n}}_1\cdot{\hat{n}}_2\big)-\|{e}_1\|^2\big({\hat{n}}_1\cdot{\hat{n}}_2\big)}{\|{n}_1\|^2}\Bigg){\hat{n}}_1{\hat{n}}_1^T \\
&+ \Bigg(\frac{2\big({e}_0\cdot{e}_2\big)\big({\hat{n}}_1\cdot{\hat{n}}_2\big)-\|{e}_0\|^2\big({\hat{n}}_1\cdot{\hat{n}}_2\big)-\|{e}_2\|^2\big({\hat{n}}_1\cdot{\hat{n}}_2\big)}{\|{n}_2\|^2}\Bigg){\hat{n}}_2{\hat{n}}_2^T \\
&+ \Bigg(\frac{\big({\hat{n}}_2\cdot{e}_1\big)\big({\hat{n}}_1\cdot{e}_2\big)-\|{e}_0\|^2\big({\hat{n}}_1\cdot{\hat{n}}_2\big)}{\|{n}_1\|\|{n}_2\|} \\
&+ \frac{\big({e}_0\cdot{e}_1\big)\big({\hat{n}}_1\cdot{\hat{n}}_2\big) + \big({e}_0\cdot{e}_2\big)\big({\hat{n}}_1\cdot{\hat{n}}_2\big) - \big({e}_1\cdot{e}_2\big)\big({\hat{n}}_1\cdot{\hat{n}}_2\big)}{\|{n}_1\|\|{n}_2\|}\Bigg)\Big({\hat{n}}_1{\hat{n}}_2^T+{\hat{n}}_2{\hat{n}}_1^T\Big)
\end{align*}
\begin{align*}
&= \frac{\big[{\hat{n}}_1,{\hat{n}}_2,{e}_0\big]}{\|{e}_0\|^2\|{n}_1\|}\Big(\big({e}_0\cdot{e}_1\big)-\|{e}_0\|^2\Big)\Bigg[-\frac{\big({e}_0\times{\hat{n}}_1\big){\hat{n}}_1^T}{\|{n}_1\|}+\frac{\big({e}_1\times{\hat{n}}_1\big){\hat{n}}_1^T}{\|{n}_1\|}-\frac{{\hat{n}}_1\big({e}_0\times{\hat{n}}_1\big)^T}{\|{n}_1\|}+\frac{{\hat{n}}_1\big({e}_1\times{\hat{n}}_1\big)^T}{\|{n}_1\|}\Bigg] \\
&+ \frac{\big[{\hat{n}}_1,{\hat{n}}_2,{e}_0\big]}{\|{e}_0\|^2\|{n}_2\|}\Big(\big({e}_0\cdot{e}_2\big)-\|{e}_0\|^2\Big)\Bigg[\frac{\big({e}_0\times{\hat{n}}_2\big){\hat{n}}_2^T}{\|{n}_2\|}-\frac{\big({e}_2\times{\hat{n}}_2\big){\hat{n}}_2^T}{\|{n}_2\|}+\frac{{\hat{n}}_2\big({e}_0\times{\hat{n}}_2\big)^T}{\|{n}_2\|}-\frac{{\hat{n}}_2\big({e}_2\times{\hat{n}}_2\big)^T}{\|{n}_2\|}\Bigg] \\
&+ \Bigg(\frac{2\big({e}_0\cdot{e}_1\big)\big({\hat{n}}_1\cdot{\hat{n}}_2\big)-\|{e}_0\|^2\big({\hat{n}}_1\cdot{\hat{n}}_2\big)-\|{e}_1\|^2\big({\hat{n}}_1\cdot{\hat{n}}_2\big)}{\|{n}_1\|^2}\Bigg){\hat{n}}_1{\hat{n}}_1^T \\
&+ \Bigg(\frac{2\big({e}_0\cdot{e}_2\big)\big({\hat{n}}_1\cdot{\hat{n}}_2\big)-\|{e}_0\|^2\big({\hat{n}}_1\cdot{\hat{n}}_2\big)-\|{e}_2\|^2\big({\hat{n}}_1\cdot{\hat{n}}_2\big)}{\|{n}_2\|^2}\Bigg){\hat{n}}_2{\hat{n}}_2^T \\
&+ \Bigg(\frac{\big({\hat{n}}_2\cdot{e}_1\big)\big({\hat{n}}_1\cdot{e}_2\big)-\|{e}_0\|^2\big({\hat{n}}_1\cdot{\hat{n}}_2\big)}{\|{n}_1\|\|{n}_2\|} \\
&+ \frac{\big({e}_0\cdot{e}_1\big)\big({\hat{n}}_1\cdot{\hat{n}}_2\big) + \big({e}_0\cdot{e}_2\big)\big({\hat{n}}_1\cdot{\hat{n}}_2\big) - \big({e}_1\cdot{e}_2\big)\big({\hat{n}}_1\cdot{\hat{n}}_2\big)}{\|{n}_1\|\|{n}_2\|}\Bigg)\Big({\hat{n}}_1{\hat{n}}_2^T+{\hat{n}}_2{\hat{n}}_1^T\Big).
\end{align*}
We thus obtain the final result,
\begin{align*}
&\nabla_{{x}_0}\big({C}_{11}-{C}_{01}+{C}_{22}-{C}_{02}\big) \\
&= \frac{\big[{\hat{n}}_1,{\hat{n}}_2,{e}_0\big]}{\|{e}_0\|^2\|{n}_1\|^3}\Big(\big({e}_0\cdot{e}_1\big)-\|{e}_0\|^2\Big)\Bigg[\Big(\|{e}_1\|^2-\big({e}_0\cdot{e}_1\big)\Big){e}_0{\hat{n}}_1^T+\Big(\|{e}_0\|^2-\big({e}_0\cdot{e}_1\big)\Big){e}_1{\hat{n}}_1^T \\
&+\Big(\|{e}_1\|^2-\big({e}_0\cdot{e}_1\big)\Big){\hat{n}}_1{e}_0^T+\Big(\|{e}_0\|^2-\big({e}_0\cdot{e}_1\big)\Big){\hat{n}}_1{e}_1^T\Bigg] \\
&+ \frac{\big[{\hat{n}}_1,{\hat{n}}_2,{e}_0\big]}{\|{e}_0\|^2\|{n}_2\|^3}\Big(\big({e}_0\cdot{e}_2\big)-\|{e}_0\|^2\Big)\Bigg[\Big(\|{e}_2\|^2-\big({e}_0\cdot{e}_2\big)\Big){e}_0{\hat{n}}_2^T+\Big(\|{e}_0\|^2-\big({e}_0\cdot{e}_2\big)\Big){e}_2{\hat{n}}_2^T \\
&+\Big(\|{e}_2\|^2-\big({e}_0\cdot{e}_2\big)\Big){\hat{n}}_2{e}_0^T+\Big(\|{e}_0\|^2-\big({e}_0\cdot{e}_2\big)\Big){\hat{n}}_2{e}_2^T\Bigg] \\
&+ \Bigg(\frac{2\big({e}_0\cdot{e}_1\big)\big({\hat{n}}_1\cdot{\hat{n}}_2\big)-\|{e}_0\|^2\big({\hat{n}}_1\cdot{\hat{n}}_2\big)-\|{e}_1\|^2\big({\hat{n}}_1\cdot{\hat{n}}_2\big)}{\|{n}_1\|^2}\Bigg){\hat{n}}_1{\hat{n}}_1^T \\
&+ \Bigg(\frac{2\big({e}_0\cdot{e}_2\big)\big({\hat{n}}_1\cdot{\hat{n}}_2\big)-\|{e}_0\|^2\big({\hat{n}}_1\cdot{\hat{n}}_2\big)-\|{e}_2\|^2\big({\hat{n}}_1\cdot{\hat{n}}_2\big)}{\|{n}_2\|^2}\Bigg){\hat{n}}_2{\hat{n}}_2^T \\
&+ \Bigg(\frac{\big({\hat{n}}_2\cdot{e}_1\big)\big({\hat{n}}_1\cdot{e}_2\big)-\|{e}_0\|^2\big({\hat{n}}_1\cdot{\hat{n}}_2\big)}{\|{n}_1\|\|{n}_2\|} \\
&+ \frac{\big({e}_0\cdot{e}_1\big)\big({\hat{n}}_1\cdot{\hat{n}}_2\big) + \big({e}_0\cdot{e}_2\big)\big({\hat{n}}_1\cdot{\hat{n}}_2\big) - \big({e}_1\cdot{e}_2\big)\big({\hat{n}}_1\cdot{\hat{n}}_2\big)}{\|{n}_1\|\|{n}_2\|}\Bigg)\Big({\hat{n}}_1{\hat{n}}_2^T+{\hat{n}}_2{\hat{n}}_1^T\Big).
\end{align*}
Note that all terms in this expression are symmetric or come in transpose pairs, validating the expected symmetry relation.

\subsubsection{\texorpdfstring{$\nabla_{{x}_1}\big({C}_{11}-{C}_{01}+{C}_{22}-{C}_{02}\big)$}{A2}}
Moving on to the next gradient term,
\begin{align*}
&\nabla_{{x}_1}\big({C}_{11}-{C}_{01}+{C}_{22}-{C}_{02}\big) \\
&= \frac{\big[{\hat{n}}_1,{\hat{n}}_2,{e}_1\big]}{\|{n}_1\|}\Big(\nabla_{{x}_1}{\hat{n}}_1 + \big[\nabla_{{x}_1}{\hat{n}}_1\big]^T\Big)+\frac{{\hat{n}}_1}{\|{n}_1\|}{e}_1^T\big({\hat{n}}_1\times\nabla_{{x}_1}{\hat{n}}_2-{\hat{n}}_2\times\nabla_{{x}_1}{\hat{n}}_1\big)+\frac{{\hat{n}}_1}{\|{n}_1\|}\big({\hat{n}}_1\times{\hat{n}}_2\big)^T\nabla_{{x}_1}{e}_1 \\
&- \frac{\big[{\hat{n}}_1,{\hat{n}}_2,{e}_0\big]}{\|{n}_1\|}\Big(\nabla_{{x}_1}{\hat{n}}_1 + \big[\nabla_{{x}_1}{\hat{n}}_1\big]^T\Big)-\frac{{\hat{n}}_1}{\|{n}_1\|}{e}_0^T\big({\hat{n}}_1\times\nabla_{{x}_1}{\hat{n}}_2-{\hat{n}}_2\times\nabla_{{x}_1}{\hat{n}}_1\big)-\frac{{\hat{n}}_1}{\|{n}_1\|}\big({\hat{n}}_1\times{\hat{n}}_2\big)^T\nabla_{{x}_1}{e}_0 \\
&+ \frac{\big[{\hat{n}}_1,{\hat{n}}_2,{e}_2\big]}{\|{n}_2\|}\Big(\nabla_{{x}_1}{\hat{n}}_2 + \big[\nabla_{{x}_1}{\hat{n}}_2\big]^T\Big)+\frac{{\hat{n}}_2}{\|{n}_2\|}{e}_2^T\big({\hat{n}}_1\times\nabla_{{x}_1}{\hat{n}}_2-{\hat{n}}_2\times\nabla_{{x}_1}{\hat{n}}_1\big)+\frac{{\hat{n}}_2}{\|{n}_2\|}\big({\hat{n}}_1\times{\hat{n}}_2\big)^T\nabla_{{x}_1}{e}_2 \\
&- \frac{\big[{\hat{n}}_1,{\hat{n}}_2,{e}_0\big]}{\|{n}_2\|}\Big(\nabla_{{x}_1}{\hat{n}}_2 + \big[\nabla_{{x}_1}{\hat{n}}_2\big]^T\Big)-\frac{{\hat{n}}_2}{\|{n}_2\|}{e}_0^T\big({\hat{n}}_1\times\nabla_{{x}_1}{\hat{n}}_2-{\hat{n}}_2\times\nabla_{{x}_1}{\hat{n}}_1\big)-\frac{{\hat{n}}_2}{\|{n}_2\|}\big({\hat{n}}_1\times{\hat{n}}_2\big)^T\nabla_{{x}_1}{e}_0
\end{align*}
\begin{align*}
&= \Bigg(\frac{\big[{\hat{n}}_1,{\hat{n}}_2,{e}_1\big]-\big[{\hat{n}}_1,{\hat{n}}_2,{e}_0\big]}{\|{n}_1\|}\Bigg)\Big(\nabla_{{x}_1}{\hat{n}}_1 + \big[\nabla_{{x}_1}{\hat{n}}_1\big]^T\Big) + \Bigg(\frac{\big[{\hat{n}}_1,{\hat{n}}_2,{e}_2\big]-\big[{\hat{n}}_1,{\hat{n}}_2,{e}_0\big]}{\|{n}_2\|}\Bigg)\Big(\nabla_{{x}_1}{\hat{n}}_2 + \big[\nabla_{{x}_1}{\hat{n}}_2\big]^T\Big) \\
&+ \frac{{\hat{n}}_1}{\|{n}_1\|}{e}_1^T\Bigg(\frac{{\hat{n}}_1\times\big({e}_2\times{\hat{n}}_2\big){\hat{n}}_2^T}{\|{n}_2\|}+\frac{{\hat{n}}_2\times\big({e}_1\times{\hat{n}}_1\big){\hat{n}}_1^T}{\|{n}_1\|}\Bigg) \\
&- \frac{{\hat{n}}_1}{\|{n}_1\|}{e}_0^T\Bigg(\frac{{\hat{n}}_1\times\big({e}_2\times{\hat{n}}_2\big){\hat{n}}_2^T}{\|{n}_2\|}+\frac{{\hat{n}}_2\times\big({e}_1\times{\hat{n}}_1\big){\hat{n}}_1^T}{\|{n}_1\|}\Bigg) - \frac{{\hat{n}}_1}{\|{n}_1\|}\big({\hat{n}}_1\times{\hat{n}}_2\big)^T \\
&+ \frac{{\hat{n}}_2}{\|{n}_2\|}{e}_2^T\Bigg(\frac{{\hat{n}}_1\times\big({e}_2\times{\hat{n}}_2\big){\hat{n}}_2^T}{\|{n}_2\|}+\frac{{\hat{n}}_2\times\big({e}_1\times{\hat{n}}_1\big){\hat{n}}_1^T}{\|{n}_1\|}\Bigg) \\
&- \frac{{\hat{n}}_2}{\|{n}_2\|}{e}_0^T\Bigg(\frac{{\hat{n}}_1\times\big({e}_2\times{\hat{n}}_2\big){\hat{n}}_2^T}{\|{n}_2\|}+\frac{{\hat{n}}_2\times\big({e}_1\times{\hat{n}}_1\big){\hat{n}}_1^T}{\|{n}_1\|}\Bigg) - \frac{{\hat{n}}_2}{\|{n}_2\|}\big({\hat{n}}_1\times{\hat{n}}_2\big)^T
\end{align*}
\begin{align*}
&= \Bigg(\frac{\big[{\hat{n}}_1,{\hat{n}}_2,{e}_1\big]-\big[{\hat{n}}_1,{\hat{n}}_2,{e}_0\big]}{\|{n}_1\|}\Bigg)\Big(\nabla_{{x}_1}{\hat{n}}_1 + \big[\nabla_{{x}_1}{\hat{n}}_1\big]^T\Big) + \Bigg(\frac{\big[{\hat{n}}_1,{\hat{n}}_2,{e}_2\big]-\big[{\hat{n}}_1,{\hat{n}}_2,{e}_0\big]}{\|{n}_2\|}\Bigg)\Big(\nabla_{{x}_1}{\hat{n}}_2 + \big[\nabla_{{x}_1}{\hat{n}}_2\big]^T\Big) \\
&+ \frac{\big({e}_1\cdot{e}_2\big)\big({\hat{n}}_1\cdot{\hat{n}}_2\big)}{\|{n}_1\|\|{n}_2\|}{\hat{n}}_1{\hat{n}}_2^T - \frac{\big({\hat{n}}_2\cdot{e}_1\big)\big({\hat{n}}_1\cdot{e}_2\big)}{\|{n}_1\|\|{n}_2\|}{\hat{n}}_1{\hat{n}}_2^T + \frac{\|{e}_1\|^2\big({\hat{n}}_1\cdot{\hat{n}}_2\big)}{\|{n}_1\|^2}{\hat{n}}_1{\hat{n}}_1^T \\
&- \frac{\big({e}_0\cdot{e}_2\big)\big({\hat{n}}_1\cdot{\hat{n}}_2\big)}{\|{n}_1\|\|{n}_2\|}{\hat{n}}_1{\hat{n}}_2^T - \frac{\big({e}_0\cdot{e}_1\big)\big({\hat{n}}_1\cdot{\hat{n}}_2\big)}{\|{n}_1\|^2}{\hat{n}}_1{\hat{n}}_1^T - \frac{{\hat{n}}_1}{\|{n}_1\|}\big({\hat{n}}_1\times{\hat{n}}_2\big)^T \\
&+ \frac{\|{e}_2\|^2\big({\hat{n}}_1\cdot{\hat{n}}_2\big)}{\|{n}_2\|^2}{\hat{n}}_2{\hat{n}}_2^T + \frac{\big({e}_1\cdot{e}_2\big)\big({\hat{n}}_1\cdot{\hat{n}}_2\big)}{\|{n}_1\|\|{n}_2\|}{\hat{n}}_2{\hat{n}}_1^T - \frac{\big({\hat{n}}_2\cdot{e}_1\big)\big({\hat{n}}_1\cdot{e}_2\big)}{\|{n}_1\|\|{n}_2\|}{\hat{n}}_2{\hat{n}}_1^T \\
&- \frac{\big({e}_0\cdot{e}_2\big)\big({\hat{n}}_1\cdot{\hat{n}}_2\big)}{\|{n}_2\|^2}{\hat{n}}_2{\hat{n}}_2^T - \frac{\big({e}_0\cdot{e}_1\big)\big({\hat{n}}_1\cdot{\hat{n}}_2\big)}{\|{n}_1\|\|{n}_2\|}{\hat{n}}_2{\hat{n}}_1^T - \frac{{\hat{n}}_2}{\|{n}_2\|}\big({\hat{n}}_1\times{\hat{n}}_2\big)^T
\end{align*}
\begin{align*}
&= \big[{\hat{n}}_1,{\hat{n}}_2,{e}_0\big]\Bigg(\frac{\big({e}_0\cdot{e}_1\big)-\|{e}_0\|^2}{\|{e}_0\|^2\|{n}_1\|}\nabla_{{x}_1}{\hat{n}}_1 + \frac{\big({e}_0\cdot{e}_1\big)-\|{e}_0\|^2}{\|{e}_0\|^2\|{n}_1\|}\big[\nabla_{{x}_1}{\hat{n}}_1\big]^T - \frac{1}{\|{e}_0\|^2\|{n}_1\|}{\hat{n}}_1{e}_0^T\\
&+ \frac{\big({e}_0\cdot{e}_2\big)-\|{e}_0\|^2}{\|{e}_0\|^2\|{n}_2\|}\nabla_{{x}_1}{\hat{n}}_2 + \frac{\big({e}_0\cdot{e}_2\big)-\|{e}_0\|^2}{\|{e}_0\|^2\|{n}_2\|}\big[\nabla_{{x}_1}{\hat{n}}_2\big]^T - \frac{1}{\|{e}_0\|^2\|{n}_2\|}{\hat{n}}_2{e}_0^T\Bigg) \\
&+ \frac{\big({e}_1\cdot{e}_2\big)\big({\hat{n}}_1\cdot{\hat{n}}_2\big)}{\|{n}_1\|\|{n}_2\|}{\hat{n}}_1{\hat{n}}_2^T - \frac{\big({\hat{n}}_2\cdot{e}_1\big)\big({\hat{n}}_1\cdot{e}_2\big)}{\|{n}_1\|\|{n}_2\|}{\hat{n}}_1{\hat{n}}_2^T + \frac{\|{e}_1\|^2\big({\hat{n}}_1\cdot{\hat{n}}_2\big)}{\|{n}_1\|^2}{\hat{n}}_1{\hat{n}}_1^T \\
&- \frac{\big({e}_0\cdot{e}_2\big)\big({\hat{n}}_1\cdot{\hat{n}}_2\big)}{\|{n}_1\|\|{n}_2\|}{\hat{n}}_1{\hat{n}}_2^T - \frac{\big({e}_0\cdot{e}_1\big)\big({\hat{n}}_1\cdot{\hat{n}}_2\big)}{\|{n}_1\|^2}{\hat{n}}_1{\hat{n}}_1^T \\
&+ \frac{\|{e}_2\|^2\big({\hat{n}}_1\cdot{\hat{n}}_2\big)}{\|{n}_2\|^2}{\hat{n}}_2{\hat{n}}_2^T + \frac{\big({e}_1\cdot{e}_2\big)\big({\hat{n}}_1\cdot{\hat{n}}_2\big)}{\|{n}_1\|\|{n}_2\|}{\hat{n}}_2{\hat{n}}_1^T - \frac{\big({\hat{n}}_2\cdot{e}_1\big)\big({\hat{n}}_1\cdot{e}_2\big)}{\|{n}_1\|\|{n}_2\|}{\hat{n}}_2{\hat{n}}_1^T \\
&- \frac{\big({e}_0\cdot{e}_2\big)\big({\hat{n}}_1\cdot{\hat{n}}_2\big)}{\|{n}_2\|^2}{\hat{n}}_2{\hat{n}}_2^T - \frac{\big({e}_0\cdot{e}_1\big)\big({\hat{n}}_1\cdot{\hat{n}}_2\big)}{\|{n}_1\|\|{n}_2\|}{\hat{n}}_2{\hat{n}}_1^T
\end{align*}
\begin{align*}
&= \big[{\hat{n}}_1,{\hat{n}}_2,{e}_0\big]\Bigg(\frac{\|{e}_0\|^2-\big({e}_0\cdot{e}_1\big)}{\|{e}_0\|^2\|{n}_1\|^2}\big({e}_1\times{\hat{n}}_1\big){\hat{n}}_1^T + \frac{\|{e}_0\|^2-\big({e}_0\cdot{e}_1\big)}{\|{e}_0\|^2\|{n}_1\|^2}{\hat{n}}_1\big({e}_1\times{\hat{n}}_1\big)^T - \frac{1}{\|{e}_0\|^2\|{n}_1\|}{\hat{n}}_1{e}_0^T\\
&+ \frac{\big({e}_0\cdot{e}_2\big)-\|{e}_0\|^2}{\|{e}_0\|^2\|{n}_2\|^2}\big({e}_2\times{\hat{n}}_2\big){\hat{n}}_2^T + \frac{\big({e}_0\cdot{e}_2\big)-\|{e}_0\|^2}{\|{e}_0\|^2\|{n}_2\|^2}{\hat{n}}_2\big({e}_2\times{\hat{n}}_2\big)^T - \frac{1}{\|{e}_0\|^2\|{n}_2\|}{\hat{n}}_2{e}_0^T\Bigg) \\
&+ \frac{\big({e}_1\cdot{e}_2\big)\big({\hat{n}}_1\cdot{\hat{n}}_2\big)}{\|{n}_1\|\|{n}_2\|}{\hat{n}}_1{\hat{n}}_2^T - \frac{\big({\hat{n}}_2\cdot{e}_1\big)\big({\hat{n}}_1\cdot{e}_2\big)}{\|{n}_1\|\|{n}_2\|}{\hat{n}}_1{\hat{n}}_2^T + \frac{\|{e}_1\|^2\big({\hat{n}}_1\cdot{\hat{n}}_2\big)}{\|{n}_1\|^2}{\hat{n}}_1{\hat{n}}_1^T \\
&- \frac{\big({e}_0\cdot{e}_2\big)\big({\hat{n}}_1\cdot{\hat{n}}_2\big)}{\|{n}_1\|\|{n}_2\|}{\hat{n}}_1{\hat{n}}_2^T - \frac{\big({e}_0\cdot{e}_1\big)\big({\hat{n}}_1\cdot{\hat{n}}_2\big)}{\|{n}_1\|^2}{\hat{n}}_1{\hat{n}}_1^T \\
&+ \frac{\|{e}_2\|^2\big({\hat{n}}_1\cdot{\hat{n}}_2\big)}{\|{n}_2\|^2}{\hat{n}}_2{\hat{n}}_2^T + \frac{\big({e}_1\cdot{e}_2\big)\big({\hat{n}}_1\cdot{\hat{n}}_2\big)}{\|{n}_1\|\|{n}_2\|}{\hat{n}}_2{\hat{n}}_1^T - \frac{\big({\hat{n}}_2\cdot{e}_1\big)\big({\hat{n}}_1\cdot{e}_2\big)}{\|{n}_1\|\|{n}_2\|}{\hat{n}}_2{\hat{n}}_1^T \\
&- \frac{\big({e}_0\cdot{e}_2\big)\big({\hat{n}}_1\cdot{\hat{n}}_2\big)}{\|{n}_2\|^2}{\hat{n}}_2{\hat{n}}_2^T - \frac{\big({e}_0\cdot{e}_1\big)\big({\hat{n}}_1\cdot{\hat{n}}_2\big)}{\|{n}_1\|\|{n}_2\|}{\hat{n}}_2{\hat{n}}_1^T
\end{align*}
\begin{align*}
&= \big[{\hat{n}}_1,{\hat{n}}_2,{e}_0\big]\Bigg(\frac{\|{e}_0\|^2\|{e}_1\|^2-\big({e}_0\cdot{e}_1\big)\|{e}_1\|^2}{\|{e}_0\|^2\|{n}_1\|^3}{e}_0{\hat{n}}_1^T + \frac{\big({e}_0\cdot{e}_1\big)^2-\big({e}_0\cdot{e}_1\big)\|{e}_0\|^2}{\|{e}_0\|^2\|{n}_1\|^3}{e}_1{\hat{n}}_1^T \\
&+ \frac{\|{e}_0\|^2\|{e}_1\|^2-\big({e}_0\cdot{e}_1\big)\|{e}_1\|^2-\|{n}_1\|^2}{\|{e}_0\|^2\|{n}_1\|^3}{\hat{n}}_1{e}_0^T + \frac{\big({e}_0\cdot{e}_1\big)^2-\big({e}_0\cdot{e}_1\big)\|{e}_0\|^2}{\|{e}_0\|^2\|{n}_1\|^3}{\hat{n}}_1{e}_1^T \\
&+ \frac{\|{e}_0\|^2\|{e}_2\|^2-\big({e}_0\cdot{e}_2\big)\|{e}_2\|^2}{\|{e}_0\|^2\|{n}_2\|^3}{e}_0{\hat{n}}_2^T + \frac{\big({e}_0\cdot{e}_2\big)^2-\big({e}_0\cdot{e}_2\big)\|{e}_0\|^2}{\|{e}_0\|^2\|{n}_2\|^3}{e}_2{\hat{n}}_2^T \\
&+ \frac{\|{e}_0\|^2\|{e}_2\|^2-\big({e}_0\cdot{e}_2\big)\|{e}_2\|^2-\|{n}_2\|^2}{\|{e}_0\|^2\|{n}_2\|^3}{\hat{n}}_2{e}_0^T + \frac{\big({e}_0\cdot{e}_2\big)^2-\big({e}_0\cdot{e}_2\big)\|{e}_0\|^2}{\|{e}_0\|^2\|{n}_2\|^3}{\hat{n}}_2{e}_2^T
\Bigg) \\
&+ \frac{\big({e}_1\cdot{e}_2\big)\big({\hat{n}}_1\cdot{\hat{n}}_2\big)}{\|{n}_1\|\|{n}_2\|}{\hat{n}}_1{\hat{n}}_2^T - \frac{\big({\hat{n}}_2\cdot{e}_1\big)\big({\hat{n}}_1\cdot{e}_2\big)}{\|{n}_1\|\|{n}_2\|}{\hat{n}}_1{\hat{n}}_2^T + \frac{\|{e}_1\|^2\big({\hat{n}}_1\cdot{\hat{n}}_2\big)}{\|{n}_1\|^2}{\hat{n}}_1{\hat{n}}_1^T \\
&- \frac{\big({e}_0\cdot{e}_2\big)\big({\hat{n}}_1\cdot{\hat{n}}_2\big)}{\|{n}_1\|\|{n}_2\|}{\hat{n}}_1{\hat{n}}_2^T - \frac{\big({e}_0\cdot{e}_1\big)\big({\hat{n}}_1\cdot{\hat{n}}_2\big)}{\|{n}_1\|^2}{\hat{n}}_1{\hat{n}}_1^T \\
&+ \frac{\|{e}_2\|^2\big({\hat{n}}_1\cdot{\hat{n}}_2\big)}{\|{n}_2\|^2}{\hat{n}}_2{\hat{n}}_2^T + \frac{\big({e}_1\cdot{e}_2\big)\big({\hat{n}}_1\cdot{\hat{n}}_2\big)}{\|{n}_1\|\|{n}_2\|}{\hat{n}}_2{\hat{n}}_1^T - \frac{\big({\hat{n}}_2\cdot{e}_1\big)\big({\hat{n}}_1\cdot{e}_2\big)}{\|{n}_1\|\|{n}_2\|}{\hat{n}}_2{\hat{n}}_1^T \\
&- \frac{\big({e}_0\cdot{e}_2\big)\big({\hat{n}}_1\cdot{\hat{n}}_2\big)}{\|{n}_2\|^2}{\hat{n}}_2{\hat{n}}_2^T - \frac{\big({e}_0\cdot{e}_1\big)\big({\hat{n}}_1\cdot{\hat{n}}_2\big)}{\|{n}_1\|\|{n}_2\|}{\hat{n}}_2{\hat{n}}_1^T,
\end{align*}
which yields the final result,
\begin{align*}
&\nabla_{{x}_1}\big({C}_{11}-{C}_{01}+{C}_{22}-{C}_{02}\big) \\
&= \big[{\hat{n}}_1,{\hat{n}}_2,{e}_0\big]\Bigg(\frac{\|{e}_0\|^2\|{e}_1\|^2-\big({e}_0\cdot{e}_1\big)\|{e}_1\|^2}{\|{e}_0\|^2\|{n}_1\|^3}{e}_0{\hat{n}}_1^T + \frac{\big({e}_0\cdot{e}_1\big)^2-\big({e}_0\cdot{e}_1\big)\|{e}_0\|^2}{\|{e}_0\|^2\|{n}_1\|^3}{e}_1{\hat{n}}_1^T \\
&+ \frac{\big({e}_0\cdot{e}_1\big)^2-\big({e}_0\cdot{e}_1\big)\|{e}_1\|^2}{\|{e}_0\|^2\|{n}_1\|^3}{\hat{n}}_1{e}_0^T + \frac{\big({e}_0\cdot{e}_1\big)^2-\big({e}_0\cdot{e}_1\big)\|{e}_0\|^2}{\|{e}_0\|^2\|{n}_1\|^3}{\hat{n}}_1{e}_1^T \\
&+ \frac{\|{e}_0\|^2\|{e}_2\|^2-\big({e}_0\cdot{e}_2\big)\|{e}_2\|^2}{\|{e}_0\|^2\|{n}_2\|^3}{e}_0{\hat{n}}_2^T + \frac{\big({e}_0\cdot{e}_2\big)^2-\big({e}_0\cdot{e}_2\big)\|{e}_0\|^2}{\|{e}_0\|^2\|{n}_2\|^3}{e}_2{\hat{n}}_2^T \\
&+ \frac{\big({e}_0\cdot{e}_2\big)^2-\big({e}_0\cdot{e}_2\big)\|{e}_2\|^2}{\|{e}_0\|^2\|{n}_2\|^3}{\hat{n}}_2{e}_0^T + \frac{\big({e}_0\cdot{e}_2\big)^2-\big({e}_0\cdot{e}_2\big)\|{e}_0\|^2}{\|{e}_0\|^2\|{n}_2\|^3}{\hat{n}}_2{e}_2^T
\Bigg) \\
&+ \frac{\big({e}_1\cdot{e}_2\big)\big({\hat{n}}_1\cdot{\hat{n}}_2\big)}{\|{n}_1\|\|{n}_2\|}{\hat{n}}_1{\hat{n}}_2^T - \frac{\big({\hat{n}}_2\cdot{e}_1\big)\big({\hat{n}}_1\cdot{e}_2\big)}{\|{n}_1\|\|{n}_2\|}{\hat{n}}_1{\hat{n}}_2^T + \frac{\|{e}_1\|^2\big({\hat{n}}_1\cdot{\hat{n}}_2\big)}{\|{n}_1\|^2}{\hat{n}}_1{\hat{n}}_1^T \\
&- \frac{\big({e}_0\cdot{e}_2\big)\big({\hat{n}}_1\cdot{\hat{n}}_2\big)}{\|{n}_1\|\|{n}_2\|}{\hat{n}}_1{\hat{n}}_2^T - \frac{\big({e}_0\cdot{e}_1\big)\big({\hat{n}}_1\cdot{\hat{n}}_2\big)}{\|{n}_1\|^2}{\hat{n}}_1{\hat{n}}_1^T \\
&+ \frac{\|{e}_2\|^2\big({\hat{n}}_1\cdot{\hat{n}}_2\big)}{\|{n}_2\|^2}{\hat{n}}_2{\hat{n}}_2^T + \frac{\big({e}_1\cdot{e}_2\big)\big({\hat{n}}_1\cdot{\hat{n}}_2\big)}{\|{n}_1\|\|{n}_2\|}{\hat{n}}_2{\hat{n}}_1^T - \frac{\big({\hat{n}}_2\cdot{e}_1\big)\big({\hat{n}}_1\cdot{e}_2\big)}{\|{n}_1\|\|{n}_2\|}{\hat{n}}_2{\hat{n}}_1^T \\
&- \frac{\big({e}_0\cdot{e}_2\big)\big({\hat{n}}_1\cdot{\hat{n}}_2\big)}{\|{n}_2\|^2}{\hat{n}}_2{\hat{n}}_2^T - \frac{\big({e}_0\cdot{e}_1\big)\big({\hat{n}}_1\cdot{\hat{n}}_2\big)}{\|{n}_1\|\|{n}_2\|}{\hat{n}}_2{\hat{n}}_1^T.
\end{align*}
By the required symmetry relations, this result should be the transpose of $\nabla_{{x}_0}\big(-{C}_{11}-{C}_{22}\big)$. We thus carry out the derivation of $\nabla_{{x}_0}\big(-{C}_{11}-{C}_{22}\big)$ here:
\begin{align*}
\nabla_{{x}_0}\big(-{C}_{11}-{C}_{22}\big) &= -\frac{\big[{\hat{n}}_1,{\hat{n}}_2,{e}_1\big]}{\|{n}_1\|}\Big(\nabla_{{x}_0}{\hat{n}}_1 + \big[\nabla_{{x}_0}{\hat{n}}_1\big]^T\Big) -\frac{\big[{\hat{n}}_1,{\hat{n}}_2,{e}_2\big]}{\|{n}_2\|}\Big(\nabla_{{x}_0}{\hat{n}}_2 + \big[\nabla_{{x}_0}{\hat{n}}_2\big]^T\Big) \\
&- \frac{{\hat{n}}_1}{\|{n}_1\|}{e}_1^T\Bigg(\frac{{\hat{n}}_1\times\big({e}_0\times{\hat{n}}_2\big){\hat{n}}_2^T}{\|{n}_2\|}-\frac{{\hat{n}}_1\times\big({e}_2\times{\hat{n}}_2\big){\hat{n}}_2^T}{\|{n}_2\|}+\frac{{\hat{n}}_2\times\big({e}_0\times{\hat{n}}_1\big){\hat{n}}_1^T}{\|{n}_1\|}-\frac{{\hat{n}}_2\times\big({e}_1\times{\hat{n}}_1\big){\hat{n}}_1^T}{\|{n}_1\|}\Bigg) \\
&+ \frac{{\hat{n}}_1}{\|{n}_1\|}\big({\hat{n}}_1\times{\hat{n}}_2\big)^T \\
&- \frac{{\hat{n}}_2}{\|{n}_2\|}{e}_2^T\Bigg(\frac{{\hat{n}}_1\times\big({e}_0\times{\hat{n}}_2\big){\hat{n}}_2^T}{\|{n}_2\|}-\frac{{\hat{n}}_1\times\big({e}_2\times{\hat{n}}_2\big){\hat{n}}_2^T}{\|{n}_2\|}+\frac{{\hat{n}}_2\times\big({e}_0\times{\hat{n}}_1\big){\hat{n}}_1^T}{\|{n}_1\|}-\frac{{\hat{n}}_2\times\big({e}_1\times{\hat{n}}_1\big){\hat{n}}_1^T}{\|{n}_1\|}\Bigg) \\
&+ \frac{{\hat{n}}_2}{\|{n}_2\|}\big({\hat{n}}_1\times{\hat{n}}_2\big)^T
\end{align*}
\begin{align*}
&= -\frac{\big[{\hat{n}}_1,{\hat{n}}_2,{e}_1\big]}{\|{n}_1\|}\Big(\nabla_{{x}_0}{\hat{n}}_1 + \big[\nabla_{{x}_0}{\hat{n}}_1\big]^T\Big) -\frac{\big[{\hat{n}}_1,{\hat{n}}_2,{e}_2\big]}{\|{n}_2\|}\Big(\nabla_{{x}_0}{\hat{n}}_2 + \big[\nabla_{{x}_0}{\hat{n}}_2\big]^T\Big) \\
&- \frac{\big({e}_0\cdot{e}_1\big)\big({\hat{n}}_1\cdot{\hat{n}}_2\big)}{\|{n}_1\|\|{n}_2\|}{\hat{n}}_1{\hat{n}}_2^T + \frac{\big({e}_1\cdot{e}_2\big)\big({\hat{n}}_1\cdot{\hat{n}}_2\big)}{\|{n}_1\|\|{n}_2\|}{\hat{n}}_1{\hat{n}}_2^T - \frac{\big({\hat{n}}_2\cdot{e}_1\big)\big({\hat{n}}_1\cdot{e}_2\big)}{\|{n}_1\|\|{n}_2\|}{\hat{n}}_1{\hat{n}}_2^T \\
&- \frac{\big({e}_0\cdot{e}_1\big)\big({\hat{n}}_1\cdot{\hat{n}}_2\big)}{\|{n}_1\|^2}{\hat{n}}_1{\hat{n}}_1^T + \frac{\|{e}_1\|^2\big({\hat{n}}_1\cdot{\hat{n}}_2\big)}{\|{n}_1\|^2}{\hat{n}}_1{\hat{n}}_1^T + \frac{{\hat{n}}_1}{\|{n}_1\|}\big({\hat{n}}_1\times{\hat{n}}_2\big)^T \\
&- \frac{\big({e}_0\cdot{e}_2\big)\big({\hat{n}}_1\cdot{\hat{n}}_2\big)}{\|{n}_2\|^2}{\hat{n}}_2{\hat{n}}_2^T + \frac{\|{e}_2\|^2\big({\hat{n}}_1\cdot{\hat{n}}_2\big)}{\|{n}_2\|^2}{\hat{n}}_2{\hat{n}}_2^T - \frac{\big({e}_0\cdot{e}_2\big)\big({\hat{n}}_1\cdot{\hat{n}}_2\big)}{\|{n}_1\|\|{n}_2\|}{\hat{n}}_2{\hat{n}}_1^T \\
&+ \frac{\big({e}_1\cdot{e}_2\big)\big({\hat{n}}_1\cdot{\hat{n}}_2\big)}{\|{n}_1\|\|{n}_2\|}{\hat{n}}_2{\hat{n}}_1^T - \frac{\big({\hat{n}}_1\cdot{e}_2\big)\big({\hat{n}}_2\cdot{e}_1\big)}{\|{n}_1\|\|{n}_2\|}{\hat{n}}_2{\hat{n}}_1^T + \frac{{\hat{n}}_2}{\|{n}_2\|}\big({\hat{n}}_1\times{\hat{n}}_2\big)^T
\end{align*}
\begin{align*}
&= \big[{\hat{n}}_1,{\hat{n}}_2,{e}_0\big]\Bigg(\frac{\big({e}_0\cdot{e}_1\big)}{\|{e}_0\|^2\|{n}_1\|^2}\big({e}_0\times{\hat{n}}_1\big){\hat{n}}_1^T - \frac{\big({e}_0\cdot{e}_1\big)}{\|{e}_0\|^2\|{n}_1\|^2}\big({e}_1\times{\hat{n}}_1\big){\hat{n}}_1^T \\
&+ \frac{\big({e}_0\cdot{e}_1\big)}{\|{e}_0\|^2\|{n}_1\|^2}{\hat{n}}_1\big({e}_0\times{\hat{n}}_1\big)^T - \frac{\big({e}_0\cdot{e}_1\big)}{\|{e}_0\|^2\|{n}_1\|^2}{\hat{n}}_1\big({e}_1\times{\hat{n}}_1\big)^T + \frac{1}{\|{e}_0\|^2\|{n}_1\|}{\hat{n}}_1{e}_0^T \\
&- \frac{\big({e}_0\cdot{e}_2\big)}{\|{e}_0\|^2\|{n}_2\|^2}\big({e}_0\times{\hat{n}}_2\big){\hat{n}}_2^T + \frac{\big({e}_0\cdot{e}_2\big)}{\|{e}_0\|^2\|{n}_2\|^2}\big({e}_2\times{\hat{n}}_2\big){\hat{n}}_2^T \\
&- \frac{\big({e}_0\cdot{e}_2\big)}{\|{e}_0\|^2\|{n}_2\|^2}{\hat{n}}_2\big({e}_0\times{\hat{n}}_2\big)^T + \frac{\big({e}_0\cdot{e}_2\big)}{\|{e}_0\|^2\|{n}_2\|^2}{\hat{n}}_2\big({e}_2\times{\hat{n}}_2\big)^T + \frac{1}{\|{e}_0\|^2\|{n}_2\|}{\hat{n}}_2{e}_0^T \Bigg) \\
&- \frac{\big({e}_0\cdot{e}_1\big)\big({\hat{n}}_1\cdot{\hat{n}}_2\big)}{\|{n}_1\|\|{n}_2\|}{\hat{n}}_1{\hat{n}}_2^T + \frac{\big({e}_1\cdot{e}_2\big)\big({\hat{n}}_1\cdot{\hat{n}}_2\big)}{\|{n}_1\|\|{n}_2\|}{\hat{n}}_1{\hat{n}}_2^T - \frac{\big({\hat{n}}_2\cdot{e}_1\big)\big({\hat{n}}_1\cdot{e}_2\big)}{\|{n}_1\|\|{n}_2\|}{\hat{n}}_1{\hat{n}}_2^T \\
&- \frac{\big({e}_0\cdot{e}_1\big)\big({\hat{n}}_1\cdot{\hat{n}}_2\big)}{\|{n}_1\|^2}{\hat{n}}_1{\hat{n}}_1^T + \frac{\|{e}_1\|^2\big({\hat{n}}_1\cdot{\hat{n}}_2\big)}{\|{n}_1\|^2}{\hat{n}}_1{\hat{n}}_1^T \\
&- \frac{\big({e}_0\cdot{e}_2\big)\big({\hat{n}}_1\cdot{\hat{n}}_2\big)}{\|{n}_2\|^2}{\hat{n}}_2{\hat{n}}_2^T + \frac{\|{e}_2\|^2\big({\hat{n}}_1\cdot{\hat{n}}_2\big)}{\|{n}_2\|^2}{\hat{n}}_2{\hat{n}}_2^T - \frac{\big({e}_0\cdot{e}_2\big)\big({\hat{n}}_1\cdot{\hat{n}}_2\big)}{\|{n}_1\|\|{n}_2\|}{\hat{n}}_2{\hat{n}}_1^T \\
&+ \frac{\big({e}_1\cdot{e}_2\big)\big({\hat{n}}_1\cdot{\hat{n}}_2\big)}{\|{n}_1\|\|{n}_2\|}{\hat{n}}_2{\hat{n}}_1^T - \frac{\big({\hat{n}}_1\cdot{e}_2\big)\big({\hat{n}}_2\cdot{e}_1\big)}{\|{n}_1\|\|{n}_2\|}{\hat{n}}_2{\hat{n}}_1^T
\end{align*}
\begin{align*}
&= \big[{\hat{n}}_1,{\hat{n}}_2,{e}_0\big]\Bigg(\frac{\big({e}_0\cdot{e}_1\big)^2-\big({e}_0\cdot{e}_1\big)\|{e}_1\|^2}{\|{e}_0\|^2\|{n}_1\|^3}{e}_0{\hat{n}}_1^T + \frac{\big({e}_0\cdot{e}_1\big)^2-\big({e}_0\cdot{e}_1\big)\|{e}_0\|^2}{\|{e}_0\|^2\|{n}_1\|^3}{e}_1{\hat{n}}_1^T \\
&+ \frac{\big({e}_0\cdot{e}_1\big)^2-\big({e}_0\cdot{e}_1\big)\|{e}_1\|^2+\|{n}_1\|^2}{\|{e}_0\|^2\|{n}_1\|^3}{\hat{n}}_1{e}_0^T + \frac{\big({e}_0\cdot{e}_1\big)^2-\big({e}_0\cdot{e}_1\big)\|{e}_0\|^2}{\|{e}_0\|^2\|{n}_1\|^3}{\hat{n}}_1{e}_1^T \\
&+ \frac{\big({e}_0\cdot{e}_2\big)^2-\big({e}_0\cdot{e}_2\big)\|{e}_2\|^2}{\|{e}_0\|^2\|{n}_2\|^3}{e}_0{\hat{n}}_2^T + \frac{\big({e}_0\cdot{e}_2\big)^2-\big({e}_0\cdot{e}_2\big)\|{e}_0\|^2}{\|{e}_0\|^2\|{n}_2\|^3}{e}_2{\hat{n}}_2^T \\
&+ \frac{\big({e}_0\cdot{e}_2\big)^2-\big({e}_0\cdot{e}_2\big)\|{e}_2\|^2+\|{n}_2\|^2}{\|{e}_0\|^2\|{n}_2\|^3}{\hat{n}}_2{e}_0^T + \frac{\big({e}_0\cdot{e}_2\big)^2-\big({e}_0\cdot{e}_2\big)\|{e}_0\|^2}{\|{e}_0\|^2\|{n}_2\|^3}{\hat{n}}_2{e}_2^T \Bigg) \\
&- \frac{\big({e}_0\cdot{e}_1\big)\big({\hat{n}}_1\cdot{\hat{n}}_2\big)}{\|{n}_1\|\|{n}_2\|}{\hat{n}}_1{\hat{n}}_2^T + \frac{\big({e}_1\cdot{e}_2\big)\big({\hat{n}}_1\cdot{\hat{n}}_2\big)}{\|{n}_1\|\|{n}_2\|}{\hat{n}}_1{\hat{n}}_2^T - \frac{\big({\hat{n}}_2\cdot{e}_1\big)\big({\hat{n}}_1\cdot{e}_2\big)}{\|{n}_1\|\|{n}_2\|}{\hat{n}}_1{\hat{n}}_2^T \\
&- \frac{\big({e}_0\cdot{e}_1\big)\big({\hat{n}}_1\cdot{\hat{n}}_2\big)}{\|{n}_1\|^2}{\hat{n}}_1{\hat{n}}_1^T + \frac{\|{e}_1\|^2\big({\hat{n}}_1\cdot{\hat{n}}_2\big)}{\|{n}_1\|^2}{\hat{n}}_1{\hat{n}}_1^T \\
&- \frac{\big({e}_0\cdot{e}_2\big)\big({\hat{n}}_1\cdot{\hat{n}}_2\big)}{\|{n}_2\|^2}{\hat{n}}_2{\hat{n}}_2^T + \frac{\|{e}_2\|^2\big({\hat{n}}_1\cdot{\hat{n}}_2\big)}{\|{n}_2\|^2}{\hat{n}}_2{\hat{n}}_2^T - \frac{\big({e}_0\cdot{e}_2\big)\big({\hat{n}}_1\cdot{\hat{n}}_2\big)}{\|{n}_1\|\|{n}_2\|}{\hat{n}}_2{\hat{n}}_1^T \\
&+ \frac{\big({e}_1\cdot{e}_2\big)\big({\hat{n}}_1\cdot{\hat{n}}_2\big)}{\|{n}_1\|\|{n}_2\|}{\hat{n}}_2{\hat{n}}_1^T - \frac{\big({\hat{n}}_1\cdot{e}_2\big)\big({\hat{n}}_2\cdot{e}_1\big)}{\|{n}_1\|\|{n}_2\|}{\hat{n}}_2{\hat{n}}_1^T,
\end{align*}
and we obtain
\begin{align*}
\nabla_{x_0}\big(-C_{11}-C_{22}\big) &= \big[{\hat{n}}_1,{\hat{n}}_2,{e}_0\big]\Bigg(\frac{\big({e}_0\cdot{e}_1\big)^2-\big({e}_0\cdot{e}_1\big)\|{e}_1\|^2}{\|{e}_0\|^2\|{n}_1\|^3}{e}_0{\hat{n}}_1^T + \frac{\big({e}_0\cdot{e}_1\big)^2-\big({e}_0\cdot{e}_1\big)\|{e}_0\|^2}{\|{e}_0\|^2\|{n}_1\|^3}{e}_1{\hat{n}}_1^T \\
&+ \frac{\|{e}_0\|^2\|{e}_1\|^2-\big({e}_0\cdot{e}_1\big)\|{e}_1\|^2}{\|{e}_0\|^2\|{n}_1\|^3}{\hat{n}}_1{e}_0^T + \frac{\big({e}_0\cdot{e}_1\big)^2-\big({e}_0\cdot{e}_1\big)\|{e}_0\|^2}{\|{e}_0\|^2\|{n}_1\|^3}{\hat{n}}_1{e}_1^T \\
&+ \frac{\big({e}_0\cdot{e}_2\big)^2-\big({e}_0\cdot{e}_2\big)\|{e}_2\|^2}{\|{e}_0\|^2\|{n}_2\|^3}{e}_0{\hat{n}}_2^T + \frac{\big({e}_0\cdot{e}_2\big)^2-\big({e}_0\cdot{e}_2\big)\|{e}_0\|^2}{\|{e}_0\|^2\|{n}_2\|^3}{e}_2{\hat{n}}_2^T \\
&+ \frac{\|{e}_0\|^2\|{e}_2\|^2-\big({e}_0\cdot{e}_2\big)\|{e}_2\|^2}{\|{e}_0\|^2\|{n}_2\|^3}{\hat{n}}_2{e}_0^T + \frac{\big({e}_0\cdot{e}_2\big)^2-\big({e}_0\cdot{e}_2\big)\|{e}_0\|^2}{\|{e}_0\|^2\|{n}_2\|^3}{\hat{n}}_2{e}_2^T \Bigg) \\
&- \frac{\big({e}_0\cdot{e}_1\big)\big({\hat{n}}_1\cdot{\hat{n}}_2\big)}{\|{n}_1\|\|{n}_2\|}{\hat{n}}_1{\hat{n}}_2^T + \frac{\big({e}_1\cdot{e}_2\big)\big({\hat{n}}_1\cdot{\hat{n}}_2\big)}{\|{n}_1\|\|{n}_2\|}{\hat{n}}_1{\hat{n}}_2^T - \frac{\big({\hat{n}}_2\cdot{e}_1\big)\big({\hat{n}}_1\cdot{e}_2\big)}{\|{n}_1\|\|{n}_2\|}{\hat{n}}_1{\hat{n}}_2^T \\
&- \frac{\big({e}_0\cdot{e}_1\big)\big({\hat{n}}_1\cdot{\hat{n}}_2\big)}{\|{n}_1\|^2}{\hat{n}}_1{\hat{n}}_1^T + \frac{\|{e}_1\|^2\big({\hat{n}}_1\cdot{\hat{n}}_2\big)}{\|{n}_1\|^2}{\hat{n}}_1{\hat{n}}_1^T \\
&- \frac{\big({e}_0\cdot{e}_2\big)\big({\hat{n}}_1\cdot{\hat{n}}_2\big)}{\|{n}_2\|^2}{\hat{n}}_2{\hat{n}}_2^T + \frac{\|{e}_2\|^2\big({\hat{n}}_1\cdot{\hat{n}}_2\big)}{\|{n}_2\|^2}{\hat{n}}_2{\hat{n}}_2^T - \frac{\big({e}_0\cdot{e}_2\big)\big({\hat{n}}_1\cdot{\hat{n}}_2\big)}{\|{n}_1\|\|{n}_2\|}{\hat{n}}_2{\hat{n}}_1^T \\
&+ \frac{\big({e}_1\cdot{e}_2\big)\big({\hat{n}}_1\cdot{\hat{n}}_2\big)}{\|{n}_1\|\|{n}_2\|}{\hat{n}}_2{\hat{n}}_1^T - \frac{\big({\hat{n}}_1\cdot{e}_2\big)\big({\hat{n}}_2\cdot{e}_1\big)}{\|{n}_1\|\|{n}_2\|}{\hat{n}}_2{\hat{n}}_1^T.
\end{align*}
Each term in this expression has its corresponding transpose that may be matched in the expression for $\nabla_{{x}_1}\big({C}_{11}-{C}_{01}+{C}_{22}-{C}_{02}\big)$; thus we have verified the symmetry condition.

\subsubsection{\texorpdfstring{$\nabla_{{x}_2}\big({C}_{11}-{C}_{01}+{C}_{22}-{C}_{02}\big)$}{A3}}
\begin{align*}
&\nabla_{{x}_2}\big({C}_{11}-{C}_{01}+{C}_{22}-{C}_{02}\big) \\
&= \frac{\big[{\hat{n}}_1,{\hat{n}}_2,{e}_1\big]}{\|{n}_1\|}\Big(\nabla_{{x}_2}{\hat{n}}_1 + \big[\nabla_{{x}_2}{\hat{n}}_1\big]^T\Big)+\frac{{\hat{n}}_1}{\|{n}_1\|}{e}_1^T\big({\hat{n}}_1\times\nabla_{{x}_2}{\hat{n}}_2-{\hat{n}}_2\times\nabla_{{x}_2}{\hat{n}}_1\big)+\frac{{\hat{n}}_1}{\|{n}_1\|}\big({\hat{n}}_1\times{\hat{n}}_2\big)^T\nabla_{{x}_2}{e}_1 \\
&- \frac{\big[{\hat{n}}_1,{\hat{n}}_2,{e}_0\big]}{\|{n}_1\|}\Big(\nabla_{{x}_2}{\hat{n}}_1 + \big[\nabla_{{x}_2}{\hat{n}}_1\big]^T\Big)-\frac{{\hat{n}}_1}{\|{n}_1\|}{e}_0^T\big({\hat{n}}_1\times\nabla_{{x}_2}{\hat{n}}_2-{\hat{n}}_2\times\nabla_{{x}_2}{\hat{n}}_1\big)-\frac{{\hat{n}}_1}{\|{n}_1\|}\big({\hat{n}}_1\times{\hat{n}}_2\big)^T\nabla_{{x}_2}{e}_0 \\
&+ \frac{\big[{\hat{n}}_1,{\hat{n}}_2,{e}_2\big]}{\|{n}_2\|}\Big(\nabla_{{x}_2}{\hat{n}}_2 + \big[\nabla_{{x}_2}{\hat{n}}_2\big]^T\Big)+\frac{{\hat{n}}_2}{\|{n}_2\|}{e}_2^T\big({\hat{n}}_1\times\nabla_{{x}_2}{\hat{n}}_2-{\hat{n}}_2\times\nabla_{{x}_2}{\hat{n}}_1\big)+\frac{{\hat{n}}_2}{\|{n}_2\|}\big({\hat{n}}_1\times{\hat{n}}_2\big)^T\nabla_{{x}_2}{e}_2 \\
&- \frac{\big[{\hat{n}}_1,{\hat{n}}_2,{e}_0\big]}{\|{n}_2\|}\Big(\nabla_{{x}_2}{\hat{n}}_2 + \big[\nabla_{{x}_2}{\hat{n}}_2\big]^T\Big)-\frac{{\hat{n}}_2}{\|{n}_2\|}{e}_0^T\big({\hat{n}}_1\times\nabla_{{x}_2}{\hat{n}}_2-{\hat{n}}_2\times\nabla_{{x}_2}{\hat{n}}_1\big)-\frac{{\hat{n}}_2}{\|{n}_2\|}\big({\hat{n}}_1\times{\hat{n}}_2\big)^T\nabla_{{x}_2}{e}_0
\end{align*}
\begin{align*}
&= \Bigg(\frac{\big[{\hat{n}}_1,{\hat{n}}_2,{e}_1\big]-\big[{\hat{n}}_1,{\hat{n}}_2,{e}_0\big]}{\|{n}_1\|}\Bigg)\Big(\nabla_{{x}_2}{\hat{n}}_1 + \big[\nabla_{{x}_2}{\hat{n}}_1\big]^T\Big) \\
&+ \frac{{\hat{n}}_1}{\|{n}_1\|}{e}_1^T\Bigg(-\frac{{\hat{n}}_2\times\big({e}_0\times{\hat{n}}_1\big){\hat{n}}_1^T}{\|{n}_1\|}\Bigg) + \frac{{\hat{n}}_1}{\|{n}_1\|}\big({\hat{n}}_1\times{\hat{n}}_2\big)^T\\
&- \frac{{\hat{n}}_1}{\|{n}_1\|}{e}_0^T\Bigg(-\frac{{\hat{n}}_2\times\big({e}_0\times{\hat{n}}_1\big){\hat{n}}_1^T}{\|{n}_1\|}\Bigg) \\
&+ \frac{{\hat{n}}_2}{\|{n}_2\|}{e}_2^T\Bigg(-\frac{{\hat{n}}_2\times\big({e}_0\times{\hat{n}}_1\big){\hat{n}}_1^T}{\|{n}_1\|}\Bigg) \\
&- \frac{{\hat{n}}_2}{\|{n}_2\|}{e}_0^T\Bigg(-\frac{{\hat{n}}_2\times\big({e}_0\times{\hat{n}}_1\big){\hat{n}}_1^T}{\|{n}_1\|}\Bigg)
\end{align*}
\begin{align*}
&= \Bigg(\frac{\big[{\hat{n}}_1,{\hat{n}}_2,{e}_1\big]-\big[{\hat{n}}_1,{\hat{n}}_2,{e}_0\big]}{\|{n}_1\|}\Bigg)\Big(\nabla_{{x}_2}{\hat{n}}_1 + \big[\nabla_{{x}_2}{\hat{n}}_1\big]^T\Big) + \frac{{\hat{n}}_1}{\|{n}_1\|}\big({\hat{n}}_1\times{\hat{n}}_2\big)^T \\
&- \frac{\big({e}_0\cdot{e}_1\big)\big({\hat{n}}_1\cdot{\hat{n}}_2\big)}{\|{n}_1\|^2}{\hat{n}}_1{\hat{n}}_1^T + \frac{\|{e}_0\|^2\big({\hat{n}}_1\cdot{\hat{n}}_2\big)}{\|{n}_1\|^2}{\hat{n}}_1{\hat{n}}_1^T \\
&- \frac{\big({e}_0\cdot{e}_2\big)\big({\hat{n}}_1\cdot{\hat{n}}_2\big)}{\|{n}_1\|\|{n}_2\|}{\hat{n}}_2{\hat{n}}_1^T + \frac{\|{e}_0\|^2\big({\hat{n}}_1\cdot{\hat{n}}_2\big)}{\|{n}_1\|\|{n}_2\|}{\hat{n}}_2{\hat{n}}_1^T
\end{align*}
\begin{align*}
&= \big[{\hat{n}}_1,{\hat{n}}_2,{e}_0\big]\Bigg(\frac{\big({e}_0\cdot{e}_1\big)-\|{e}_0\|^2}{\|{e}_0\|^2\|{n}_1\|}\nabla_{{x}_2}{\hat{n}}_1+\frac{\big({e}_0\cdot{e}_1\big)-\|{e}_0\|^2}{\|{e}_0\|^2\|{n}_1\|}\big[\nabla_{{x}_2}{\hat{n}}_1\big]^T+\frac{1}{\|{e}_0\|^2\|{n}_1\|}{\hat{n}}_1{e}_0^T\Bigg) \\
&- \frac{\big({e}_0\cdot{e}_1\big)\big({\hat{n}}_1\cdot{\hat{n}}_2\big)}{\|{n}_1\|^2}{\hat{n}}_1{\hat{n}}_1^T + \frac{\|{e}_0\|^2\big({\hat{n}}_1\cdot{\hat{n}}_2\big)}{\|{n}_1\|^2}{\hat{n}}_1{\hat{n}}_1^T \\
&- \frac{\big({e}_0\cdot{e}_2\big)\big({\hat{n}}_1\cdot{\hat{n}}_2\big)}{\|{n}_1\|\|{n}_2\|}{\hat{n}}_2{\hat{n}}_1^T + \frac{\|{e}_0\|^2\big({\hat{n}}_1\cdot{\hat{n}}_2\big)}{\|{n}_1\|\|{n}_2\|}{\hat{n}}_2{\hat{n}}_1^T
\end{align*}
\begin{align*}
&= \big[{\hat{n}}_1,{\hat{n}}_2,{e}_0\big]\Bigg(\frac{\big({e}_0\cdot{e}_1\big)-\|{e}_0\|^2}{\|{e}_0\|^2\|{n}_1\|^2}\big({e}_0\times{\hat{n}}_1\big){\hat{n}}_1^T+\frac{\big({e}_0\cdot{e}_1\big)-\|{e}_0\|^2}{\|{e}_0\|^2\|{n}_1\|}{\hat{n}}_1\big({e}_0\times{\hat{n}}_1\big)^T+\frac{1}{\|{e}_0\|^2\|{n}_1\|}{\hat{n}}_1{e}_0^T\Bigg) \\
&- \frac{\big({e}_0\cdot{e}_1\big)\big({\hat{n}}_1\cdot{\hat{n}}_2\big)}{\|{n}_1\|^2}{\hat{n}}_1{\hat{n}}_1^T + \frac{\|{e}_0\|^2\big({\hat{n}}_1\cdot{\hat{n}}_2\big)}{\|{n}_1\|^2}{\hat{n}}_1{\hat{n}}_1^T \\
&- \frac{\big({e}_0\cdot{e}_2\big)\big({\hat{n}}_1\cdot{\hat{n}}_2\big)}{\|{n}_1\|\|{n}_2\|}{\hat{n}}_2{\hat{n}}_1^T + \frac{\|{e}_0\|^2\big({\hat{n}}_1\cdot{\hat{n}}_2\big)}{\|{n}_1\|\|{n}_2\|}{\hat{n}}_2{\hat{n}}_1^T
\end{align*}
\begin{align*}
&= \big[{\hat{n}}_1,{\hat{n}}_2,{e}_0\big]\Bigg(\frac{\big({e}_0\cdot{e}_1\big)^2-\big({e}_0\cdot{e}_1\big)\|{e}_0\|^2}{\|{e}_0\|^2\|{n}_1\|^3}{e}_0{\hat{n}}_1^T + \frac{\|{e}_0\|^2-\big({e}_0\cdot{e}_1\big)}{\|{n}_1\|^3}{e}_1{\hat{n}}_1^T \\
&+\frac{\big({e}_0\cdot{e}_1\big)^2-\big({e}_0\cdot{e}_1\big)\|{e}_0\|^2+\|{n}_1\|^2}{\|{e}_0\|^2\|{n}_1\|^3}{\hat{n}}_1{e}_0^T + \frac{\|{e}_0\|^2-\big({e}_0\cdot{e}_1\big)}{\|{n}_1\|^3}{\hat{n}}_1{e}_1^T\Bigg)\\
&- \frac{\big({e}_0\cdot{e}_1\big)\big({\hat{n}}_1\cdot{\hat{n}}_2\big)}{\|{n}_1\|^2}{\hat{n}}_1{\hat{n}}_1^T + \frac{\|{e}_0\|^2\big({\hat{n}}_1\cdot{\hat{n}}_2\big)}{\|{n}_1\|^2}{\hat{n}}_1{\hat{n}}_1^T \\
&- \frac{\big({e}_0\cdot{e}_2\big)\big({\hat{n}}_1\cdot{\hat{n}}_2\big)}{\|{n}_1\|\|{n}_2\|}{\hat{n}}_2{\hat{n}}_1^T + \frac{\|{e}_0\|^2\big({\hat{n}}_1\cdot{\hat{n}}_2\big)}{\|{n}_1\|\|{n}_2\|}{\hat{n}}_2{\hat{n}}_1^T,
\end{align*}
which yields
\begin{align*}
\nabla_{{x}_2}\big({C}_{11}-{C}_{01}+{C}_{22}-{C}_{02}\big) &= \big[{\hat{n}}_1,{\hat{n}}_2,{e}_0\big]\Bigg(\frac{\big({e}_0\cdot{e}_1\big)^2-\big({e}_0\cdot{e}_1\big)\|{e}_0\|^2}{\|{e}_0\|^2\|{n}_1\|^3}{e}_0{\hat{n}}_1^T + \frac{\|{e}_0\|^2-\big({e}_0\cdot{e}_1\big)}{\|{n}_1\|^3}{e}_1{\hat{n}}_1^T \\
&+\frac{\|{e}_1\|^2-\big({e}_0\cdot{e}_1\big)}{\|{n}_1\|^3}{\hat{n}}_1{e}_0^T + \frac{\|{e}_0\|^2-\big({e}_0\cdot{e}_1\big)}{\|{n}_1\|^3}{\hat{n}}_1{e}_1^T\Bigg)\\
&- \frac{\big({e}_0\cdot{e}_1\big)\big({\hat{n}}_1\cdot{\hat{n}}_2\big)}{\|{n}_1\|^2}{\hat{n}}_1{\hat{n}}_1^T + \frac{\|{e}_0\|^2\big({\hat{n}}_1\cdot{\hat{n}}_2\big)}{\|{n}_1\|^2}{\hat{n}}_1{\hat{n}}_1^T \\
&- \frac{\big({e}_0\cdot{e}_2\big)\big({\hat{n}}_1\cdot{\hat{n}}_2\big)}{\|{n}_1\|\|{n}_2\|}{\hat{n}}_2{\hat{n}}_1^T + \frac{\|{e}_0\|^2\big({\hat{n}}_1\cdot{\hat{n}}_2\big)}{\|{n}_1\|\|{n}_2\|}{\hat{n}}_2{\hat{n}}_1^T.
\end{align*}
Next we can compare this result with its transpose, $\nabla_{{x}_0}{C}_{01}$:
\begin{align*}
\nabla_{{x}_0}{C}_{01} &= \frac{\big[{\hat{n}}_1,{\hat{n}}_2,{e}_0\big]}{\|{n}_1\|}\Big(\nabla_{{x}_0}{\hat{n}}_1 + \big[\nabla_{{x}_0}{\hat{n}}_1\big]^T\Big)+\frac{{\hat{n}}_1}{\|{n}_1\|}{e}_0^T\big({\hat{n}}_1\times\nabla_{{x}_0}{\hat{n}}_2-{\hat{n}}_2\times\nabla_{{x}_0}{\hat{n}}_1\big)+\frac{{\hat{n}}_1}{\|{n}_1\|}\big({\hat{n}}_1\times{\hat{n}}_2\big)^T\nabla_{{x}_0}{e}_0, \\
&= \frac{\big[{\hat{n}}_1,{\hat{n}}_2,{e}_0\big]}{\|{n}_1\|}\Big(\nabla_{{x}_0}{\hat{n}}_1 + \big[\nabla_{{x}_0}{\hat{n}}_1\big]^T\Big) \\
&+\frac{{\hat{n}}_1}{\|{n}_1\|}{e}_0^T\Bigg(\frac{{\hat{n}}_1\times\big({e}_0\times{\hat{n}}_2\big){\hat{n}}_2^T}{\|{n}_2\|}-\frac{{\hat{n}}_1\times\big({e}_2\times{\hat{n}}_2\big){\hat{n}}_2^T}{\|{n}_2\|}+\frac{{\hat{n}}_2\times\big({e}_0\times{\hat{n}}_1\big){\hat{n}}_1^T}{\|{n}_1\|}-\frac{{\hat{n}}_2\times\big({e}_1\times{\hat{n}}_1\big){\hat{n}}_1^T}{\|{n}_1\|}\Bigg) \\
&- \frac{{\hat{n}}_1}{\|{n}_1\|}\big({\hat{n}}_1\times{\hat{n}}_2\big)^T
\end{align*}
\begin{align*}
&= \frac{\big[{\hat{n}}_1,{\hat{n}}_2,{e}_0\big]}{\|{n}_1\|}\Big(\nabla_{{x}_0}{\hat{n}}_1 + \big[\nabla_{{x}_0}{\hat{n}}_1\big]^T\Big) - \frac{{\hat{n}}_1}{\|{n}_1\|}\big({\hat{n}}_1\times{\hat{n}}_2\big)^T\\
&+ \frac{\|{e}_0\|^2\big({\hat{n}}_1\cdot{\hat{n}}_2\big)}{\|{n}_1\|\|{n}_2\|}{\hat{n}}_1{\hat{n}}_2^T - \frac{\big({e}_0\cdot{e}_2\big)\big({\hat{n}}_1\cdot{\hat{n}}_2\big)}{\|{n}_1\|\|{n}_2\|}{\hat{n}}_1{\hat{n}}_2^T \\
&+ \frac{\|{e}_0\|^2\big({\hat{n}}_1\cdot{\hat{n}}_2\big)}{\|{n}_1\|^2}{\hat{n}}_1{\hat{n}}_1^T - \frac{\big({e}_0\cdot{e}_1\big)\big({\hat{n}}_1\cdot{\hat{n}}_2\big)}{\|{n}_1\|^2}{\hat{n}}_1{\hat{n}}_1^T
\end{align*}
\begin{align*}
&= \big[{\hat{n}}_1,{\hat{n}}_2,{e}_0\big]\Bigg(-\frac{\big({e}_0\times{\hat{n}}_1\big){\hat{n}}_1^T}{\|{n}_1\|^2}+\frac{\big({e}_1\times{\hat{n}}_1\big){\hat{n}}_1^T}{\|{n}_1\|^2}-\frac{{\hat{n}}_1\big({e}_0\times{\hat{n}}_1\big)^T}{\|{n}_1\|^2}+\frac{{\hat{n}}_1\big({e}_1\times{\hat{n}}_1\big)^T}{\|{n}_1\|^2}-\frac{1}{\|{e}_0\|^2\|{n}_1\|}{\hat{n}}_1{e}_0^T\Bigg) \\
&+ \frac{\|{e}_0\|^2\big({\hat{n}}_1\cdot{\hat{n}}_2\big)}{\|{n}_1\|\|{n}_2\|}{\hat{n}}_1{\hat{n}}_2^T - \frac{\big({e}_0\cdot{e}_2\big)\big({\hat{n}}_1\cdot{\hat{n}}_2\big)}{\|{n}_1\|\|{n}_2\|}{\hat{n}}_1{\hat{n}}_2^T \\
&+ \frac{\|{e}_0\|^2\big({\hat{n}}_1\cdot{\hat{n}}_2\big)}{\|{n}_1\|^2}{\hat{n}}_1{\hat{n}}_1^T - \frac{\big({e}_0\cdot{e}_1\big)\big({\hat{n}}_1\cdot{\hat{n}}_2\big)}{\|{n}_1\|^2}{\hat{n}}_1{\hat{n}}_1^T
\end{align*}
\begin{align*}
&= \big[{\hat{n}}_1,{\hat{n}}_2,{e}_0\big]\Bigg(\frac{\|{e}_1\|^2-\big({e}_0\cdot{e}_1\big)}{\|{n}_1\|^3}{e}_0{\hat{n}}_1^T + \frac{\|{e}_0\|^2-\big({e}_0\cdot{e}_1\big)}{\|{n}_1\|^3}{e}_1{\hat{n}}_1^T\\
&+ \frac{\|{e}_0\|^2\|{e}_1\|^2-\big({e}_0\cdot{e}_1\big)\|{e}_0\|^2-\|{n}_1\|^2}{\|{e}_0\|^2\|{n}_1\|^3}{\hat{n}}_1{e}_0^T + \frac{\|{e}_0\|^2-\big({e}_0\cdot{e}_1\big)}{\|{n}_1\|^3}{\hat{n}}_1{e}_1^T\Bigg)\\
&+ \frac{\|{e}_0\|^2\big({\hat{n}}_1\cdot{\hat{n}}_2\big)}{\|{n}_1\|\|{n}_2\|}{\hat{n}}_1{\hat{n}}_2^T - \frac{\big({e}_0\cdot{e}_2\big)\big({\hat{n}}_1\cdot{\hat{n}}_2\big)}{\|{n}_1\|\|{n}_2\|}{\hat{n}}_1{\hat{n}}_2^T \\
&+ \frac{\|{e}_0\|^2\big({\hat{n}}_1\cdot{\hat{n}}_2\big)}{\|{n}_1\|^2}{\hat{n}}_1{\hat{n}}_1^T - \frac{\big({e}_0\cdot{e}_1\big)\big({\hat{n}}_1\cdot{\hat{n}}_2\big)}{\|{n}_1\|^2}{\hat{n}}_1{\hat{n}}_1^T,
\end{align*}
and we obtain
\begin{align*}
\nabla_{{x}_0}{C}_{01} &= \big[{\hat{n}}_1,{\hat{n}}_2,{e}_0\big]\Bigg(\frac{\|{e}_1\|^2-\big({e}_0\cdot{e}_1\big)}{\|{n}_1\|^3}{e}_0{\hat{n}}_1^T + \frac{\|{e}_0\|^2-\big({e}_0\cdot{e}_1\big)}{\|{n}_1\|^3}{e}_1{\hat{n}}_1^T\\
&+ \frac{\big({e}_0\cdot{e}_1\big)^2-\big({e}_0\cdot{e}_1\big)\|{e}_0\|^2}{\|{e}_0\|^2\|{n}_1\|^3}{\hat{n}}_1{e}_0^T + \frac{\|{e}_0\|^2-\big({e}_0\cdot{e}_1\big)}{\|{n}_1\|^3}{\hat{n}}_1{e}_1^T\Bigg)\\
&+ \frac{\|{e}_0\|^2\big({\hat{n}}_1\cdot{\hat{n}}_2\big)}{\|{n}_1\|\|{n}_2\|}{\hat{n}}_1{\hat{n}}_2^T - \frac{\big({e}_0\cdot{e}_2\big)\big({\hat{n}}_1\cdot{\hat{n}}_2\big)}{\|{n}_1\|\|{n}_2\|}{\hat{n}}_1{\hat{n}}_2^T \\
&+ \frac{\|{e}_0\|^2\big({\hat{n}}_1\cdot{\hat{n}}_2\big)}{\|{n}_1\|^2}{\hat{n}}_1{\hat{n}}_1^T - \frac{\big({e}_0\cdot{e}_1\big)\big({\hat{n}}_1\cdot{\hat{n}}_2\big)}{\|{n}_1\|^2}{\hat{n}}_1{\hat{n}}_1^T.
\end{align*}
Once again we arrive at a set of terms that are transposed with respect to $\nabla_{x_2}\big(C_{11}-C_{01}+C_{22}-C_{02}\big)$. As all derivations rely on applying similar principles of vector calculus, we present only the solutions from here on.

\subsubsection{\texorpdfstring{$\nabla_{{x}_3}\big({C}_{11}-{C}_{01}+{C}_{22}-{C}_{02}\big)$}{A4}}
\begin{align*}
\nabla_{{x}_3}\big({C}_{11}-{C}_{01}+{C}_{22}-{C}_{02}\big) &= \big[{\hat{n}}_1,{\hat{n}}_2,{e}_0\big]\Bigg(\frac{\big({e}_0\cdot{e}_2\big)^2-\big({e}_0\cdot{e}_2\big)\|{e}_0\|^2}{\|{e}_0\|^2\|{n}_2\|^3}{e}_0{\hat{n}}_2^T + \frac{\|{e}_0\|^2-\big({e}_0\cdot{e}_2\big)}{\|{n}_2\|^3}{e}_2{\hat{n}}_2^T \\
&+\frac{\|{e}_2\|^2-\big({e}_0\cdot{e}_2\big)}{\|{n}_2\|^3}{\hat{n}}_2{e}_0^T + \frac{\|{e}_0\|^2-\big({e}_0\cdot{e}_2\big)}{\|{n}_2\|^3}{\hat{n}}_2{e}_2^T\Bigg)\\
&- \frac{\big({e}_0\cdot{e}_1\big)\big({\hat{n}}_1\cdot{\hat{n}}_2\big)}{\|{n}_1\|\|{n}_2\|}{\hat{n}}_1{\hat{n}}_2^T + \frac{\|{e}_0\|^2\big({\hat{n}}_1\cdot{\hat{n}}_2\big)}{\|{n}_1\|\|{n}_2\|}{\hat{n}}_1{\hat{n}}_2^T \\
&- \frac{\big({e}_0\cdot{e}_2\big)\big({\hat{n}}_1\cdot{\hat{n}}_2\big)}{\|{n}_2\|^2}{\hat{n}}_2{\hat{n}}_2^T + \frac{\|{e}_0\|^2\big({\hat{n}}_1\cdot{\hat{n}}_2\big)}{\|{n}_2\|^2}{\hat{n}}_2{\hat{n}}_2^T.
\end{align*}
Comparing this result with its transpose, $\nabla_{{x}_0}{C}_{02}$:
\begin{align*}
\nabla_{{x}_0}{C}_{02} &= \big[{\hat{n}}_1,{\hat{n}}_2,{e}_0\big]\Bigg(\frac{\|{e}_2\|^2-\big({e}_0\cdot{e}_2\big)}{\|{n}_2\|^3}{e}_0{\hat{n}}_2^T + \frac{\|{e}_0\|^2-\big({e}_0\cdot{e}_2\big)}{\|{n}_2\|^3}{e}_2{\hat{n}}_2^T\\
&+ \frac{\big({e}_0\cdot{e}_2\big)^2-\big({e}_0\cdot{e}_2\big)\|{e}_0\|^2}{\|{e}_0\|^2\|{n}_2\|^3}{\hat{n}}_2{e}_0^T + \frac{\|{e}_0\|^2-\big({e}_0\cdot{e}_2\big)}{\|{n}_2\|^3}{\hat{n}}_2{e}_2^T\Bigg)\\
&+ \frac{\|{e}_0\|^2\big({\hat{n}}_1\cdot{\hat{n}}_2\big)}{\|{n}_2\|^2}{\hat{n}}_2{\hat{n}}_2^T - \frac{\big({e}_0\cdot{e}_2\big)\big({\hat{n}}_1\cdot{\hat{n}}_2\big)}{\|{n}_2\|^2}{\hat{n}}_2{\hat{n}}_2^T \\
&+ \frac{\|{e}_0\|^2\big({\hat{n}}_1\cdot{\hat{n}}_2\big)}{\|{n}_1\|\|{n}_2\|}{\hat{n}}_2{\hat{n}}_1^T - \frac{\big({e}_0\cdot{e}_1\big)\big({\hat{n}}_1\cdot{\hat{n}}_2\big)}{\|{n}_1\|\|{n}_2\|}{\hat{n}}_2{\hat{n}}_1^T.
\end{align*}

\subsubsection{\texorpdfstring{$\nabla_{{x}_0}\big(-{C}_{11}-{C}_{22}\big)$}{A5}}
This is the transpose of $\nabla_{{x}_1}\big({C}_{11}-{C}_{01}+{C}_{22}-{C}_{02}\big) $, which we have shown:
\begin{align*}
\nabla_{{x}_0}\big(-{C}_{11}-{C}_{22}\big) &= \big[{\hat{n}}_1,{\hat{n}}_2,{e}_0\big]\Bigg(\frac{\big({e}_0\cdot{e}_1\big)^2-\big({e}_0\cdot{e}_1\big)\|{e}_1\|^2}{\|{e}_0\|^2\|{n}_1\|^3}{e}_0{\hat{n}}_1^T + \frac{\big({e}_0\cdot{e}_1\big)^2-\big({e}_0\cdot{e}_1\big)\|{e}_0\|^2}{\|{e}_0\|^2\|{n}_1\|^3}{e}_1{\hat{n}}_1^T \\
&+ \frac{\|{e}_0\|^2\|{e}_1\|^2-\big({e}_0\cdot{e}_1\big)\|{e}_1\|^2}{\|{e}_0\|^2\|{n}_1\|^3}{\hat{n}}_1{e}_0^T + \frac{\big({e}_0\cdot{e}_1\big)^2-\big({e}_0\cdot{e}_1\big)\|{e}_0\|^2}{\|{e}_0\|^2\|{n}_1\|^3}{\hat{n}}_1{e}_1^T \\
&+ \frac{\big({e}_0\cdot{e}_2\big)^2-\big({e}_0\cdot{e}_2\big)\|{e}_2\|^2}{\|{e}_0\|^2\|{n}_2\|^3}{e}_0{\hat{n}}_2^T + \frac{\big({e}_0\cdot{e}_2\big)^2-\big({e}_0\cdot{e}_2\big)\|{e}_0\|^2}{\|{e}_0\|^2\|{n}_2\|^3}{e}_2{\hat{n}}_2^T \\
&+ \frac{\|{e}_0\|^2\|{e}_2\|^2-\big({e}_0\cdot{e}_2\big)\|{e}_2\|^2}{\|{e}_0\|^2\|{n}_2\|^3}{\hat{n}}_2{e}_0^T + \frac{\big({e}_0\cdot{e}_2\big)^2-\big({e}_0\cdot{e}_2\big)\|{e}_0\|^2}{\|{e}_0\|^2\|{n}_2\|^3}{\hat{n}}_2{e}_2^T \Bigg) \\
&- \frac{\big({e}_0\cdot{e}_1\big)\big({\hat{n}}_1\cdot{\hat{n}}_2\big)}{\|{n}_1\|\|{n}_2\|}{\hat{n}}_1{\hat{n}}_2^T + \frac{\big({e}_1\cdot{e}_2\big)\big({\hat{n}}_1\cdot{\hat{n}}_2\big)}{\|{n}_1\|\|{n}_2\|}{\hat{n}}_1{\hat{n}}_2^T - \frac{\big({\hat{n}}_2\cdot{e}_1\big)\big({\hat{n}}_1\cdot{e}_2\big)}{\|{n}_1\|\|{n}_2\|}{\hat{n}}_1{\hat{n}}_2^T \\
&- \frac{\big({e}_0\cdot{e}_1\big)\big({\hat{n}}_1\cdot{\hat{n}}_2\big)}{\|{n}_1\|^2}{\hat{n}}_1{\hat{n}}_1^T + \frac{\|{e}_1\|^2\big({\hat{n}}_1\cdot{\hat{n}}_2\big)}{\|{n}_1\|^2}{\hat{n}}_1{\hat{n}}_1^T \\
&- \frac{\big({e}_0\cdot{e}_2\big)\big({\hat{n}}_1\cdot{\hat{n}}_2\big)}{\|{n}_2\|^2}{\hat{n}}_2{\hat{n}}_2^T + \frac{\|{e}_2\|^2\big({\hat{n}}_1\cdot{\hat{n}}_2\big)}{\|{n}_2\|^2}{\hat{n}}_2{\hat{n}}_2^T - \frac{\big({e}_0\cdot{e}_2\big)\big({\hat{n}}_1\cdot{\hat{n}}_2\big)}{\|{n}_1\|\|{n}_2\|}{\hat{n}}_2{\hat{n}}_1^T \\
&+ \frac{\big({e}_1\cdot{e}_2\big)\big({\hat{n}}_1\cdot{\hat{n}}_2\big)}{\|{n}_1\|\|{n}_2\|}{\hat{n}}_2{\hat{n}}_1^T - \frac{\big({\hat{n}}_1\cdot{e}_2\big)\big({\hat{n}}_2\cdot{e}_1\big)}{\|{n}_1\|\|{n}_2\|}{\hat{n}}_2{\hat{n}}_1^T.
\end{align*}

\subsubsection{\texorpdfstring{$\nabla_{{x}_1}\big(-{C}_{11}-{C}_{22}\big)$}{A6}}
\begin{align*}
\nabla_{{x}_1}\big(-{C}_{11}-{C}_{22}\big) &= \frac{\big[{\hat{n}}_1,{\hat{n}}_2,{e}_0\big]}{\|{e}_0\|^2\|{n}_1\|^3}\big({e}_0\cdot{e}_1\big)\Bigg[\|{e}_1\|^2{e}_0{\hat{n}}_1^T-\big({e}_0\cdot{e}_1\big){e}_1{\hat{n}}_1^T+\|{e}_1\|^2{\hat{n}}_1{e}_0^T-\big({e}_0\cdot{e}_1\big){\hat{n}}_1{e}_1^T\Bigg] \\ 
&+\frac{\big[{\hat{n}}_1,{\hat{n}}_2,{e}_0\big]}{\|{e}_0\|^2\|{n}_2\|^3}\big({e}_0\cdot{e}_2\big)\Bigg[\|{e}_2\|^2{e}_0{\hat{n}}_2^T-\big({e}_0\cdot{e}_2\big){e}_2{\hat{n}}_2^T+\|{e}_2\|^2{\hat{n}}_2{e}_0^T-\big({e}_0\cdot{e}_2\big){\hat{n}}_2{e}_2^T\Bigg] \\
&- \frac{\big({e}_1\cdot{e}_2\big)\big({\hat{n}}_1\cdot{\hat{n}}_2\big)}{\|{n}_1\|\|{n}_2\|}\Big({\hat{n}}_1{\hat{n}}_2^T+{\hat{n}}_2{\hat{n}}_1^T\Big) + \frac{\big({\hat{n}}_2\cdot{e}_1\big)\big({\hat{n}}_1\cdot{e}_2\big)}{\|{n}_1\|\|{n}_2\|}\Big({\hat{n}}_1{\hat{n}}_2^T + {\hat{n}}_2{\hat{n}}_1^T\Big) \\
&- \frac{\|{e}_1\|^2\big({\hat{n}}_1\cdot{\hat{n}}_2\big)}{\|{n}_1\|^2}{\hat{n}}_1{\hat{n}}_1^T - \frac{\|{e}_2\|^2\big({\hat{n}}_1\cdot{\hat{n}}_2\big)}{\|{n}_2\|^2}{\hat{n}}_2{\hat{n}}_2^T.
\end{align*}
This expression is likewise symmetric, as all terms are either symmetric or come in transpose pairs.

\subsubsection{\texorpdfstring{$\nabla_{{x}_2}\big(-{C}_{11}-{C}_{22}\big)$}{A7}}
\begin{align*}
\nabla_{{x}_2}\big(-{C}_{11}-{C}_{22}\big) &= -\frac{\big[{\hat{n}}_1,{\hat{n}}_2,{e}_0\big]}{\|{e}_0\|^2\|{n}_1\|^3}\Bigg[\big({e}_0\cdot{e}_1\big)^2{e}_0{\hat{n}}_1^T+\|{e}_0\|^2\|{e}_1\|^2{\hat{n}}_1{e}_0^T-\big({e}_0\cdot{e}_1\big)\|{e}_0\|^2{e}_1{\hat{n}}_1^T-\big({e}_0\cdot{e}_1\big)\|{e}_0\|^2{\hat{n}}_1{e}_1^T\Bigg] \\
&+ \frac{\big({e}_0\cdot{e}_1\big)\big({\hat{n}}_1\cdot{\hat{n}}_2\big)}{\|{n}_1\|^2}{\hat{n}}_1{\hat{n}}_1^T + \frac{\big({e}_0\cdot{e}_2\big)\big({\hat{n}}_1\cdot{\hat{n}}_2\big)}{\|{n}_1\|\|{n}_2\|}{\hat{n}}_2{\hat{n}}_1^T.
\end{align*}
Comparing this result with its transpose, $\nabla_{{x}_1}{C}_{01}$, we obtain
\begin{align*}
\nabla_{{x}_1}{C}_{01} &= -\frac{\big[{\hat{n}}_1,{\hat{n}}_2,{e}_0\big]}{\|{e}_0\|^2\|{n}_1\|^3}\Bigg[\|{e}_0\|^2\|{e}_1\|^2{e}_0{\hat{n}}_1^T+\big({e}_0\cdot{e}_1\big)^2{\hat{n}}_1{e}_0^T-\big({e}_0\cdot{e}_1\big)\|{e}_0\|^2{e}_1{\hat{n}}_1^T-\big({e}_0\cdot{e}_1\big)\|{e}_0\|^2{\hat{n}}_1{e}_1^T\Bigg] \\
&+ \frac{\big({e}_0\cdot{e}_1\big)\big({\hat{n}}_1\cdot{\hat{n}}_2\big)}{\|{n}_1\|^2}{\hat{n}}_1{\hat{n}}_1^T + \frac{\big({e}_0\cdot{e}_2\big)\big({\hat{n}}_1\cdot{\hat{n}}_2\big)}{\|{n}_1\|\|{n}_2\|}{\hat{n}}_1{\hat{n}}_2^T 
\end{align*}
in agreement with our result.

\subsubsection{\texorpdfstring{$\nabla_{{x}_3}\big(-{C}_{11}-{C}_{22}\big)$}{A8}}
\begin{align*}
\nabla_{{x}_3}\big(-{C}_{11}-{C}_{22}\big) &= \frac{\big[{\hat{n}}_1,{\hat{n}}_2,{e}_0\big]}{\|{e}_0\|^2\|{n}_2\|^3}\Bigg[-\big({e}_0\cdot{e}_2\big)^2{e}_0{\hat{n}}_2^T-\|{e}_0\|^2\|{e}_2\|^2{\hat{n}}_2{e}_0^T+\big({e}_0\cdot{e}_2\big)\|{e}_0\|^2{e}_2{\hat{n}}_2^T+\big({e}_0\cdot{e}_2\big)\|{e}_0\|^2{\hat{n}}_2{e}_2^T\Bigg] \\
&+ \frac{\big({e}_0\cdot{e}_2\big)\big({\hat{n}}_1\cdot{\hat{n}}_2\big)}{\|{n}_2\|^2}{\hat{n}}_2{\hat{n}}_2^T + \frac{\big({e}_0\cdot{e}_1\big)\big({\hat{n}}_1\cdot{\hat{n}}_2\big)}{\|{n}_1\|\|{n}_2\|}{\hat{n}}_1{\hat{n}}_2^T.
\end{align*}
Comparing this result with its transpose, $\nabla_{{x}_1}{C}_{02}$:
\begin{align*}
\nabla_{{x}_1}{C}_{02} &= \frac{\big[{\hat{n}}_1,{\hat{n}}_2,{e}_0\big]}{\|{e}_0\|^2\|{n}_2\|^3}\Bigg[-\|{e}_0\|^2\|{e}_2\|^2{e}_0{\hat{n}}_2^T-\big({e}_0\cdot{e}_2\big)^2{\hat{n}}_2{e}_0^T+\big({e}_0\cdot{e}_2\big)\|{e}_0\|^2{e}_2{\hat{n}}_2^T+\big({e}_0\cdot{e}_2\big)\|{e}_0\|^2{\hat{n}}_2{e}_2^T\Bigg] \\
&+ \frac{\big({e}_0\cdot{e}_2\big)\big({\hat{n}}_1\cdot{\hat{n}}_2\big)}{\|{n}_2\|^2}{\hat{n}}_2{\hat{n}}_2^T + \frac{\big({e}_0\cdot{e}_1\big)\big({\hat{n}}_1\cdot{\hat{n}}_2\big)}{\|{n}_1\|\|{n}_2\|}{\hat{n}}_2{\hat{n}}_1^T.
\end{align*}

\subsubsection{\texorpdfstring{$\nabla_{{x}_0}\big({C}_{01}\big)$}{A9}}
This is the transpose of $\nabla_{{x}_2}\big({C}_{11}-{C}_{01}+{C}_{22}-{C}_{02}\big)$, which we have shown:
\begin{align*}
\nabla_{{x}_0}\big({C}_{01}\big) &= \big[{\hat{n}}_1,{\hat{n}}_2,{e}_0\big]\Bigg(\frac{\|{e}_1\|^2-\big({e}_0\cdot{e}_1\big)}{\|{n}_1\|^3}{e}_0{\hat{n}}_1^T + \frac{\|{e}_0\|^2-\big({e}_0\cdot{e}_1\big)}{\|{n}_1\|^3}{e}_1{\hat{n}}_1^T\\
&+ \frac{\big({e}_0\cdot{e}_1\big)^2-\big({e}_0\cdot{e}_1\big)\|{e}_0\|^2}{\|{e}_0\|^2\|{n}_1\|^3}{\hat{n}}_1{e}_0^T + \frac{\|{e}_0\|^2-\big({e}_0\cdot{e}_1\big)}{\|{n}_1\|^3}{\hat{n}}_1{e}_1^T\Bigg)\\
&+ \frac{\|{e}_0\|^2\big({\hat{n}}_1\cdot{\hat{n}}_2\big)}{\|{n}_1\|\|{n}_2\|}{\hat{n}}_1{\hat{n}}_2^T - \frac{\big({e}_0\cdot{e}_2\big)\big({\hat{n}}_1\cdot{\hat{n}}_2\big)}{\|{n}_1\|\|{n}_2\|}{\hat{n}}_1{\hat{n}}_2^T \\
&+ \frac{\|{e}_0\|^2\big({\hat{n}}_1\cdot{\hat{n}}_2\big)}{\|{n}_1\|^2}{\hat{n}}_1{\hat{n}}_1^T - \frac{\big({e}_0\cdot{e}_1\big)\big({\hat{n}}_1\cdot{\hat{n}}_2\big)}{\|{n}_1\|^2}{\hat{n}}_1{\hat{n}}_1^T.
\end{align*}

\subsubsection{\texorpdfstring{$\nabla_{{x}_1}\big({C}_{01}\big)$}{A10}}
This is the transpose of $\nabla_{{x}_2}\big(-{C}_{11}-{C}_{22}\big)$, which we have shown:
\begin{align*}
\nabla_{{x}_1}\big({C}_{01}\big) &= -\frac{\big[{\hat{n}}_1,{\hat{n}}_2,{e}_0\big]}{\|{e}_0\|^2\|{n}_1\|^3}\Bigg[\|{e}_0\|^2\|{e}_1\|^2{e}_0{\hat{n}}_1^T+\big({e}_0\cdot{e}_1\big)^2{\hat{n}}_1{e}_0^T-\big({e}_0\cdot{e}_1\big)\|{e}_0\|^2{e}_1{\hat{n}}_1^T-\big({e}_0\cdot{e}_1\big)\|{e}_0\|^2{\hat{n}}_1{e}_1^T\Bigg] \\
&+ \frac{\big({e}_0\cdot{e}_1\big)\big({\hat{n}}_1\cdot{\hat{n}}_2\big)}{\|{n}_1\|^2}{\hat{n}}_1{\hat{n}}_1^T + \frac{\big({e}_0\cdot{e}_2\big)\big({\hat{n}}_1\cdot{\hat{n}}_2\big)}{\|{n}_1\|\|{n}_2\|}{\hat{n}}_1{\hat{n}}_2^T.
\end{align*}

\subsubsection{\texorpdfstring{$\nabla_{{x}_2}\big({C}_{01}\big)$}{A11}}
\begin{align*}
\nabla_{{x}_2}\big({C}_{01}\big) &= \frac{\big[{\hat{n}}_1,{\hat{n}}_2,{e}_0\big]}{\|{n}_1\|^3}\Bigg[\big({e}_0\cdot{e}_1\big){e}_0{\hat{n}}_1^T-\|{e}_0\|^2{e}_1{\hat{n}}_1^T+\big({e}_0\cdot{e}_1\big){\hat{n}}_1{e}_0^T-\|{e}_0\|^2{\hat{n}}_1{e}_1^T\Bigg] - \frac{\|{e}_0\|^2\big({\hat{n}}_1\cdot{\hat{n}}_2\big)}{\|{n}_1\|^2}{\hat{n}}_1{\hat{n}}_1^T,
\end{align*}
which is symmetric.

\subsubsection{\texorpdfstring{$\nabla_{{x}_3}\big({C}_{01}\big)$}{A12}}
\begin{align*}
\nabla_{{x}_3}\big({C}_{01}\big) &= - \frac{\|{e}_0\|^2\big({\hat{n}}_1\cdot{\hat{n}}_2\big)}{\|{n}_1\|\|{n}_2\|}{\hat{n}}_1{\hat{n}}_2^T.
\end{align*}
Verifying with its transpose, $\nabla_{{x}_2}{C}_{02}$:
\begin{align*}
\nabla_{{x}_2}{C}_{02} &= - \frac{\|{e}_0\|^2\big({\hat{n}}_1\cdot{\hat{n}}_2\big)}{\|{n}_1\|\|{n}_2\|}{\hat{n}}_2{\hat{n}}_1^T.
\end{align*}

\subsubsection{\texorpdfstring{$\nabla_{{x}_0}\big({C}_{02}\big)$}{A13}}
This is the transpose of $\nabla_{{x}_3}\big({C}_{11}-{C}_{01}+{C}_{22}-{C}_{02}\big)$, which we have shown:
\begin{align*}
\nabla_{{x}_0}\big({C}_{02}\big) &= \big[{\hat{n}}_1,{\hat{n}}_2,{e}_0\big]\Bigg(\frac{\|{e}_2\|^2-\big({e}_0\cdot{e}_2\big)}{\|{n}_2\|^3}{e}_0{\hat{n}}_2^T + \frac{\|{e}_0\|^2-\big({e}_0\cdot{e}_2\big)}{\|{n}_2\|^3}{e}_2{\hat{n}}_2^T\\
&+ \frac{\big({e}_0\cdot{e}_2\big)^2-\big({e}_0\cdot{e}_2\big)\|{e}_0\|^2}{\|{e}_0\|^2\|{n}_2\|^3}{\hat{n}}_2{e}_0^T + \frac{\|{e}_0\|^2-\big({e}_0\cdot{e}_2\big)}{\|{n}_2\|^3}{\hat{n}}_2{e}_2^T\Bigg)\\
&+ \frac{\|{e}_0\|^2\big({\hat{n}}_1\cdot{\hat{n}}_2\big)}{\|{n}_2\|^2}{\hat{n}}_2{\hat{n}}_2^T - \frac{\big({e}_0\cdot{e}_2\big)\big({\hat{n}}_1\cdot{\hat{n}}_2\big)}{\|{n}_2\|^2}{\hat{n}}_2{\hat{n}}_2^T \\
&+ \frac{\|{e}_0\|^2\big({\hat{n}}_1\cdot{\hat{n}}_2\big)}{\|{n}_1\|\|{n}_2\|}{\hat{n}}_2{\hat{n}}_1^T - \frac{\big({e}_0\cdot{e}_1\big)\big({\hat{n}}_1\cdot{\hat{n}}_2\big)}{\|{n}_1\|\|{n}_2\|}{\hat{n}}_2{\hat{n}}_1^T.
\end{align*}

\subsubsection{\texorpdfstring{$\nabla_{{x}_1}\big({C}_{02}\big)$}{A14}}
This is the transpose of $\nabla_{{x}_3}\big(-{C}_{11}-{C}_{22}\big)$, which we have shown:
\begin{align*}
\nabla_{{x}_1}\big({C}_{02}\big) &= \frac{\big[{\hat{n}}_1,{\hat{n}}_2,{e}_0\big]}{\|{e}_0\|^2\|{n}_2\|^3}\Bigg[-\|{e}_0\|^2\|{e}_2\|^2{e}_0{\hat{n}}_2^T-\big({e}_0\cdot{e}_2\big)^2{\hat{n}}_2{e}_0^T+\big({e}_0\cdot{e}_2\big)\|{e}_0\|^2{e}_2{\hat{n}}_2^T+\big({e}_0\cdot{e}_2\big)\|{e}_0\|^2{\hat{n}}_2{e}_2^T\Bigg] \\
&+ \frac{\big({e}_0\cdot{e}_2\big)\big({\hat{n}}_1\cdot{\hat{n}}_2\big)}{\|{n}_2\|^2}{\hat{n}}_2{\hat{n}}_2^T + \frac{\big({e}_0\cdot{e}_1\big)\big({\hat{n}}_1\cdot{\hat{n}}_2\big)}{\|{n}_1\|\|{n}_2\|}{\hat{n}}_2{\hat{n}}_1^T.
\end{align*}

\subsubsection{\texorpdfstring{$\nabla_{{x}_2}\big({C}_{02}\big)$}{A15}}
This is the transpose of $\nabla_{{x}_3}\big({C}_{01}\big)$, which we have shown:
\begin{align*}
&\nabla_{{x}_2}\big({C}_{02}\big) = - \frac{\|{e}_0\|^2\big({\hat{n}}_1\cdot{\hat{n}}_2\big)}{\|{n}_1\|\|{n}_2\|}{\hat{n}}_2{\hat{n}}_1^T.
\end{align*}

\subsubsection{\texorpdfstring{$\nabla_{{x}_3}\big({C}_{02}\big)$}{A16}}
Finally, we have
\begin{align*}
\nabla_{{x}_3}\big({C}_{02}\big)
&= \frac{\big[{\hat{n}}_1,{\hat{n}}_2,{e}_0\big]}{\|{n}_2\|^3}\Bigg[\big({e}_0\cdot{e}_2\big){e}_0{\hat{n}}_2^T-\|{e}_0\|^2{e}_2{\hat{n}}_2^T+\big({e}_0\cdot{e}_2\big){\hat{n}}_2{e}_0^T-\|{e}_0\|^2{\hat{n}}_2{e}_2^T\Bigg] - \frac{\|{e}_0\|^2\big({\hat{n}}_1\cdot{\hat{n}}_2\big)}{\|{n}_2\|^2}{\hat{n}}_2{\hat{n}}_2^T,
\end{align*}
which is symmetric.

\end{document}